\pdfoutput=1
\documentclass{iopjournal}

\usepackage{amsmath} %
\usepackage{amsfonts} %
\usepackage{amssymb} %
\usepackage{amsthm} %
\usepackage{bm}
\usepackage{tikz}
\usepackage{tensor}
\usepackage{graphicx} %
\usepackage{hyperref} %
\usepackage[numbers,square,sort&compress]{natbib}
\usepackage{cleveref}
\usepackage{verbatim}
\usepackage{physics}
\usepackage{ragged2e}
\justifying

\makeatletter
\renewcommand{\@makecaption}[2]{\vskip\abovecaptionskip
\sbox\@tempboxa{\fontsize{8}{10}\selectfont {\bfseries #1.} #2}%
\ifdim \wd\@tempboxa >\hsize
\justifying \fontsize{8}{10}\selectfont {\bfseries #1.} #2\par
\else
\global \@minipagefalse
\hb@xt@\hsize{\hfil\box\@tempboxa\hfil}%
\fi
\vskip\belowcaptionskip}
\makeatother

\newcommand{\fs}[1]{\left< #1\right>}
\newcommand{\favre}[1]{\left\langle\!\left\langle  #1\right\rangle\!\right\rangle}
\newcommand{\eqnref}[1]{Eq.~(\ref{#1})}

\makeatletter
\renewcommand{\@makecaption}[2]{\vskip\abovecaptionskip
\sbox\@tempboxa{\fontsize{8}{10}\selectfont {\bfseries #1.} {\itshape #2}}%
\ifdim \wd\@tempboxa >\hsize
\raggedright \fontsize{8}{10}\selectfont {\bfseries #1.} {\itshape #2}\par
\else
\global \@minipagefalse
\hb@xt@\hsize{\hfil\box\@tempboxa\hfil}%
\fi
\vskip\belowcaptionskip}
\makeatother

\begin{document}

\articletype{Paper} %

\title{Turbulent Origin of the Edge Radial Electric Field Asymmetry between Favourable and Unfavourable Drift Configurations}

\author{B. J. Frei$^{1,*}$\orcid{0000-0003-2177-8068}, R. Bilato$^1$, W. Zholobenko$^1$\orcid{0000-0002-2624-0251}, O. Grover$^1$, C. Angioni$^1$, M. Bergmann$^1$, P. Ulbl$^1$\orcid{0000-0002-6592-2298}, F. Jenko$^1$\orcid{0000-0001-6686-1469}, S. Rien\"acker$^2$, L. Vermare$^2$\orcid{0000-0002-3090-2713}, P. Hennequin$^2$\orcid{0000-0002-4848-4898} and B. Labit$^3$\orcid{0000-0002-0751-8182}}

\affil{$^1$Max-Planck-Institut f\"ur Plasmaphysik, 85748 Garching, Germany}

\affil{$^2$Laboratoire de Physique des Plasmas, CNRS-\'Ecole Polytechnique, 91128 Palaiseau cedex, France}

\affil{$^3$\'Ecole Polytechnique F\'ed\'erale de Lausanne (EPFL), Swiss Plasma Center (SPC), CH-1015 Lausanne, Switzerland}

\affil{$^*$Author to whom any correspondence should be addressed.}

\email{baptiste.frei@ipp.mpg.de}

\keywords{gyrokinetic simulation, edge and scrape-off layer turbulence, radial electric field, ion $\nabla B$ drift direction, poloidal flow, ASDEX Upgrade, TCV}

\begin{abstract}
\justifying
We study the influence of the ion magnetic-drift direction relative to the active X-point (favourable and unfavourable configurations) on edge turbulence, transport, and the radial electric field $E_r$ in ASDEX Upgrade and TCV. First-principles gyrokinetic simulations of the edge and scrape-off layer are performed in diverted X-point geometry using the spectral full-$f$ \texttt{GENE-X} code. Two matched pairs of L-mode discharges are considered, and the simulations are validated against the available measurements in all four cases, showing good agreement, in particular for the $E_r$ profiles. A radial force-balance analysis shows that the deeper $E_r$ well in the favourable configuration is determined by the larger poloidal ion flow near the separatrix, which cannot be reproduced by neoclassical solvers and is identified in axisymmetric simulations as turbulence-driven. We then derive an exact evolution equation for the mean poloidal-flow acceleration from the full-$f$ gyrokinetic vorticity equation, which identifies the radial gradient of the total velocity stress as its nonlinear drive. The observed asymmetry is found to arise from stronger nonlinear energy transfer into the mean flow in the favourable configuration, associated with enhanced nonlinear $\bm{E}\times\bm{B}$ advection of vorticity near the separatrix and X-point. Flux-driven TCV simulations further show that the asymmetry persists at matched power more than six times the L-mode value. Assuming a critical $E_r$ at the L--H transition, this yields an estimated ratio of approximately $1.7$ between the unfavourable and favourable L--H power thresholds, close to the experimentally observed factor of two. Overall, the unfavourable configuration exhibits larger fluctuation amplitudes, a weaker $E_r$ well, stronger geodesic acoustic modes, and shorter energy confinement times for the same profiles. Taken together, these effects may explain the observed difference in access to high-confinement regimes. The present paper is the companion paper to B. J. Frei \textit{et al.}, Phys. Rev. Lett. \textbf{137}, 035103 (2026).
\end{abstract}

\section{Introduction} \label{sec:intro}

The direction of the ion magnetic drift relative to the position of the active X-point defines the two drift configurations in diverted geometry in tokamaks: the \textit{favourable} (\textit{fav}) configuration, where the drift points towards the active X-point, and the \textit{unfavourable} (\textit{unfav}) one, where it points away from it. Despite the apparent smallness of this operational change, its consequences on the plasma performance are large and have been documented for more than three decades. The most prominent of them is the power threshold ($P_{\mathrm{LH}}$) for the transition from the low- (L-mode) to the high-confinement mode (H-mode), which is typically larger by a factor of about two in \textit{unfav} than in \textit{fav} \cite{ryter1996,labombard2005,sauter2011,plank2023b,manz2025}. In \textit{unfav}, the plasma typically evolves from L-mode to H-mode through an intermediate improved energy confinement regime, the I-mode, which combines an H-mode-like temperature pedestal with an L-mode-like density profile \cite{hubbard2012,ryter2016,happel2017}. Related asymmetries are reported in the scrape-off layer (SOL) parallel flows \cite{labombard2005}, in the edge fluctuation levels and in the turbulence dynamics preceding the transition \cite{cziegler2017,yan2023,bielajew2023}, and in the sustainment of H-mode itself \cite{grover2024}. Identifying their origin is of direct practical relevance, since $P_{\mathrm{LH}}$ sets the auxiliary heating required to operate a fusion device in H-mode and thus enters the design of the operational scenarios of ITER and of future fusion reactors \cite{ryter1996,manz2025}.

Among the key quantities that differ between the two configurations, the edge radial electric field $E_r$ has emerged as the most promising candidate to explain the asymmetry. Indeed, the $\bm E \times \bm B$ flow shear associated with the $E_r$ well, which develops just inside the separatrix, is widely considered to regulate and eventually quench the turbulent transport at the L--H transition \cite{diamond1994,terry2000,carlstrom2002,conway2011}. Dedicated experiments at ASDEX Upgrade (AUG), comparing matched pairs of \textit{fav} and \textit{unfav} L-mode discharges, show that, at the same input power and for nearly identical kinetic profiles, the measured $E_r$ well is systematically deeper in \textit{fav} than in \textit{unfav} \cite{schirmer2006,viezzer2013,viezzer2014,plank2023}, the access to H-mode being experimentally linked to the local value of $E_r$ \cite{cavedon2020,cavedon2024}. A threshold value of $E_r$ at the L--H transition has indeed been identified in \textit{fav} at AUG \cite{plank2023b}, while it remains unclear in \textit{unfav}. In \textit{unfav}, the $E_r$ well is found to deepen progressively from L-mode to I-mode, and further in H-mode \cite{happel2017}. Similar surveys of the edge $E_r$ were performed on WEST \cite{vermare2021} and, more recently, on TCV \cite{rienacker2025,rienacker2026phd,rienacker2026}, showing qualitatively consistent results across the three devices. Since the kinetic profiles are matched, the deeper $E_r$ well in \textit{fav} must originate in the contributions to the radial force balance that are not set by the equilibrium pressure gradient, i.e. in the ion flows.

Explaining the depth of the $E_r$ well from first principles has nevertheless remained elusive. In particular, neoclassical (NC) theory relates the poloidal ion flow to the equilibrium gradients and to the collisionality through the parallel viscosity \cite{hinton1976,kim1991,houlberg1997,belli2008,landreman2014}. These quantities being identical in a matched \textit{fav} and \textit{unfav} pair, the NC poloidal flow changes sign upon field reversal but keeps a similar amplitude, such that NC predictions cannot, by themselves, produce the observed asymmetry \cite{plank2023}. Mechanisms that break this symmetry, such as the collisional loss of ion orbits near the separatrix \cite{brzozowski2019,zhu2023} or the momentum losses by charge exchange with neutrals \cite{versloot2011}, have been considered, but their quantitative assessment in realistic diverted geometry is difficult. On the other hand, the generation of mean (zonal) poloidal flows by turbulence through the radial gradient of the Reynolds stress (RS) \cite{diamond1991,hidalgo1999,smolyakov2000,diamond2005} provides a natural symmetry-breaking mechanism since the tilting of the turbulent eddies, together with the poloidal localization of the turbulent envelope and its truncation near the X-point, depends on the drift configuration \cite{fedorczak2012,fedorczak2013,schmid2017,cziegler2017,manz2018,grover2024}. The applicability of these models to the tokamak edge is, however, questionable, since they are often local, assume small fluctuation amplitudes and rely on reduced fluid descriptions, such that they lack the kinetic and full-$f$ effects known to modify the RS and, therefore, the nonlinear generation of mean flows by turbulence \cite{scott2003,scott2005,held2018,sarazin2021}. A first-principles description of the edge and SOL dynamics that retains, simultaneously, the X-point geometry, the full-$f$ turbulence and the self-consistent electric field is therefore needed to assess the origin of the \textit{fav} and \textit{unfav} asymmetry.

In the present work, leveraging the spectral full-$f$ formulation \cite{frei2020,frei2025vspec} of the gyrokinetic (GK) \texttt{GENE-X} code \cite{michels2021}, we perform first-principles simulations of the edge and SOL turbulence in diverted X-point geometry, and assess the influence of the drift configuration on turbulence, transport and radial electric field in two devices. In particular, we consider two matched pairs of lower single-null (LSN) L-mode discharges, performed at AUG \cite{plank2023} and at TCV \cite{rienacker2025,rienacker2026}, which differ by the direction of the ion $\grad B$ drift, obtained by reversing the toroidal magnetic field (together with the plasma current in TCV), and we validate the simulations against the available measurements in all four cases. The present work is the companion of Ref.~\cite{frei2026prl}, where a first account of the AUG results is reported, and extends the analysis to TCV, whose smaller size, different edge collisionality and different dominant instability provide an independent test of the mechanism proposed there. We note that the present simulations are full-$f$, such that they retain the turbulent and NC components of $E_r$ on the same footing.

We show that \texttt{GENE-X} yields nearly identical outboard midplane (OMP) profiles in \textit{fav} and \textit{unfav}, consistent with experiments, while reproducing the deeper $E_r$ well measured in \textit{fav} in both devices. Decomposing $E_r$ through the radial force balance, we find the depth of the well to be set almost entirely by the poloidal ion flow, whose \textit{fav} and \textit{unfav} asymmetry is retrieved neither by the NC codes \texttt{NCLASS} and \texttt{PERFECT} nor by dedicated axisymmetric simulations, pointing to a turbulence-driven origin, while the geodesic acoustic mode (GAM) activity is found to be stronger in \textit{unfav}. We then show that the same linear instabilities, i.e. a resistive electron drift wave (eDW) and a trapped-electron mode (TEM) in AUG and a dissipative trapped-electron mode (DTEM) in TCV, drive the turbulence in \textit{fav} and \textit{unfav}, and derive from the GK vorticity equation an exact, full-$f$ evolution equation for the flux-surface averaged (FSA) poloidal momentum density, whose nonlinear drive is the total electrostatic velocity stress $\Pi$, i.e. the full-$f$ generalization of the RS based on the Favre decomposition \cite{favre1965,held2018}. The asymmetry is traced to the work done by $\Pi$, $\Pi_W$, which measures the nonlinear energy transfer from turbulence into the mean poloidal flow near the separatrix and is substantially weaker in \textit{unfav}. By means of flux-driven simulations at matched power, we finally show that the $E_r$ asymmetry, as well as the fluctuation and GAM asymmetries, persist in TCV when the power crossing the separatrix is raised to more than six times its L-mode value, and we estimate from the resulting offset in the depth of the $E_r$ well a ratio of about $1.7$ between the \textit{unfav} and \textit{fav} L--H power thresholds, close to the experimental factor of two.

The structure of the paper is the following. The \textit{fav} and \textit{unfav} experiments are described in Section~\ref{sec:experiments} and the spectral full-$f$ GK model in Section~\ref{sec:model}, while Section~\ref{sec:results} reports the validation of the simulations. The validation and the decomposition of $E_r$ are given in Section~\ref{sec:er}, the turbulence being characterized in Section~\ref{sec:turbulence}. Section~\ref{sec:meanflow} contains the full-$f$ theory of the poloidal mean flow acceleration and the analysis of the nonlinear energy transfer, while the persistence of the asymmetry at high and matched power is assessed in Section~\ref{sec:fluxdriven}, together with its implications for the access to H-mode. Finally, we summarize our findings and conclude in Section~\ref{sec:conclusions}. The appendices report the coordinates and magnetic geometry (Appendix~\ref{appendix:signs}), the boundary and initial conditions, numerical resolution and computational cost (Appendix~\ref{appendix:A}), the derivation of the FSA poloidal momentum balance (Appendix~\ref{appendix:momentum}), the GAM dispersion relation with pressure anisotropy (Appendix~\ref{appendix:gam}) and the local plasma parameters together with the linear parameter scans (Appendix~\ref{appendix:localparameters}).

\section{Favourable and Unfavourable Experiments at AUG and TCV} \label{sec:experiments}

\subsection{Experiments at AUG} \label{subsec:aug}

The AUG discharges considered in this work are the same as those reported in \cite{plank2023}, where dedicated deuterium experiments investigated the effect of the ion $\nabla B$ drift direction on the radial electric field $E_r$ by comparing a pair of LSN discharges, \#36983 (\textit{fav}) and \#37375 (\textit{unfav}). Both discharges were performed under closely matched conditions: a line-averaged electron density of $n_e \simeq 2.7\times10^{19}$~m$^{-3}$ (lower density branch) and $600$~kW of electron cyclotron resonance heating (ECRH) power in L-mode. The on-axis toroidal magnetic field and plasma current were $\lvert B_\phi\rvert = 2.5$~T and $\lvert I_p\rvert = 0.8$~MA, respectively. The radial electric field $E_r$ was measured using Doppler backscattering (DBS), while the electron temperature ($T_e$) and density ($n_e$) profiles were reconstructed through integrated data analysis (IDA) \cite{fischer2010}. The main ion temperature ($T_i$) was assumed equal to the impurity ion temperature measured by charge-exchange recombination spectroscopy (CXRS). The deuterium toroidal ($U_{i \phi}$) and poloidal ($U_{i \theta}$) velocities were inferred from impurity flow measurements using the neoclassical code \texttt{NEOART} \cite{peeters2000}.

\subsection{Experiments at TCV} \label{subsec:tcv}

The TCV discharges analysed in this work consist of a pair of LSN Ohmic L-mode deuterium plasmas, \#86593 (\textit{fav}) and \#86594 (\textit{unfav}), presented in \cite{rienacker2026}. Both discharges feature identical magnetic configurations, with $\lvert B_\phi\rvert = 1.44$~T, $\lvert I_p\rvert = 210$~kA, and a line-averaged density of $n_e \simeq 3.6\times10^{19}$~m$^{-3}$ over the $1.3 \leq t \leq 1.4$~s time interval. The total input power is approximately $250$~kW in both cases. These closely matched conditions enable a direct comparison between the \textit{fav} and \textit{unfav} configurations, with minimal differences in the kinetic profiles. The radial electric field $E_r$ was inferred from measurements of the perpendicular velocity using DBS. The electron density and temperature profiles were obtained from Thomson scattering (TS), while the carbon ion temperature (a proxy for $T_i$) was measured via CXRS.

\section{Gyrokinetic Edge and SOL Turbulence Modelling} \label{sec:model}

To model self-consistently the edge and SOL turbulence, we use in this work the spectral formulation of the GK full-$f$ \texttt{GENE-X} code \cite{frei2025vspec}. The spectral approach solves for the full gyroaveraged part of the gyrocenter distribution function, denoted by $ f_\alpha = f_\alpha(\bm{R}, v_\parallel, \mu,t)$, for particle species $\alpha$ (mass $m_\alpha$ and charge $q_\alpha$) in the gyrocenter phase-space described by the coordinates $(\bm{R}, v_\parallel, \mu, \theta)$. Here, $\bm{R}$ is the gyrocenter position, $v_\parallel = \bm{b} \cdot \bm{v}$ the velocity parallel to the equilibrium magnetic field $\bm B$ (with $\bm{b} = \bm{B} / B$), and $\mu = m_\alpha v_\perp^2 / (2B)$ the magnetic moment, being $v_\perp = \left|\bm{v} - \bm{b} v_\parallel\right|$. The self-consistent evolution of $f_\alpha$ is obtained using the long-wavelength electromagnetic and collisional GK Vlasov equation. In particular, within the spectral approach and for computational efficiency, $f_\alpha$ is expanded onto a set of orthogonal and scaled Hermite and Laguerre polynomials, such that $f_\alpha$ is approximated by \cite{frei2020}

\begin{equation} \label{eq:faspectral}
f_\alpha \simeq \sum_{p =0}^{N_{v_\parallel}-1} \sum_{j =0}^{N_\mu-1} \mathcal{N}_\alpha^{pj} \hat H_p(\hat v_{\parallel \alpha}) L_j(\hat \mu_\alpha) F_{\mathcal{M} \alpha}.
\end{equation}
Here, $\hat H_p(\hat v_{\parallel \alpha}) = H_p(\hat v_{\parallel \alpha})  / \sqrt{2^p p !} $ and $L_j(\hat \mu_\alpha)$ are the Hermite and Laguerre polynomials respectively \cite{gradshteyn2014}, with the scaled velocity-space coordinates $\hat v_{\parallel \alpha} = v_\parallel \sqrt{2 \tau_\alpha / m_\alpha}$ and $\hat \mu_\alpha = \mu B  / \tau_\alpha$ as arguments. $\tau_\alpha$ is a constant reference temperature introduced to adjust the velocity-space basis \cite{frei2025vspec}. In \eqnref{eq:faspectral}, the spectral coefficients, $\mathcal{N}_\alpha^{pj}$, are defined as Hermite-Laguerre weighted moments of $f_\alpha$,

\begin{equation} \label{eq:npj}
  \mathcal{N}_\alpha^{pj}  = \int d W  \hat H_p(\hat v_{\parallel \alpha}) L_j(\hat \mu_\alpha) f_\alpha,
\end{equation}
and are related to fluid quantities \cite{frei2025vspec}. In the spectral formulation, the dynamics is entirely described by the spectral coefficients, $\mathcal{N}_\alpha^{pj}$, which are evolved according to

\begin{align} \label{eq:spectralvlasov}
    & \frac{\partial}{\partial t } \mathcal{N}^{pj}_\alpha  + \nabla  \cdot \bm \Gamma_\alpha^{pj} + \mathcal{F}_{\alpha \ell k}^{pj}\mathcal{N}^{\ell k}_\alpha + \mathcal{D}_{\alpha \ell k}^{pj}\mathcal{N}^{\ell k}_\alpha \frac{\partial A_{1 \parallel} }{\partial t}  = \sum_{\beta} \mathcal{C}_{\alpha \beta}^{pj} + \mathcal{S}_{n_\alpha}^{pj}.
\end{align}
On the right-hand side of \eqnref{eq:spectralvlasov}, collisional effects ($\mathcal{C}_{\alpha \beta}^{pj}$) are modeled using a full-$f$ Lenard-Bernstein-Daugherty (LBD) collision operator \cite{ulbl2022,frei2025vspec}. The analytical expressions of $\bm \Gamma_\alpha^{pj}$, $\mathcal{F}_{\alpha \ell k}^{pj}$, $\mathcal{D}_{\alpha \ell k}^{pj}$, and $\mathcal{C}_{\alpha \beta}^{pj}$ can be found in \cite{frei2025vspec}.  The term $\mathcal{S}_{n_\alpha}^{pj}$ is the spectral representation associated with a localized edge density source \cite{frei2026ppcf}, which we detail in the next section. The spectral QN condition, Amp\`ere's law and Ohm's law, which determine the electrostatic potential $\phi_1$ and the parallel vector potential $A_{\parallel 1}$, are given in \cite{frei2025vspec}.

\subsection{Self-consistent Electric Field and Vorticity Equation} \label{subsec:efield}

In \texttt{GENE-X}, the electric field $\bm E_1$, central to the present work, is determined directly by the QN condition

\begin{align} \label{eq:qn}
\omega = \sum_{\alpha } q_\alpha n_\alpha,
\end{align}
with $\omega = \grad \cdot \bm \Omega$ the vorticity variable, with $\bm \Omega =  c^2 M  \bm E_1 /  B^2 $, $\bm E_1 = - \grad \phi_1$ the electric field, and $M = \sum_\alpha n_\alpha m_\alpha$ the total mass density. Here, the plasma density $n_\alpha$ is given by $n_\alpha = \mathcal{N}^{00}_\alpha$. We remark that $\bm E_1$ contains both short-wavelength, rapid (turbulent) and long-wavelength, slow (neoclassical) components.

An alternative way of determining $\bm E_1$ is from the GK vorticity equation, which can be derived from the QN condition given in \eqnref{eq:qn}. While yielding the same self-consistent electric field, the vorticity equation allows gaining physical insight about the drive of $\bm E_1$, relating the time evolution of the latter to parallel and perpendicular gyrocenter current densities. More precisely, taking the time derivative of \eqnref{eq:qn} and using the continuity equation, the GK vorticity equation is given by

\begin{align} \label{eq:vorticity}
      \frac{\partial \omega}{\partial t}  +   \grad \cdot \bm J = 0.
\end{align}
with $\bm J =  J_\parallel \bm b + \bm J_\perp = \sum_\alpha \bm J_\alpha = \sum_\alpha q_\alpha \int d W  \dot{\bm R} f_\alpha$ the total current density. The parallel current density is given by $J_{\parallel  \alpha}= q_\alpha   n_\alpha U_{\parallel \alpha}$, while the perpendicular current density can be separated into different contributions, such that $\bm J_{\perp  \alpha} = \bm J^{es}_{\perp  \alpha} + \bm J^{em}_{\perp  \alpha} + \bm J^{dia}_{\perp  \alpha} + \bm J_{\perp \alpha}^{pol}$, with $\bm J^{es}_{\perp  \alpha} = q_\alpha n_\alpha \bm u$, $ \bm J^{em}_{\perp  \alpha} = J_{\parallel \alpha} \bm B_1 /B$, $\bm J^{dia}_{\perp  \alpha} =  c P_{\parallel \alpha} \grad \times \bm b  / B   + c  P_{\perp \alpha} \bm b \times \grad B / B^2  $, $\bm J_{\perp \alpha}^{pol} =  - c m_\alpha n_\alpha  \bm b \times \grad \left| \bm u \right|^2/ (2 B)$. Here, $\bm u$ denotes the electrostatic $\bm E \times \bm B$ velocity, i.e. $\bm u = c \bm E_1 \times \bm B / B^2$. We remark that $\bm J^{dia}_{\perp \alpha}$ differs from the fluid diamagnetic current, $c \bm b \times \grad P_\alpha / B$, by the classical magnetization current $-c \grad \times (P_{\perp \alpha} \bm b / B)$, the two being equal in the limit of isotropic pressure once this correction is added \cite{hazeltine2013,frei2025}. Since the magnetization current is divergence free, it does not contribute to \eqnref{eq:vorticity}. At quasi-steady state (QSS), \eqnref{eq:vorticity} imposes a divergence-free gyrocenter current, $\grad \cdot \bm J = 0$, and it serves as the basis of the turbulence--mean flow analysis of Section~\ref{sec:meanflow}.

\subsection{Numerical Setup} \label{subsec:numsetup}

In each device, the \textit{fav} and \textit{unfav} simulations use the same magnetic equilibrium, reconstructed from AUG \#36983 and TCV \#86594 (see Appendix~\ref{appendix:A}). The two configurations are obtained by reversing the toroidal magnetic field only, the plasma current being kept fixed, such that the \textit{fav} and \textit{unfav} simulations share the same magnetic geometry and differ only by the direction of the ion $\grad B$ drift. This differs from the AUG and TCV experiments, where the toroidal magnetic field and the plasma current are reversed together, such that the field-line helicity is preserved \cite{plank2023,rienacker2026}. As a consequence, the \textit{unfav} simulations have a helicity opposite to that of the \textit{unfav} discharges (see Appendix~\ref{appendix:signs}). We expect this difference to have a minor impact on the present results since reversing the helicity at a fixed drift direction was found to have no sizeable effect on the edge $E_r$ \cite{rienacker2025}.

The present GK turbulence simulations evolve a deuterium ion species ($m_i = 2 m_{\mathrm{ref}}$, $q_i = e$) and an electron species with realistic mass ($m_e = 5.486 \times 10^{-4} m_{\mathrm{ref}}$, $q_e = - e$), with $m_{\mathrm{ref}}$ the proton mass. At the boundaries of the numerical domain, the distribution functions are assumed to be Maxwellian \cite{frei2025vspec}, with density and temperatures fixed at values close to experiments and identical in \textit{fav} and \textit{unfav}. The boundary values, the initial conditions, the numerical resolution, the reference quantities and the computational cost of the simulations are detailed in Appendix~\ref{appendix:A} and collected in Table~\ref{tab:2}.

To mimic the effects of neutral ionization in the edge, we consider a poloidally homogeneous, axisymmetric, radially localized Gaussian energy-conserving density source centred at $\rho_{\mathrm{pol}}=0.99$, with a constant amplitude of $1.5\times10^{22} \mathrm{s}^{-1}$ for both \textit{fav} and \textit{unfav}. The implementation of the density source in \texttt{GENE-X} is detailed in Ref.~\cite{frei2026ppcf}. In the present setup, particle and energy sources and sinks arise from a combination of Dirichlet boundary conditions at the edge and SOL boundaries—where the profiles are fixed to values close to experimental measurements—and diffusive relaxation applied within a narrow buffer region between ghost points and the active numerical domain. Upon discretization, this approach is equivalent to a Krook-like operator relaxing the distribution function towards a Maxwellian with fixed density and temperature near the boundary, which injects (or removes) particles and energy as required, such that the present simulations are effectively flux-driven with adaptive sources. 

\section{Simulation Results} \label{sec:results}

\begin{figure}
\centering
\includegraphics[scale=0.55]{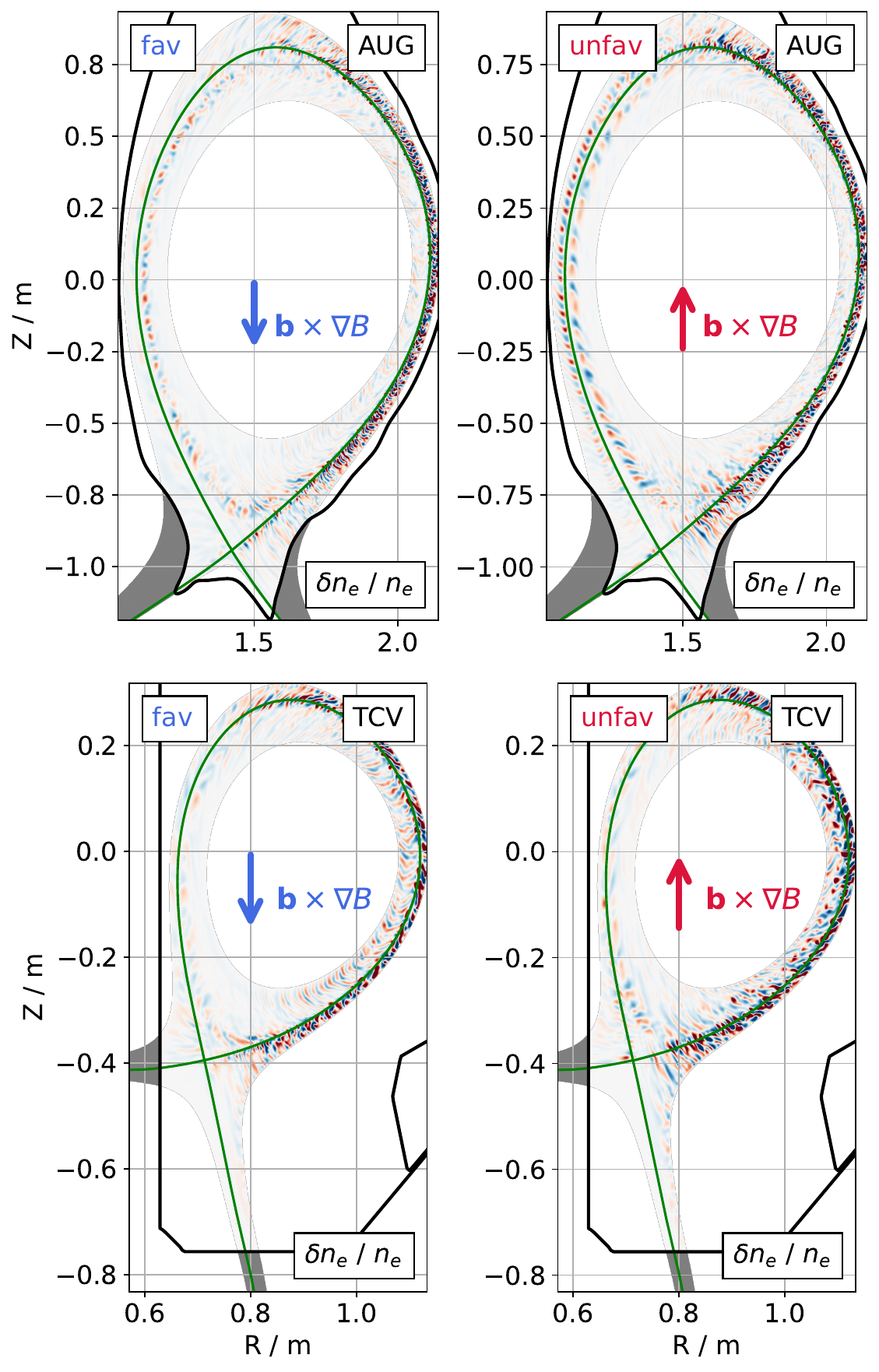}
\caption{\label{fig:fig1} Snapshots of (top) AUG and (bottom) TCV simulations at QSS with \texttt{GENE-X}. \textit{Fav} and \textit{unfav} results are shown in the left and right columns, respectively.}
\end{figure}

We now report the results of the \texttt{GENE-X} simulations, snapshots of which are shown at quasi-steady state (QSS) in Fig.~\ref{fig:fig1} for both devices and both drift configurations. We validate the OMP profiles against the AUG and TCV measurements and compare the relative fluctuation levels between \textit{fav} and \textit{unfav} in Section~\ref{subsec:ompvalidation}, while the particle and energy edge confinement times and the total energy balance are assessed in Section~\ref{subsec:particleandenergy}.

\subsection{Validation of Outboard Midplane Profiles} \label{subsec:ompvalidation}

\begin{figure}
\includegraphics[scale=0.55]{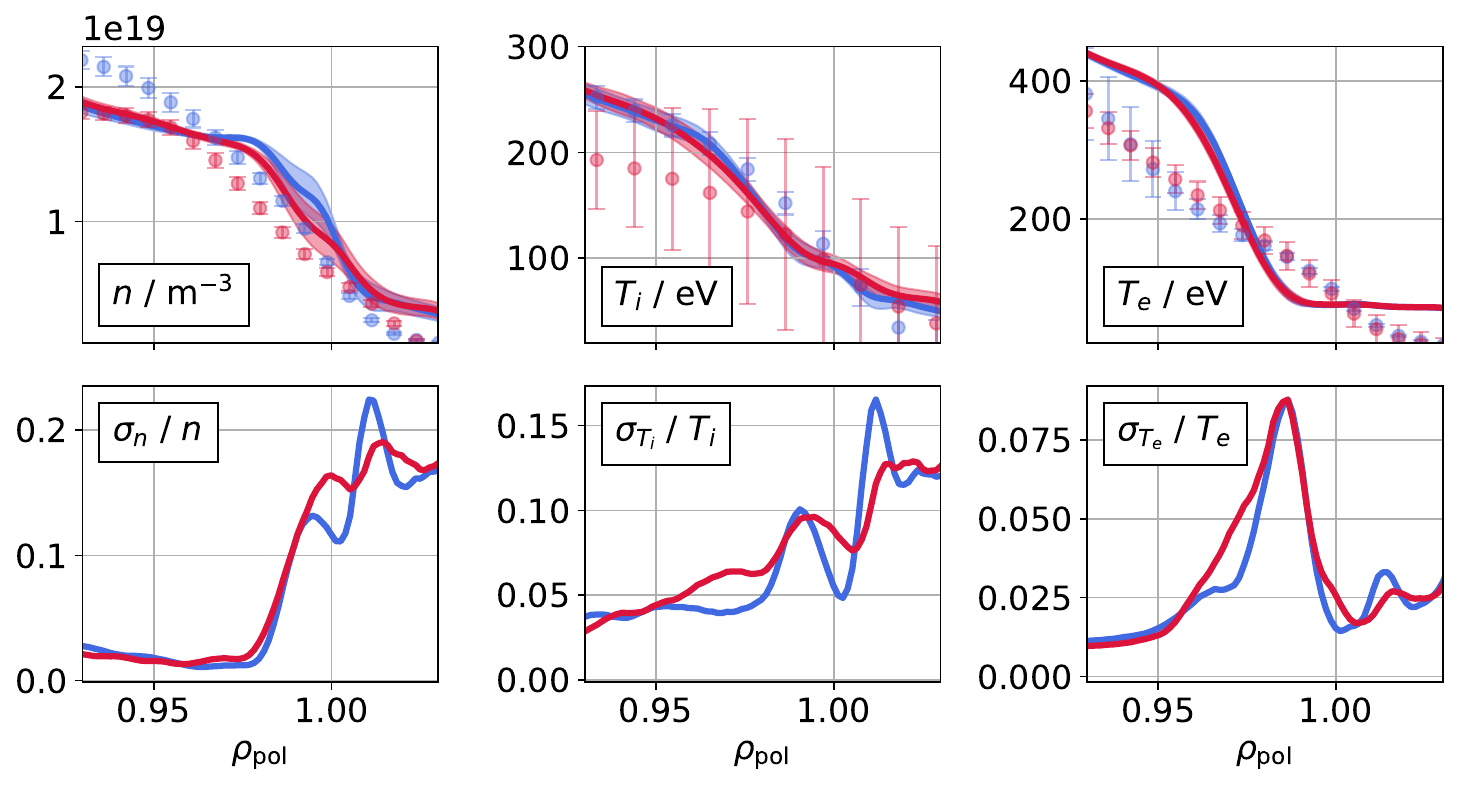}
\caption{\label{fig:fig2} Top row: OMP density ($n$), ion ($T_i$) and electron ($T_e$) temperature profiles (from left to right) obtained from \texttt{GENE-X} (solid coloured lines) in \textit{fav} (blue) and \textit{unfav} (red) at AUG. The experimental data (coloured circles) are taken from \cite{plank2023}. The coloured shaded areas are the standard deviations ($\sigma_f$) of $f$ associated with the toroidal and time averages. Bottom row: density, ion and electron temperature normalized fluctuation levels, $\sigma_f / f $,  (from left to right) associated with the OMP profiles.}
\end{figure}

We note that the validations of the density and ion temperature OMP profiles for AUG are reported in Ref.~\cite{frei2026prl}. At the top of Fig.~\ref{fig:fig2}, the OMP profiles obtained from \texttt{GENE-X} are shown for AUG and compared with experimental data from \cite{plank2023}. Good agreement (within the error bar) is obtained for the density and ion temperature profiles in particular near the separatrix. On the other hand, the electron temperature profile is found to depart the most from experiments. In particular, lower $T_e$ (and gradient) near the separatrix are observed in \texttt{GENE-X} compared to experiments because of the presence of the edge density source. In the lower row of Fig.~\ref{fig:fig2}, the relative fluctuations $\sigma_f / f$, with $\sigma_f$ the standard deviation associated with the toroidal and temporal averages, are shown at the OMP. While the relative fluctuation levels are similar in the edge, $\sigma_f / f$ is found to be slightly larger in \textit{unfav} than in \textit{fav} near the separatrix and in the region of steep gradients.

\begin{figure}
\centering
\includegraphics[scale=0.55]{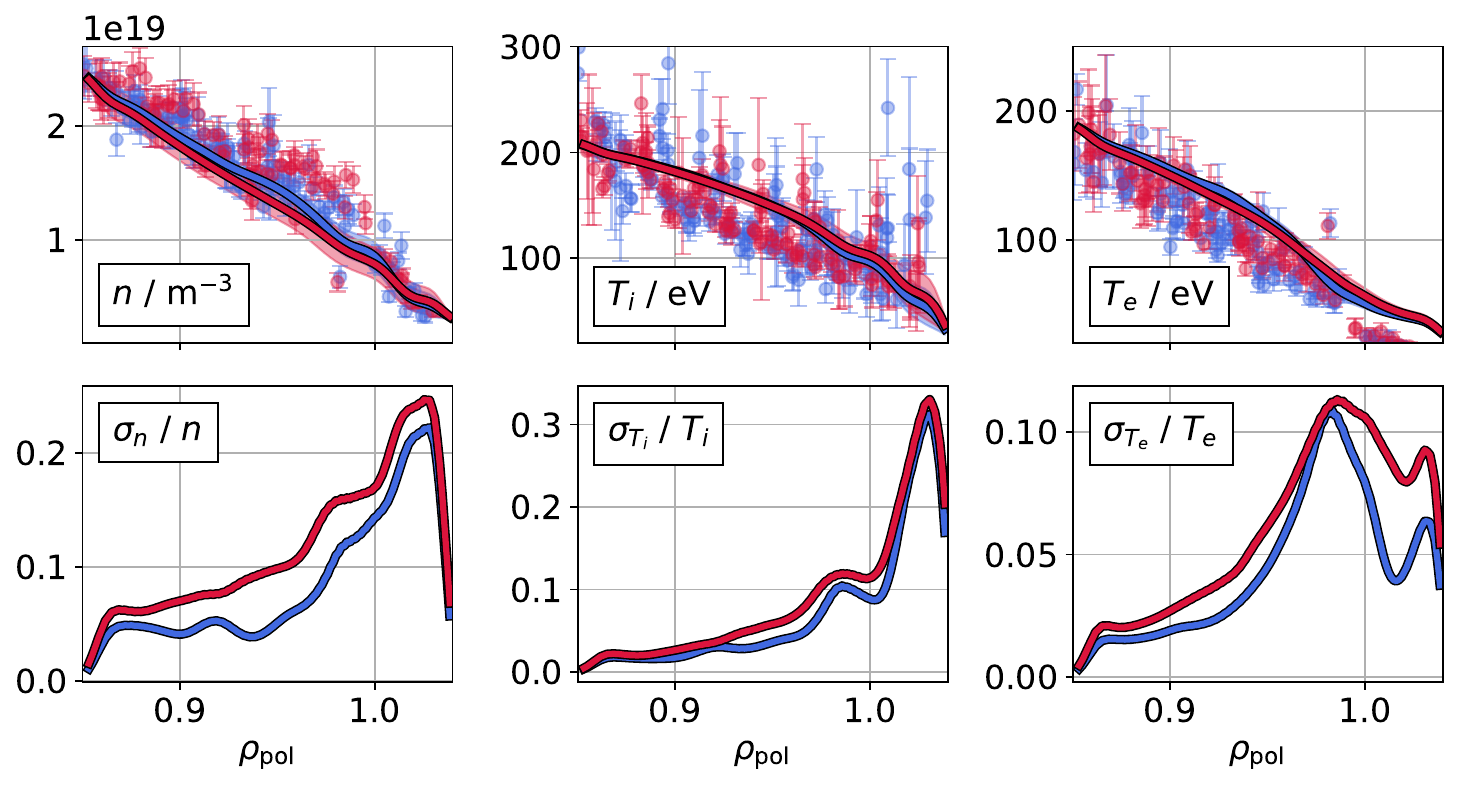}
\caption{\label{fig:fig3} Same as Fig.~\ref{fig:fig2} for TCV \textit{fav} (blue) and \textit{unfav} (red). The experimental data are taken from \cite{rienacker2026}.}
\end{figure}

Fig.~\ref{fig:fig3} presents the same validation exercise for the TCV discharges, the experimental profiles being taken from \cite{rienacker2026}. Similar to AUG, the simulations achieve reasonable agreement with experiments, although the temperatures are slightly overestimated by \texttt{GENE-X}. Despite the large scatter in the measurements, the \textit{fav} and \textit{unfav} profiles are similar in both simulations and experiments. We nonetheless remark that the density and electron temperature profiles are slightly higher in \textit{fav} than in \textit{unfav}, a larger $T_e$ gradient near the separatrix being observed in the former case. Finally, the lower plots of Fig.~\ref{fig:fig3} also confirm that larger relative fluctuation levels are observed in \textit{unfav} compared to \textit{fav}, similar to AUG, the difference between the two configurations being enhanced in TCV. We remark, however, that the measured asymmetry depends on the fluctuating quantity. In the same TCV Ohmic pair, the DBS backscattered power is larger in \textit{fav} near the $E_r$ well, larger density fluctuations in \textit{unfav} being recovered only with sufficient auxiliary heating \cite{rienacker2026}. At AUG, on the other hand, the edge electron temperature fluctuations are larger in \textit{unfav} \cite{bielajew2023}, as in the present simulations.

\subsection{Particle and Energy Confinement Time} \label{subsec:particleandenergy}

The particle and energy (edge) confinement time, $\tau_p$ and $\tau_E$, can be calculated at quasi-steady state. More precisely, we define the particle (ion) confinement time as $\tau_p =  \int d V n_i /  \oint d S\Gamma_i$, with $\int d V$ the volume integral evaluated over the edge region (without the core region) and $\oint d S\Gamma_i$ the surface-integrated total radial particle flux ($\Gamma_i$) across the separatrix. Similarly, the energy confinement time is defined by $\tau_E =  \sum_\alpha \int d V 3 n_\alpha T_\alpha / 2 / \oint d S Q$, with $3 n_\alpha T_\alpha / 2$ the thermal energy and $\oint d S Q$ the surface integrated total radial heat flux $Q$.

\begin{table}[h!]
\centering
\begin{tabular}{ccccc}
\hline
Machine& Conf. & $\tau_p$ / ms & $\tau_E$ / ms & $P$ / MW \\
\hline
\hline
AUG  & \textit{fav}  &  $18.7$ & $14.9$ & $0.50$ \\
AUG  & \textit{unfav} & $15.4$ & $12.7$ & $0.50$ \\
TCV  & \textit{fav}  &  $5.6$ &  $3.7$ & $0.19$ \\
TCV  & \textit{unfav} & $5.3$ &  $2.1$ & $0.25$ \\
\hline
\end{tabular}
\caption{Particle ($\tau_p$) and energy ($\tau_E$) edge confinement times (in ms) and total power $P$ (in MW) across the edge evaluated at $\rho_{\mathrm{pol}} = 0.98$, obtained in \texttt{GENE-X} simulations for AUG and TCV in \textit{fav} and \textit{unfav}. The power $P$ is composed of both the $\bm E \times \bm B$ and diamagnetic fluxes and is averaged over $0.2$~ms at quasi-steady state.}
\label{tab:1}
\end{table}

In Table~\ref{tab:1}, the particle and energy confinement times obtained in the AUG and TCV simulations are shown and compared between \textit{fav} and \textit{unfav}. First, $\tau_p$ and $\tau_E$ are of the same order for \textit{fav} and \textit{unfav} at AUG, with slightly larger values for \textit{fav}. Although the particle transport (and sources) and heat transport are similar in \textit{fav} and \textit{unfav}, the larger $\tau_p$ and $\tau_E$ values are a consequence of the higher density and temperature profiles observed in \textit{fav} (see Fig.~\ref{fig:fig2}).

In the case of TCV, while the particle confinement $\tau_p$ is similar in both \textit{fav} and \textit{unfav}, the energy confinement time $\tau_E$ is found to be enhanced in \textit{fav} (by a factor $\sim 1.7$). While such an enhancement is not observed in AUG, this larger asymmetry in $\tau_E$ results from two effects: (i) the higher density and temperature profiles observed in \textit{fav} as shown in Fig.~\ref{fig:fig3} and (ii) the lower total heat flux in \textit{fav}. At quasi-steady state, the total energy balance is given, in the absence of volumetric sources and at leading order, by

\begin{equation} \label{eq:nrgtotbalance}
 \frac{1}{V^\prime} \frac{d}{d \psi} \left(V'\fs{Q^\psi } \right) \simeq  0,
\end{equation}
with $Q^\psi  = \bm Q \cdot \grad \psi$ the radial component of the total energy flux $\bm Q_\alpha = \int d W m_\alpha v^2 \dot{\bm R} f_\alpha / 2 $. Here, $\fs{(\cdot)} =  (V')^{-1} \oint d S (\cdot) / \left|\grad \psi \right|$ ($V(\psi)$ is the volume enclosed by the $\psi$ flux-surface with $V' = d V(\psi) / d \psi$, and $dS$ is the surface element of the flux-surface $\psi$) denotes the FSA operator. A time average is assumed in \eqnref{eq:nrgtotbalance}.

We remark that, in the present case, the total particle and energy fluxes are dominated by the $\bm E \times \bm B$ turbulent and diamagnetic fluxes. In the absence of a volumetric energy source in the present simulations, \eqnref{eq:nrgtotbalance} imposes the total energy flux to be constant. This is illustrated in Table~\ref{tab:1} where the total power, $P =  \oint d S  Q^\psi / \left|\grad \psi \right| $, evaluated at $\rho_{\mathrm{pol}} = 0.98$ is reported for the different simulations. While the total power across the edge at AUG is approximately $P \simeq 0.5$~MW for both configurations, a noticeable difference is observed in TCV, where $P \simeq 0.25$~MW is required in \textit{unfav} against $P \simeq 0.19$~MW in \textit{fav}. This asymmetry results from the Dirichlet boundary conditions at the inner edge, which act as an adaptive source forcing similar profiles there, such that a larger heat flux is required in \textit{unfav} to compensate for its larger transport, hence its smaller $\tau_E$. Such a difference in energy confinement is not observed in the TCV experiments \cite{rienacker2026}. In Section~\ref{sec:fluxdriven}, flux-driven TCV simulations show that the \textit{fav} and \textit{unfav} asymmetry remains also at matched power. Finally, we note that the confinement times are larger for AUG than for TCV due to the machine size.

\section{Validation and Decomposition of the Radial Electric Field} \label{sec:er}

Having validated the equilibrium profiles and characterized the confinement, we now turn to the radial electric field $E_r$. We first show that \texttt{GENE-X} reproduces the deeper $E_r$ well measured near the separatrix in \textit{fav} in both AUG and TCV (Section~\ref{subsec:ervalidation}), its depth being set almost entirely by the poloidal ion flow according to the radial force balance (Section~\ref{subsec:forcebalance}). We then analyse the finite-frequency GAM component of $E_r$, larger in \textit{unfav} near the well (Section~\ref{subsec:gam}), and quantify its nonlinear coupling with the turbulence by a bispectral analysis (Section~\ref{subsec:gamcoupling}). Finally, we compare the toroidal and poloidal flows with the NC predictions (Section~\ref{subsec:flows}) and test the role of turbulence in driving the asymmetry with dedicated axisymmetric simulations (Section~\ref{subsec:n0}).

\subsection{Radial Electric Field Validation} \label{subsec:ervalidation}

\begin{figure}
\centering
\includegraphics[scale=0.55]{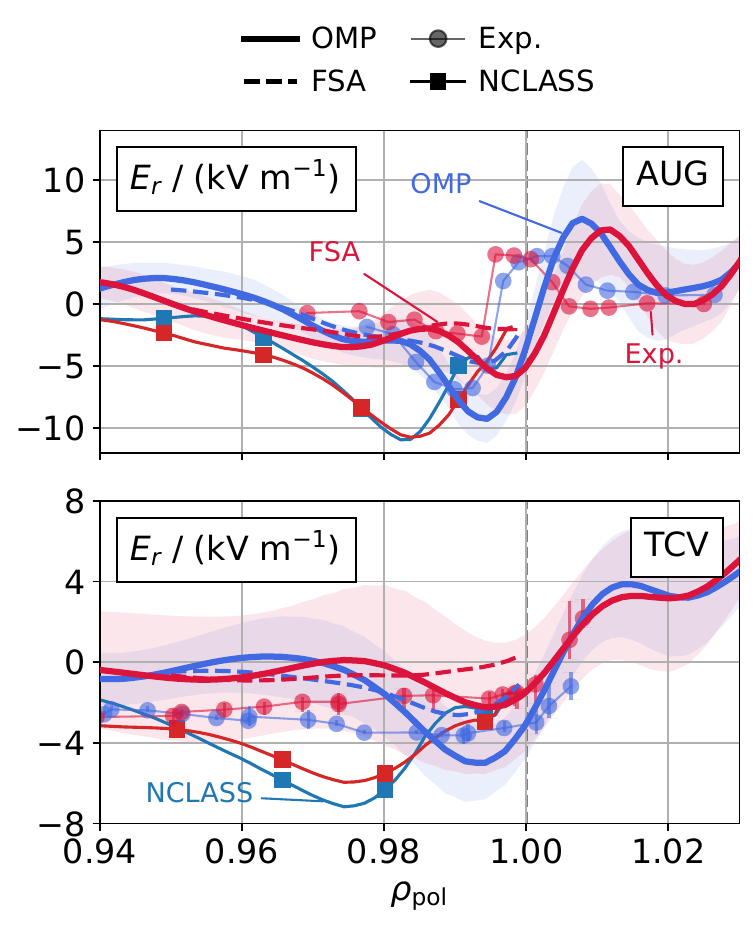}
\caption{\label{fig:fig4} $E_r$ profiles in AUG (top) and TCV (bottom) for \textit{fav} (blue) and \textit{unfav} (red): \texttt{GENE-X} at the OMP (solid lines, shaded areas giving the standard deviation over toroidal angle and time) and FSA (dashed lines), experiment (circles joined by thin lines; AUG \cite{plank2023}, TCV \cite{rienacker2026}), and radial force balance with the \texttt{NCLASS} NC poloidal flow (thin lines with squares). The grey dashed line marks the separatrix ($\rho_{\mathrm{pol}} = 1$). The AUG data and \texttt{GENE-X} results are those of \cite{frei2026prl}.}
\end{figure}

While the matched experimental parameters yield similar OMP profiles between \textit{fav} and \textit{unfav} (Figs.~\ref{fig:fig2} and \ref{fig:fig3}), a clear asymmetry in $E_r$ is observed in both AUG and TCV: a deeper $E_r$ well is found close to the separatrix in \textit{fav}. Fig.~\ref{fig:fig4} compares the OMP $E_r$ profiles from \texttt{GENE-X} with the experimental measurements \cite{plank2023,rienacker2026}. The simulations reproduce the \textit{fav} and \textit{unfav} asymmetry with the $E_r$ well, located near $\rho_{\mathrm{pol}} \sim 0.98$, deeper in \textit{fav} than in \textit{unfav}. Further inside the edge ($\rho_{\mathrm{pol}} \lesssim 0.98$) and in the SOL, $E_r$ is very similar between the two configurations. For both AUG and TCV, the depth of the $E_r$ well agrees closely with experiments, within the statistical uncertainties. We note that no measurements are available for $\rho_{\mathrm{pol}} \lesssim 0.98$ in AUG, while in TCV the simulations overestimate $E_r$ in this region. The simulations also feature the rapid radial increase of $E_r$ (from negative to positive) across the separatrix, associated with the transition to the SOL region where the fast parallel electron dynamics dominates \cite{frei2025}. As a consequence, both the inner and outer $E_r$ shear layers are more pronounced in \textit{fav}. The similarity of $E_r$ in the edge and SOL indicates that the origin of the \textit{fav} and \textit{unfav} asymmetry is confined radially just inside the separatrix. However, this asymmetry is not limited to the OMP: it persists in the FSA profiles shown in Fig.~\ref{fig:fig4}, indicating that the $E_r$ asymmetry does not average out over a flux surface and thus reflects an asymmetry in the balance that sets $E_r$.

\subsection{Radial Force Balance Analysis} \label{subsec:forcebalance}

To gain physical insight into the origin of the $E_r$ asymmetry between \textit{fav} and \textit{unfav}, we perform a radial force balance analysis, decomposing $E_r$ at the OMP into its different contributions. The radial force balance reads \cite{frei2025}

\begin{align}
\label{eq:radialforcebalance}
E_r & = \frac{\bm e_\psi \cdot \nabla \cdot \bm \Pi_i }{q_i n_e} - (\bm U_i \times \bm B)\cdot \bm e_\psi = \frac{\partial_\psi P_i}{q_i n_e} + U_{i \phi} B_\theta - U_{i \theta} B_\phi
\end{align}
with $\partial_\psi = \bm e_\psi \cdot  \grad $ the radial gradient, and $U_{i \theta}$ ($U_{i \phi}$) the poloidal (toroidal) components of the main ion velocity $\bm U_i$, defined as in \cite{frei2025}. In deriving \eqnref{eq:radialforcebalance}, we assume that the ion pressure anisotropy is negligible, which is a good approximation in the present cases. Since the radial force balance is satisfied identically in the simulations, it provides an exact decomposition of $E_r$ into physically distinct contributions, which we use here as a diagnostic. From left to right in \eqnref{eq:radialforcebalance}, these are the diamagnetic ($\partial_\psi P_i / (q_i n_e)$), toroidal ($+ U_{i \phi } B_\theta$), and poloidal ($- U_{i \theta } B_\phi$) contributions to the Lorentz force. 

\begin{figure}
\centering
\includegraphics[scale=0.55]{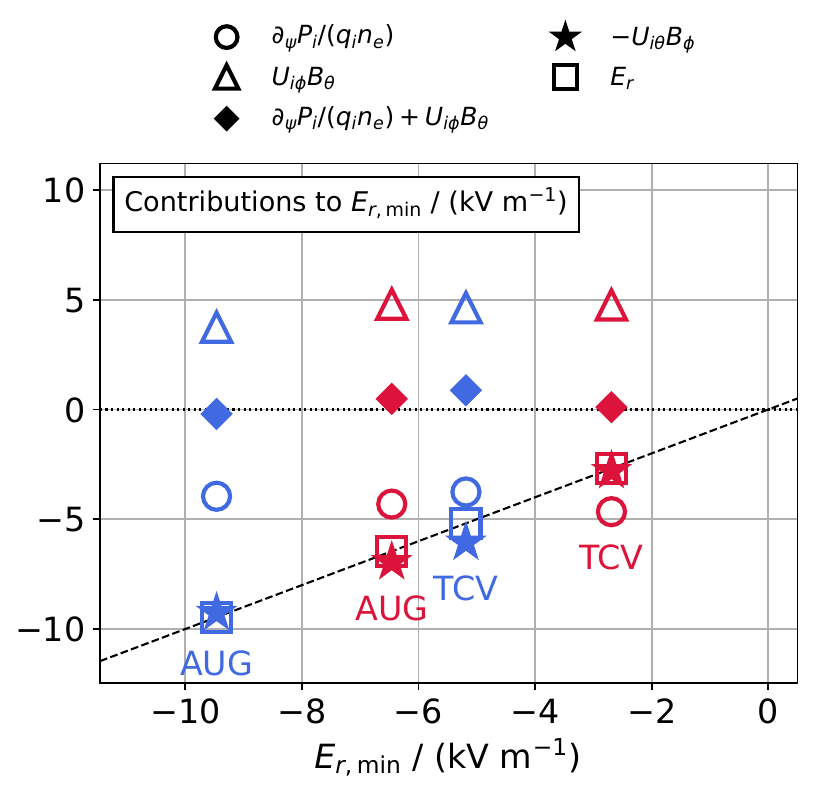}
\caption{\label{fig:fig5} Contributions to $E_{r, \mathrm{min}}$ at the OMP from the radial force balance, \eqnref{eq:radialforcebalance}, vs. $E_{r, \mathrm{min}}$ for AUG and TCV obtained in \texttt{GENE-X}: diamagnetic (open circles) and toroidal (open triangles) contributions, their sum (filled diamonds), the poloidal contribution (filled stars), and $E_{r, \mathrm{min}}$ itself (open squares), in \textit{fav} (blue) and \textit{unfav} (red). All quantities are averaged over the last $0.2$~ms. In all cases, $E_{r, \mathrm{min}} \simeq  - U_{i \theta} B_\phi$.}
\end{figure}

Fig.~\ref{fig:fig5} shows these contributions to the minimum of the $E_r$ well ($E_{r, \textrm{min}}$) evaluated at the OMP, plotted against $E_{r, \textrm{min}}$ itself. In all cases, the diamagnetic and toroidal contributions nearly cancel each other at the location of $E_{r, \textrm{min}}$, leaving only the poloidal contribution, such that $E_{r, \textrm{min}} \simeq  - U_{i \theta} B_\phi$. This implies that the asymmetry in the depth of the $E_r$ well between \textit{fav} and \textit{unfav} is induced by an asymmetry in the dynamics that sets the poloidal flow $U_{i \theta}$. This is consistent with the AUG and TCV measurements, where $\partial_\psi P_i / (q_i n_e)$ is matched between the two configurations and deviate from the measured $E_r$ near the separatrix \cite{plank2023,rienacker2026}. This also indicates that the parallel flows are similar (in amplitude) in both \textit{fav} and \textit{unfav}. A similar decomposition is obtained for the FSA but not shown here, demonstrating that the dominance of $U_{i \theta}$ in $E_r$ is not restricted to the OMP. Finally but not shown, $E_r$ follows closely $\partial_\psi P_i /(q_i n_e) + U_{i \phi} B_\theta$ in the edge (with $U_{i \phi} \simeq U_{i \parallel}$), where the poloidal contribution is much smaller, while $E_r$ follows $- \grad T_{e \parallel} /e$ in the SOL \cite{frei2025}.

\subsection{Edge Geodesic Acoustic Oscillations} \label{subsec:gam}

In addition to the zero-frequency zonal $E_r$ flow, $E_r$ can also carry a finite-frequency component associated with geodesic acoustic modes (GAMs), which is an axisymmetric $n = m = 0$ oscillation of the $\bm E \times \bm B$ flow arising from the $\bm E \times \bm B$ compressibility in toroidal geometry and coupling to $m=1$ pressure sidebands \cite{conway2011,conway2021}. Turbulence-driven GAMs are of particular interest for their possible role in regulating turbulent transport in the edge region and in the L--H transition \cite{conway2011}. Their amplitude is highly sensitive to the magnetic-field direction: experiments show the GAMs to be suppressed inside the $E_r$ well (or even absent) in \textit{fav} and larger in \textit{unfav} \cite{cziegler2017,vermare2026eps,rienacker2026}, a contrast reproduced qualitatively by the present simulations.

\begin{figure}
\centering
\includegraphics[scale=0.65]{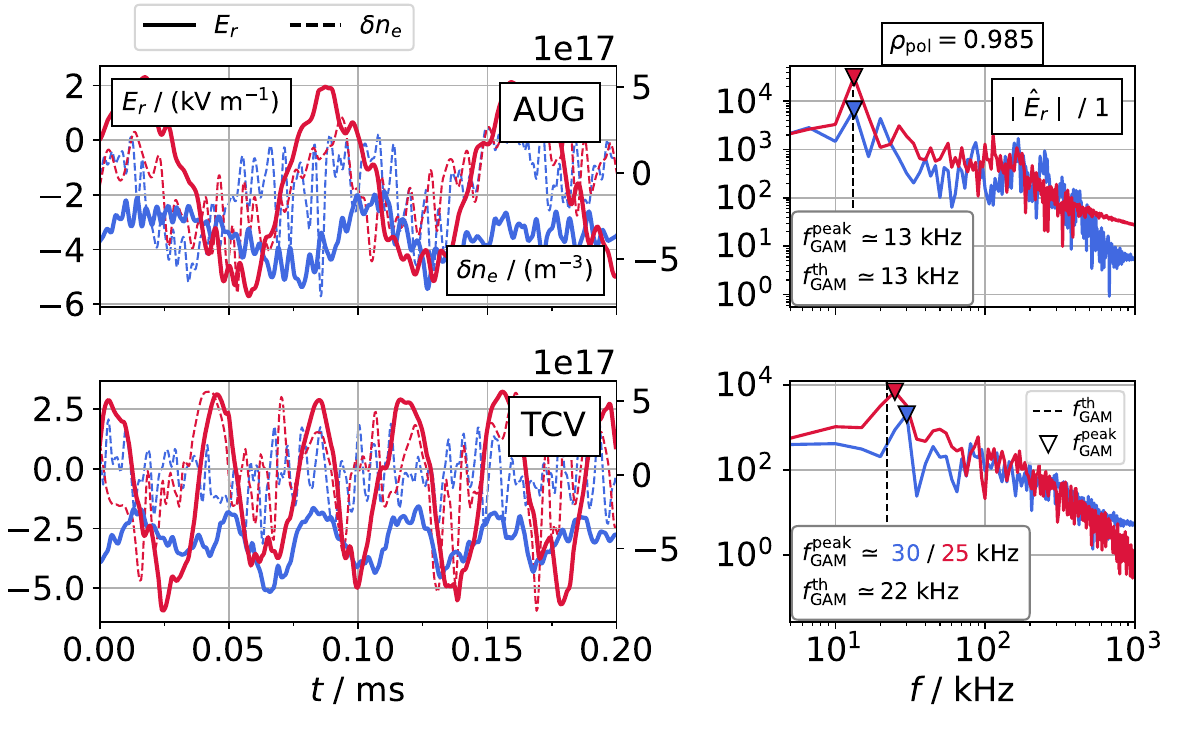}
\caption{\label{fig:fig6} Left: time traces of $E_r$ (solid lines plotted on the left y-axis) and density fluctuations $\delta n_e$ (dashed lines plotted on the right y-axis) evaluated at the OMP on the $\rho_{\mathrm{pol}} = 0.985$ flux-surface. Right: temporal spectra of $E_r$ measured at $\rho_{\mathrm{pol}} = 0.985$ at the OMP for AUG (top) and TCV (bottom). The vertical black dashed line marks the linear estimate of the GAM frequency, $f_{\mathrm{GAM}}^{\mathrm{th}}$, calculated from the local parameters using the dispersion relation given in \eqnref{eq:omegagam}, while the triangles mark the spectral maxima within $10 \leq f \leq 35$~kHz, $f_{\mathrm{GAM}}^{\mathrm{peak}}$, in \textit{fav} (blue) and \textit{unfav} (red).}
\end{figure}

\begin{figure}
\centering
\includegraphics[scale=0.55]{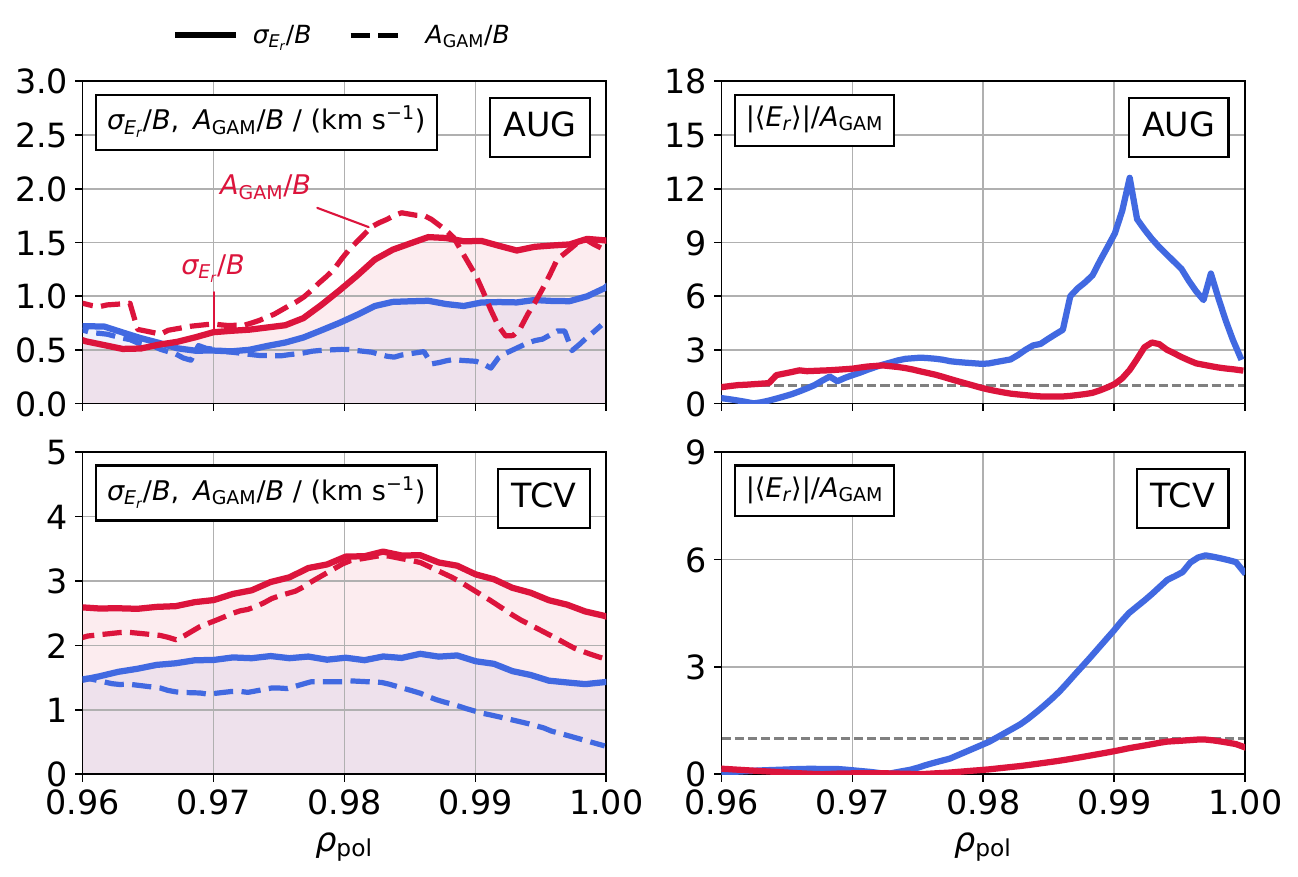}
\caption{\label{fig:fig7} Left: radial profiles, at the OMP for AUG (top) and TCV (bottom), of the GAM amplitude $A_{\mathrm{GAM}}$ (dashed lines) and of the $E_r$ standard deviation $\sigma_{E_r}$ (solid lines, with the area below shaded), both normalized to $B$, the standard deviation being taken over the toroidal angle and time. The amplitude $A_{\mathrm{GAM}}$ (half the peak-to-peak amplitude) is obtained by applying to the toroidally averaged $E_r$ a Butterworth band-pass filter of $\pm 5$~kHz width centred on the local spectral peak in the range $10 \leq f \leq 35$~kHz. Right: mean (zonal and zero-frequency) component of $E_r$, normalized to $A_{\mathrm{GAM}}$, the grey dashed line marking $\left| \langle E_r \rangle \right| = A_{\mathrm{GAM}}$.}
\end{figure}

The GAM oscillations appear in the OMP time traces of $E_r$, shown in Fig.~\ref{fig:fig6} for AUG and TCV at $\rho_{\mathrm{pol}} = 0.985$, as low-frequency oscillations at the GAM frequency, $f_{\mathrm{GAM}} \sim c_s / R$ (a linear estimate for $f_{\mathrm{GAM}}$ is given in Appendix~\ref{appendix:gam}, based on the \texttt{GENE-X} model and retaining the pressure anisotropy). The associated temporal Fourier spectra (right column of Fig.~\ref{fig:fig6}) peak at $f_{\mathrm{GAM}}^{\mathrm{peak}} \simeq 13$~kHz in both configurations in AUG, and at $f_{\mathrm{GAM}}^{\mathrm{peak}} \simeq 30$~kHz in \textit{fav} and $25$~kHz in \textit{unfav} in TCV, the frequency resolution of the spectra being $3.3$~kHz and $5$~kHz, respectively. The linear estimate of \eqnref{eq:omegagam}, $f_{\mathrm{GAM}}^{\mathrm{th}}$, yields $13$~kHz in AUG, in very good agreement with the measured peak, and $22$~kHz in TCV, below the measured peak in \textit{fav} and within the frequency resolution in \textit{unfav}. This machine-dependent agreement may be ascribed to the effects neglected in \eqnref{eq:omegagam}, i.e. radial profile variation, finite orbit width and wave-particle resonance, whose importance can differ between the two devices. Note that $f_{\mathrm{GAM}}$ is lower in AUG, principally because of the $1/R$ scaling of $\omega_{\mathrm{GAM}}$. Since $f_{\mathrm{GAM}}^{\mathrm{th}}$ depends only on the equilibrium profiles, which are matched between \textit{fav} and \textit{unfav}, comparable peak frequencies are expected where both configurations carry a coherent GAM, as in AUG at $\rho_{\mathrm{pol}} = 0.985$, while they can differ where the GAM is weak, as in TCV \textit{fav} inside the $E_r$ well (see Fig.~\ref{fig:fig7}). In \textit{unfav}, $f_{\mathrm{GAM}}$ is nearly radially constant across the edge in both devices, forming a plateau-like structure rather than following the $\sqrt{\fs{P_\parallel} / \fs{M}}$ scaling of \eqnref{eq:omegagam}, as also reported experimentally \cite{conway2011,conway2021}.

In contrast to the frequency, the GAM amplitude depends strongly on the drift configuration, being sensitive to the presence of an $E_r$ well, a behaviour recently reported experimentally in TCV \cite{vermare2026eps,rienacker2026}. The GAM amplitude $A_{\mathrm{GAM}}$ is extracted from the toroidally averaged $E_r$ time traces as half the peak-to-peak amplitude of the signal band-pass filtered with a zero-phase, fourth-order Butterworth filter of $\pm 5$~kHz width, centred on the local spectral peak in the range $10 \leq f \leq 35$~kHz, which encloses $f_{\mathrm{GAM}}$ in both devices. The resulting radial profiles of $A_{\mathrm{GAM}}$ are shown in Fig.~\ref{fig:fig7} with the total $E_r$ fluctuation level, given by the standard deviation, $\sigma_{E_r}$. Consistent with the time traces of Fig.~\ref{fig:fig6}, $A_{\mathrm{GAM}}$ is larger in \textit{unfav}, being comparable to $\sigma_{E_r}$, such that the GAM dominates the temporal modulation of $E_r$. On the other hand, $A_{\mathrm{GAM}}$ is reduced in \textit{fav} near the $E_r$ well, by a factor of about $2$ in AUG and $3$ in TCV with respect to \textit{unfav}, $\sigma_{E_r}$ being there increasingly carried by higher-frequency fluctuations.

These configuration-dependent amplitudes translate into a GAM-driven $\bm E \times \bm B$ flow shear, quantified by the mean (zero-frequency) $E_r$ normalized to $A_{\mathrm{GAM}}$ displayed in the right column of Fig.~\ref{fig:fig7}. Since the mean and the oscillating components of $E_r$ are found to vary on comparable radial scales, this amplitude ratio also measures the relative magnitude of the two contributions to the $\bm E \times \bm B$ shear. In \textit{unfav}, the low-frequency oscillating $E_r$ associated with the GAM is comparable to the mean one in both AUG and TCV. Its effect is also visible as a clear modulation at the GAM frequency of the density fluctuations $\delta n_e$ in Fig.~\ref{fig:fig6}. On the other hand, in \textit{fav} the mean component dominates, exceeding $A_{\mathrm{GAM}}$ by up to a factor of about $12$ in AUG and $6$ in TCV near the $E_r$ well. Whether this larger GAM-driven shear in \textit{unfav} can regulate turbulent transport, however, remains an open question that we leave to future work.

\subsection{Nonlinear GAM--Turbulence Coupling} \label{subsec:gamcoupling}

The amplitude of the GAM, discussed above, does not by itself indicate whether the axisymmetric oscillation exchanges energy with the turbulence. To quantify the nonlinear coupling between GAMs and turbulence, we evaluate the bispectrum of the density and $E_r$ fluctuations \cite{holland2007,ulbl2025}. Two channels separated poloidally by $3$~cm are placed at the OMP on each analysed flux-surface, from which the ion density $n_i$, its binormal derivative and $E_r$ are extracted, Fourier transformed and combined into the weighted and normalized bispectrum $\mathcal{B}(f, f^\prime)$ defined in \cite{holland2007,ulbl2025}. The ensemble average is taken over different time segments and over the $32$ toroidal planes. A GAM nonlinearly coupled to the turbulence appears in $\mathcal{B}$ as an enhancement along the sideband lines $f = f^\prime \pm f_{\mathrm{GAM}}$, which is the signature of the three-wave interactions exchanging energy between the turbulent fluctuations and the GAM oscillation. We denote by $\left| \mathcal{B} \right|_{\mathrm{sb}}$ the average of $\left| \mathcal{B} \right|$ along these two sideband (sb) lines, with $f_{\mathrm{GAM}}$ measured on the flux-surface considered. To assess the significance of this nonlinear coupling, a phase-randomized surrogate of the same signals, which by construction carries no coupling, is used as reference. Note that $\left| \mathcal{B} \right|_{\mathrm{sb}}$ measures the degree of phase locking between the GAM and the turbulence and not the direction of the energy flow, such that the present analysis establishes that the two are nonlinearly coupled without stating which of them gains energy.

\begin{figure}
\centering
\includegraphics[scale=0.65]{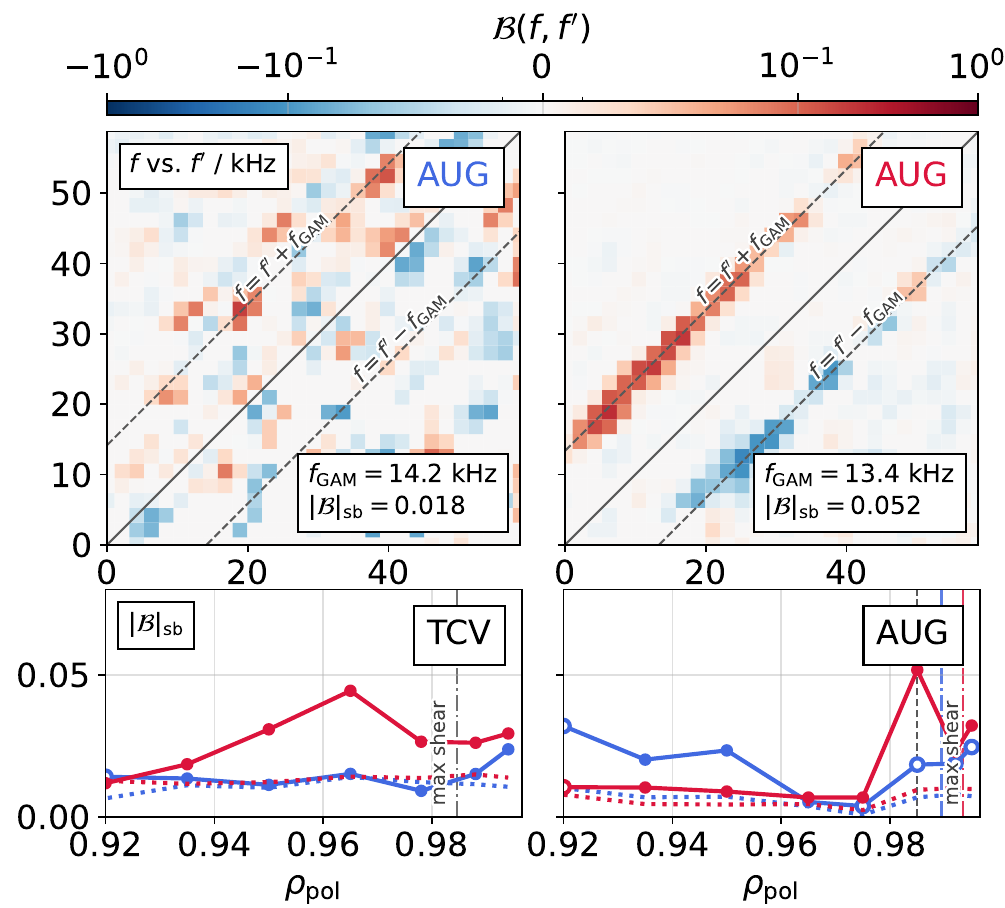}
\caption{\label{fig:figbicoh} Top: bispectrum $\mathcal{B}(f, f^\prime)$ at the OMP on the $\rho_{\mathrm{pol}} = 0.985$ flux-surface in AUG, in \textit{fav} (left) and \textit{unfav} (right), with $f = f^\prime$ (solid lines) and the GAM sidebands (dashed lines). Bottom: radial profiles of the sideband transfer $\left| \mathcal{B} \right|_{\mathrm{sb}}$ in TCV (left) and AUG (right), for \textit{fav} (blue) and \textit{unfav} (red), with their phase-randomized surrogates (dotted lines) and open markers where the measured $f_{\mathrm{GAM}}$ differs from the run value used for the surrogate. The vertical dashed line in AUG marks the flux-surface of the top row, the dash-dotted lines the maximum $E_r$ shear, in grey where the two configurations coincide.}
\end{figure}

Figure~\ref{fig:figbicoh} displays the bispectrum obtained in AUG and the radial profiles of $\left| \mathcal{B} \right|_{\mathrm{sb}}$ obtained in both AUG and TCV. First, in TCV, $\left| \mathcal{B} \right|_{\mathrm{sb}}$ remains at the level of its surrogate across the $E_r$ well region in \textit{fav}, while in \textit{unfav} it exceeds the surrogate from $\rho_{\mathrm{pol}} = 0.935$ outwards and peaks at $\rho_{\mathrm{pol}} = 0.965$, where it is $3$ times larger than the \textit{fav} value on the same flux-surface. This indicates a stronger nonlinear GAM--turbulence coupling in \textit{unfav}.
Second, AUG shows the coupling to be localized where the GAM is radially coherent (see Fig.~\ref{fig:fig7}). Indeed, in \textit{unfav} the GAM is a narrow band at $f_{\mathrm{GAM}} \simeq 13.4$~kHz that extends radially from $\rho_{\mathrm{pol}} \simeq 0.97$ outwards, while in \textit{fav} it lies further inside. Similar to TCV, $\left| \mathcal{B} \right|_{\mathrm{sb}}$ is the largest in \textit{unfav} approaching the $E_r$ well, where it reaches $0.052$ at $\rho_{\mathrm{pol}} = 0.985$ against $0.014$ in \textit{fav}.

The localization of significant GAM-mediated energy transfer does not follow the mean $\bm E \times \bm B$ shear. Indeed, $\left| \mathcal{B} \right|_{\mathrm{sb}}$ decreases at the flux-surface of maximal $E_r$ shear (indicated in Fig.~\ref{fig:figbicoh}) and peaks instead radially inside it, at $\rho_{\mathrm{pol}} = 0.965$ in TCV and $0.985$ in AUG, reaching a minimum at the $E_r$ well and increasing towards the separatrix. This profile is consistent with the reduced $A_{\mathrm{GAM}}$ inside the \textit{fav} $E_r$ well. We remark, however, that $A_{\mathrm{GAM}}$ measures the GAM amplitude and $\left| \mathcal{B} \right|_{\mathrm{sb}}$ its phase coupling to the turbulence, such that even a weak GAM may be strongly coupled.

\subsection{Toroidal and Poloidal Flows} \label{subsec:flows}

The decomposition of $E_r$ reported in Section~\ref{subsec:forcebalance} reveals the role of flows, and in particular of the poloidal flow, in the asymmetry between \textit{fav} and \textit{unfav}. In the present simulations, the toroidal and poloidal flows arise self-consistently from the balance between damping (by NC viscosity and collisional friction) and nonlinear interaction with turbulence.

\begin{figure}
\centering
\includegraphics[scale=0.55]{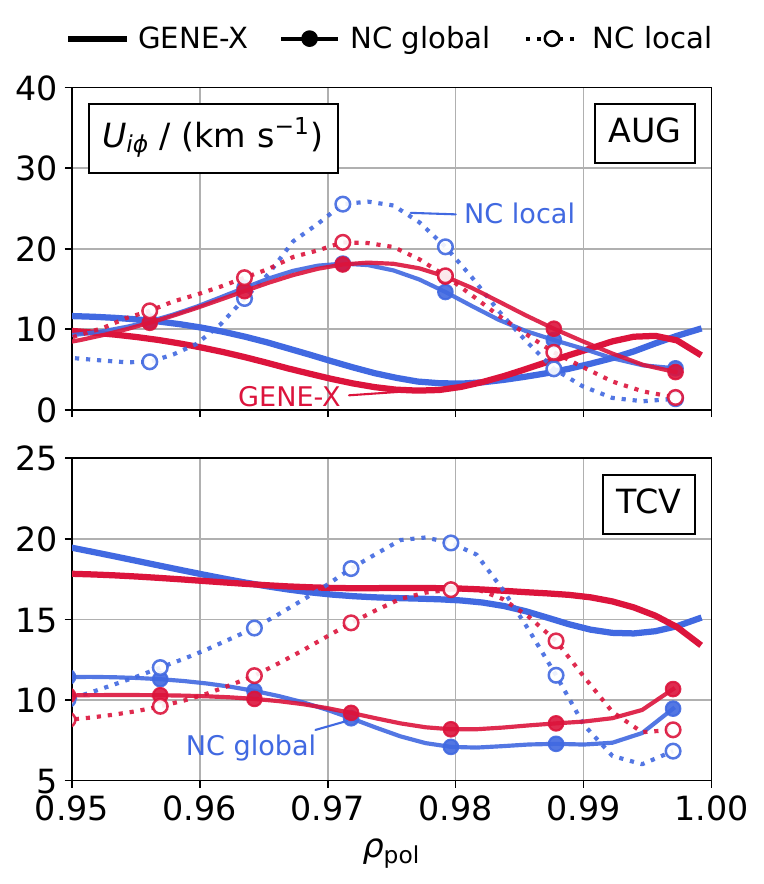}
\caption{\label{fig:fig8} FSA profiles of the toroidal main ion velocity $U_{i \phi}$ obtained from \texttt{GENE-X} (thick lines) for AUG (top) and TCV (bottom) for \textit{fav} (blue) and \textit{unfav} (red). NC predictions of the parallel main ion flow, $\fs{V_{\parallel i} B} / B_0$, from \texttt{PERFECT} \cite{landreman2014} with the diamagnetic $E_r$ prescription, are shown for comparison by the markers for radially global (filled circles) and radially local (open circles) calculations. To facilitate comparison between the two configurations, the velocities are multiplied by $\sigma_B = \mathrm{sign}(B_\phi)$, such that positive values of $U_{i \phi}$ indicate a toroidal flow in the direction of the toroidal magnetic field (see Appendix~\ref{appendix:signs}). The \texttt{GENE-X} data are averaged over $0.2$~ms.}
\end{figure}

The OMP and FSA profiles of $U_{i \phi}$ obtained from the AUG and TCV simulations are shown in Fig.~\ref{fig:fig8}. In both machines, the profiles of $U_{i \phi}$ are very similar in the edge and in the SOL between \textit{fav} and \textit{unfav}. As there is no external momentum source (e.g. from neutral beam injection, NBI), the toroidal velocity is set by turbulent and neoclassical transport, the parallel force, and collisional friction. Parallel momentum can also be generated directly by turbulence, through the residual stress \cite{diamond2008} and the parallel nonlinearity \cite{mcdevitt2009}. The similar $U_{i \phi}$ between \textit{fav} and \textit{unfav} therefore shows that these drives are not the origin of the asymmetry. This is consistent with the AUG and TCV measurements, where the edge carbon toroidal rotation shows no systematic dependence on the drift configuration \cite{plank2023,rienacker2026}. We remark that the SOL profiles are expected to be sensitive to the sheath boundary conditions at the divertor targets, which are set to zero in the present setup. More realistic conditions could change the SOL $U_{i \phi}$ profiles and the edge rotation at the separatrix.

\begin{figure}
\centering
\includegraphics[scale=0.55]{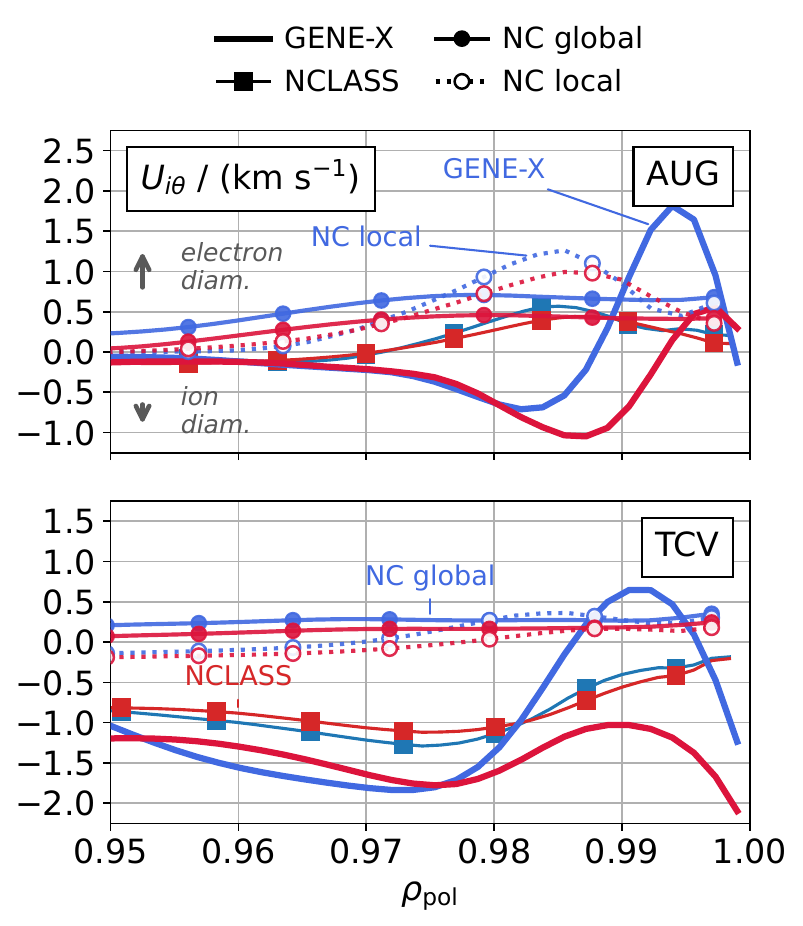}
\caption{\label{fig:fig9} FSA profiles of the poloidal main ion velocity $U_{i \theta}$ obtained from \texttt{GENE-X} (thick lines) for AUG (top) and TCV (bottom), for \textit{fav} (blue) and \textit{unfav} (red). NC estimates are shown for comparison: \texttt{NCLASS} \cite{houlberg1997} (square markers) and \texttt{PERFECT} \cite{landreman2014} with the diamagnetic $E_r$, both radially global (filled circles) and radially local (open circles). To facilitate comparison between the two configurations, the velocities are multiplied by $\sigma_B = \mathrm{sign}(B_\phi)$, such that positive (negative) values of $U_{i \theta}$ indicate the electron (ion) diamagnetic direction (see Appendix~\ref{appendix:signs}). The \texttt{GENE-X} data are averaged over $0.2$~ms.}
\end{figure}

Contrary to the toroidal velocity, the FSA poloidal velocity ($U_{i \theta}$) profiles in Fig.~\ref{fig:fig9} show a noticeable asymmetry near the separatrix, around the $E_r$ well: it is much more pronounced in \textit{fav} than in \textit{unfav}, consistent with the decomposition of $E_{r,\textrm{min}}$ shown in Section~\ref{subsec:forcebalance}. This asymmetry is also present at the OMP, but is not shown for simplicity. Away from the separatrix, $U_{i \theta}$ is in the ion diamagnetic direction and of the order of $-0.5$~km\,s$^{-1}$ ($-1$~km\,s$^{-1}$) in AUG (TCV). Close to the separatrix, the flows deviate strongly. In AUG, both the \textit{fav} and \textit{unfav} $U_{i \theta}$ flip sign to the electron diamagnetic direction, with $U_{i \theta}$ much larger in \textit{fav}, reaching $+1.5$~km\,s$^{-1}$. In TCV, $U_{i \theta}$ stays negative in \textit{unfav}, while it turns to the electron diamagnetic direction in \textit{fav}.

To understand the nature of these flows, we compare the self-consistent \texttt{GENE-X} toroidal and poloidal velocities with dedicated NC calculations. In particular, we use the drift-kinetic (DK) solver \texttt{PERFECT} \cite{landreman2014}, which computes the NC ion flows by solving the radially global $\delta f$ DK equation with the linearised Fokker-Planck collision operator, as well as the reduced NC model \texttt{NCLASS} \cite{houlberg1997}, which solves a moment hierarchy equation. The \texttt{PERFECT} code retains the effects of finite orbit width (FOW), which are important in the edge region where the equilibrium pressure gradient scale length can become comparable to the ion poloidal Larmor radius. To assess the role of FOW effects, we also perform \texttt{PERFECT} simulations in the local limit neglecting the coupling between neighbouring flux-surfaces. In all cases, we use a circular geometry and the \texttt{GENE-X} equilibrium profiles as input in both codes. In \texttt{NCLASS}, we compute the poloidal flow assuming $E_r \simeq 0$, consistent with the lowest-order NC theory, while in \texttt{PERFECT} we prescribe the diamagnetic $E_r$, so that the $\bm E \times \bm B$ contribution to the flow is retained.

The NC predictions from \texttt{PERFECT} (global and local) are shown in Fig.~\ref{fig:fig8}. While both global and local NC predictions agree in AUG, indicating weak FOW effects, they deviate in TCV. More precisely, in AUG, the NC predictions yield a non-monotonic radial dependence of $U_{i\phi}$ opposite to the one observed in \texttt{GENE-X}. On the other hand, in TCV, only the global calculation reproduces the nearly consistent $U_{i\phi}$ profiles, the deviations being related to the higher collisionality than in AUG. Although the comparison between global and local calculations indicates that FOW might play a role in the shape of $U_{i\phi}$ near the edge, the absolute values observed in \texttt{GENE-X} cannot be explained by the NC predictions, indicating the possible presence of other non-NC mechanisms. We nonetheless remark that the weak difference between the \textit{fav} and \textit{unfav} results is consistent with the NC calculations. Finally, we note that the approximation $U_{i \phi} \simeq U_{i \parallel}$ yields a variation of only about $10 \%$ in the results.

The self-consistent $U_{i \theta}$, on the other hand, deviates from the NC predictions towards the separatrix. At the lower AUG collisionality, both the NC predictions and \texttt{GENE-X} yield a negligible poloidal flow away from the separatrix. While the NC predictions correctly yield a poloidal flow in the electron diamagnetic direction towards the separatrix, their amplitude remains well below the \textit{fav} \texttt{GENE-X} results, $U_{i \theta}$ reaching $+1.5$~km\,s$^{-1}$ in AUG \textit{fav} against only $0.5$~km\,s$^{-1}$ from the NC calculations, whereas in \textit{unfav} it remains closer to the NC prediction (also around $0.5$~km\,s$^{-1}$). Finally, we remark that both \texttt{NCLASS} and \texttt{PERFECT} (local and global) yield similar results for the poloidal flows, indicating weak effects from FOW and negligible effects from deviations in the NC models.

At the higher TCV collisionality, deviations between \texttt{NCLASS} and \texttt{PERFECT} are instead clearly visible. While \texttt{PERFECT} (global and local) yields small $U_{i \theta}$ in the edge near $\rho_{\mathrm{pol}} \lesssim 0.96$, the results from \texttt{GENE-X} are quantitatively similar to \texttt{NCLASS}, predicting $U_{i \theta}$ near $-1$~km\,s$^{-1}$. The decrease of $U_{i \theta}$ in amplitude towards the edge is also quantitatively captured by \texttt{NCLASS}. However, the latter predicts $U_{i \theta} \simeq -0.5$~km\,s$^{-1}$ near $\rho_{\mathrm{pol}} \lesssim 0.99$, while \texttt{GENE-X} produces an electron diamagnetic direction poloidal flow of about $U_{i \theta} \simeq + 0.5$~km\,s$^{-1}$. On the other hand, all the \texttt{PERFECT} results yield positive $U_{i \theta}$, but smaller in amplitude than \texttt{GENE-X}.

Finally, to investigate the effect of these poloidal-flow deviations on $E_r$, we reconstruct $E_r$ from the radial force balance, \eqnref{eq:radialforcebalance}, using the \texttt{NCLASS} poloidal rotation (instead of the self-consistent $U_{i \theta}$) together with the \texttt{GENE-X} diamagnetic and toroidal contributions. This is shown in Fig.~\ref{fig:fig4}. Replacing the self-consistent $U_{i \theta}$ by the NC prediction removes the asymmetry in the $E_r$ well and shifts the well inwards, where it is then dominated by the diamagnetic term. Overall, the mechanism setting $U_{i \theta}$ near the separatrix is therefore nonlinear and turbulence-driven, no clear asymmetry being seen in the toroidal flow or in the diamagnetic drive of $E_r$.

The deviations between the NC calculations and \texttt{GENE-X} cannot be attributed to FOW effects, as the comparison between global and local \texttt{PERFECT} calculations shows, the deviations between \texttt{PERFECT} and \texttt{NCLASS} indicating instead the effect of inherent differences between the NC models, in particular in TCV. We note that one possible explanation for the difference between the NC calculations and \texttt{GENE-X} is the use of the LBD collision operator in the latter. Indeed, although the thermal relaxation rate of the LBD collision operator is tuned to match the one of the linearized Fokker-Planck operator, the NC damping of the parallel flow can be severely overestimated, which can in turn modify $U_{i \theta}$. More dedicated NC calculations to assess these deviations are the subject of future publications. The present comparison therefore remains qualitative, but none of the NC models, which rely on the equilibrium gradients only, retrieves the asymmetry in $U_{i \theta}$, pointing to non-NC effects such as its nonlinear drive by turbulence \cite{plank2023,rienacker2026}.

\subsection{Axisymmetric Simulations} \label{subsec:n0}

To assess directly the role of turbulence in setting the $E_r$ well asymmetry, we perform a dedicated axisymmetric numerical experiment in TCV. More precisely, we restart the \textit{fav} and \textit{unfav} TCV (less computationally demanding than AUG, see Table~\ref{tab:2}) simulations from their saturated turbulent states, and we apply at each time step a toroidal average (axisymmetric filter) on the electromagnetic fields, such that $\phi_1 \to \left< \phi_1 \right>_\phi = \int_0^{2 \pi} d \phi \phi_1 / (2 \pi)$ (and similarly for $A_{\parallel 1}$). This filter removes all $n \neq 0$ components of $\phi_1$, and hence the turbulence. The simulations are continued for approximately $0.5$~ms in both configurations. The axisymmetric simulations retain the NC dynamics as well as the GAM and zonal flow dynamics, the latter being $n = 0$ modes, thus enabling us to decouple the contributions from turbulence and from the competition between NC effects and magnetic geometry in driving the \textit{fav} and \textit{unfav} asymmetry. We however do not attempt to simulate a true NC equilibrium, since it would require a simulation time of the order of several ion collisional time scales (see Fig.~\ref{fig:fig10}).

\begin{figure}
\centering
\includegraphics[scale=0.55]{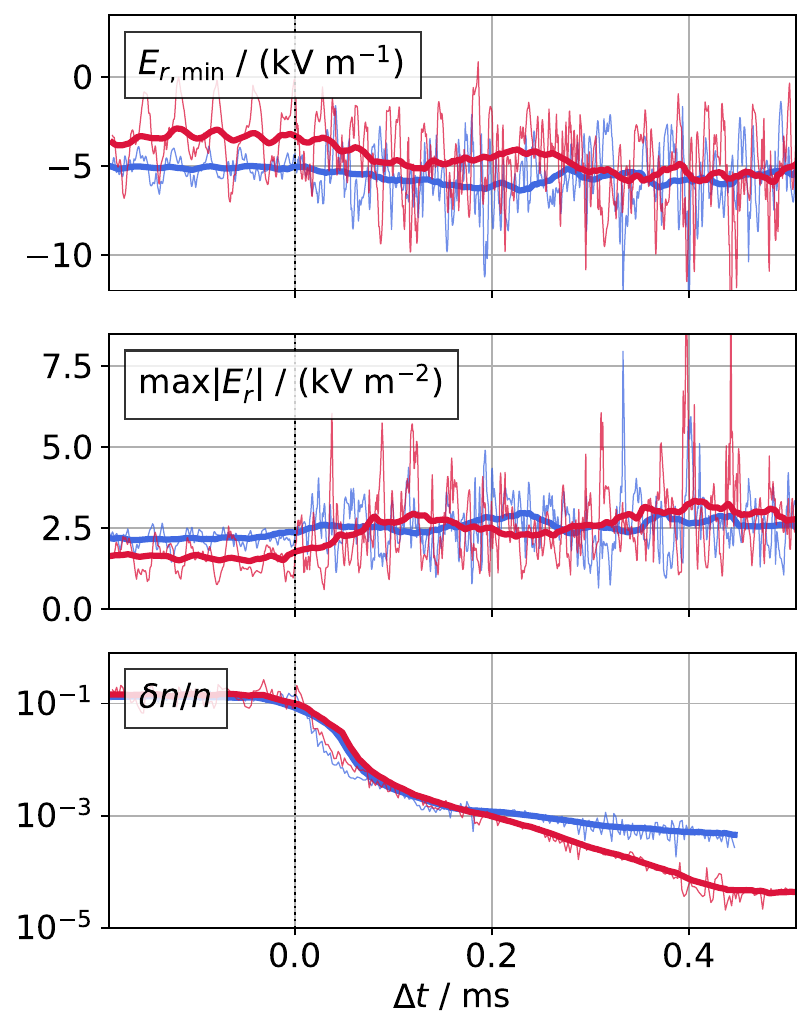}
\caption{\label{fig:fign0} Time traces across the removal of the $n \neq 0$ toroidal mode numbers for \textit{fav} and \textit{unfav} in TCV, against $\Delta t = t - t_{\mathrm{filter}}$. The dotted vertical black line marks the time $t_{\mathrm{filter}}$ at which the axisymmetric filter is applied. From top to bottom: the depth of the $E_r$ well, $E_{r, \mathrm{min}}$, and the maximum $E_r$ shear, $\max \left| E_r^\prime \right|$ with $E_r^\prime = d_r E_r $ (both measured at the OMP over $0.96 < \rho_{\mathrm{pol}} < 1.01$) and the normalized density fluctuation $\delta n / n$ (with $\delta n$ associated with the toroidal average) at the location of $E_{r, \mathrm{min}}$. The thin coloured lines are the instantaneous time traces and thick lines are their $93$~$\mu$s window average.}
\end{figure}

The time traces of $E_{r, \mathrm{min}}$, of the maximum $E_r$ shear across the well and of the normalized density fluctuation level evaluated at the location of $E_{r, \mathrm{min}}$ are shown at the OMP in Fig.~\ref{fig:fign0}, before the axisymmetric filter is applied ($\Delta t < 0$) and after ($\Delta t > 0$). First of all, when $\Delta t < 0$, the \textit{fav} and \textit{unfav} asymmetry is clearly visible, with a deeper $E_r$ well and a larger $E_r$ shear in \textit{fav}. Once the filter is applied, the turbulence decays rapidly, the fluctuation level dropping by more than two orders of magnitude within about $0.1$~ms in both configurations. Consistently, the Fourier spectrum of $\phi_1$ (not shown here) loses several orders of magnitude at the scales carrying the turbulent transport (see Section~\ref{subsec:fourier}), while the long-wavelength axisymmetric structures (e.g. poloidal sidebands) remain essentially unaffected. In such a case, the evolution of the mean (FSA) $E_r$ is then set by the NC viscous damping (see \eqnref{eq:vort5}).

The character of the $E_{r, \mathrm{min}}$ time traces also changes once the filter is applied. While in the presence of turbulence the oscillations of $E_{r, \mathrm{min}}$ are considerably larger in \textit{unfav} than in \textit{fav}, consistently with the GAM amplitudes reported in Section~\ref{subsec:gam}, both configurations develop oscillations of comparable amplitude for $\Delta t > 0$, reaching a sizeable fraction of $E_{r, \mathrm{min}}$ itself. These oscillations are associated with GAM relaxation coupled with higher order harmonics. This suggests that the turbulence contributes to the asymmetry observed in the amplitude of the $E_r$ oscillations associated with GAMs. 

Most importantly, the asymmetry in $E_{r, \mathrm{min}}$ that is clearly visible in Fig.~\ref{fig:fig5} when turbulence is present is progressively reduced once the axisymmetric filter is applied. For $\Delta t \gtrsim 0.3$~ms, the mean $E_{r, \mathrm{min}}$ settles at comparable values in \textit{fav} and \textit{unfav}, the residual difference being within the uncertainty associated with the large $E_r$ oscillations. Removing the turbulence therefore removes the asymmetry in the depth of the $E_r$ well, on the same time scale as the decay of the fluctuations. The same holds for the maximum $E_r$ shear, larger in \textit{fav} in the presence of turbulence, the \textit{unfav} shear rising to the \textit{fav} level for $\Delta t > 0.3$~ms.

We finally emphasize that removing the turbulent transport yields the relaxation of the equilibrium profiles. As a consequence, the contribution of the diamagnetic term in the radial force balance, \eqnref{eq:radialforcebalance}, grows during the axisymmetric simulations. This affects the absolute value of $E_{r, \mathrm{min}}$ in both configurations, but, since both the \textit{fav} and \textit{unfav} simulations relax in the same manner, it affects much more weakly the difference between them, which is the quantity of interest here. A controlled determination of the relaxation of $U_{i \theta}$ towards its NC value would require fixing the equilibrium profiles during the axisymmetric simulations and following the poloidal flow relaxation over several NC damping times. This is left to future work.

\section{Turbulence Characterization} \label{sec:turbulence}

Before analyzing the mechanism underlying the poloidal flow asymmetry, we characterize the turbulence properties in the edge of AUG and TCV: the flux-surface Fourier analysis of the nonlinear spectra is reported in Section~\ref{subsec:fourier}, the statistics of the fluctuations at the OMP are analyzed in Section~\ref{subsec:pdf}, the linear instabilities driving the turbulent transport are identified in Section~\ref{subsec:linear} and, finally, the radial density correlation length is computed in Section~\ref{subsec:corrlength}. The local plasma parameters extracted from the simulations and the collisionality and $\beta_e$ scans of the linear growth rates are reported in Appendix~\ref{appendix:localparameters}.

\subsection{Nonlinear Flux-Surface Fourier Analysis} \label{subsec:fourier}

\begin{figure}[h]
\centering
\includegraphics[scale=0.5]{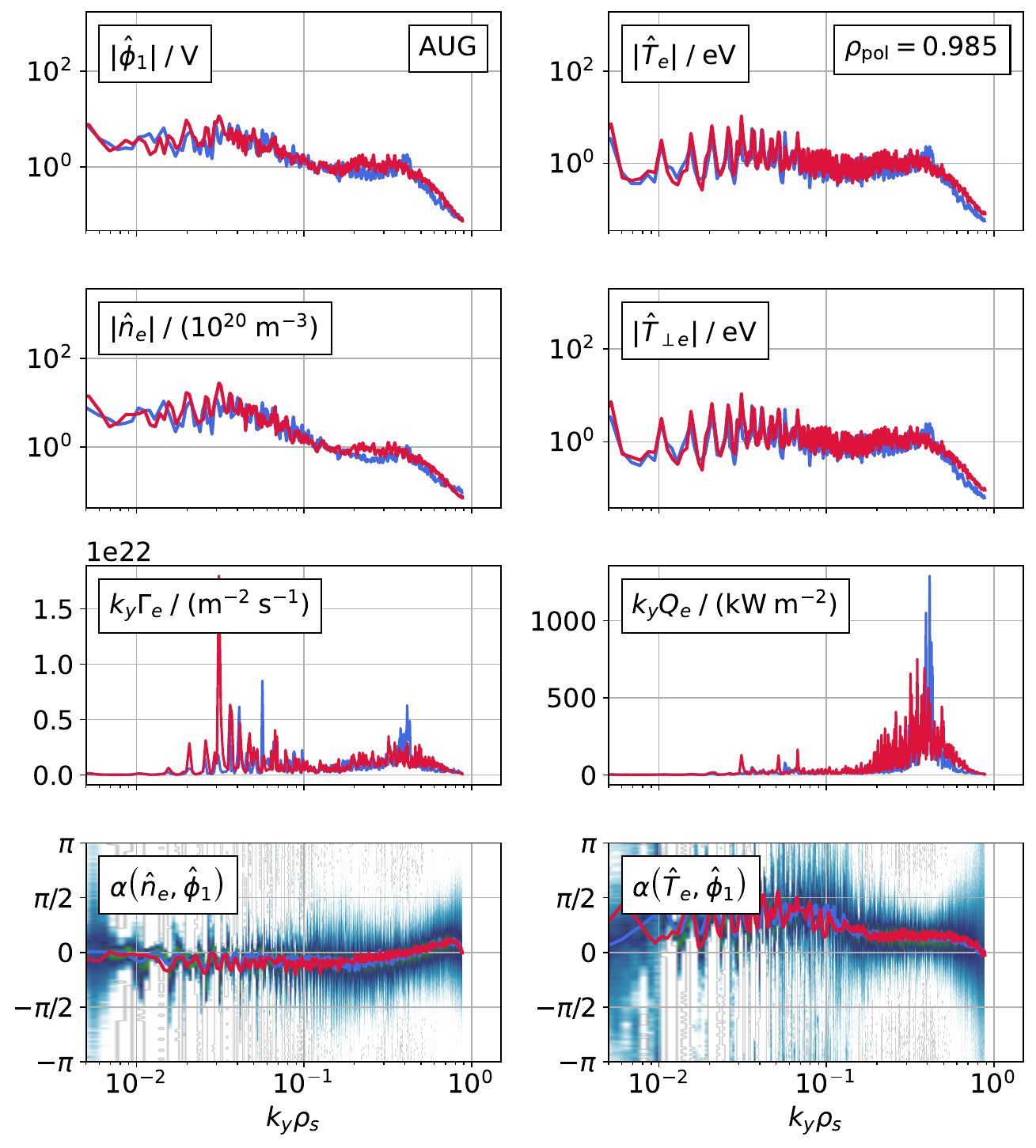}
\caption{\label{fig:fig11} Fourier spectra at $\rho_{\mathrm{pol}} = 0.985$ in AUG, for \textit{fav} (blue) and \textit{unfav} (red). From top to bottom: amplitude spectra of the electrostatic potential $\hat \phi_1$ and of the total electron temperature $\hat T_e$; of the electron density $\hat n_e$ and of the perpendicular electron temperature $\hat T_{\perp e}$; spectra of the electron particle and heat fluxes, $k_y \Gamma_e$ and $k_y Q_e$; and cross-phases $\alpha(\hat n_e, \hat \phi_1)$ and $\alpha(\hat T_e, \hat \phi_1)$, the coloured lines being the mean cross-phase and the shading its distribution.}
\end{figure}

\begin{figure}[h]
\centering
\includegraphics[scale=0.5]{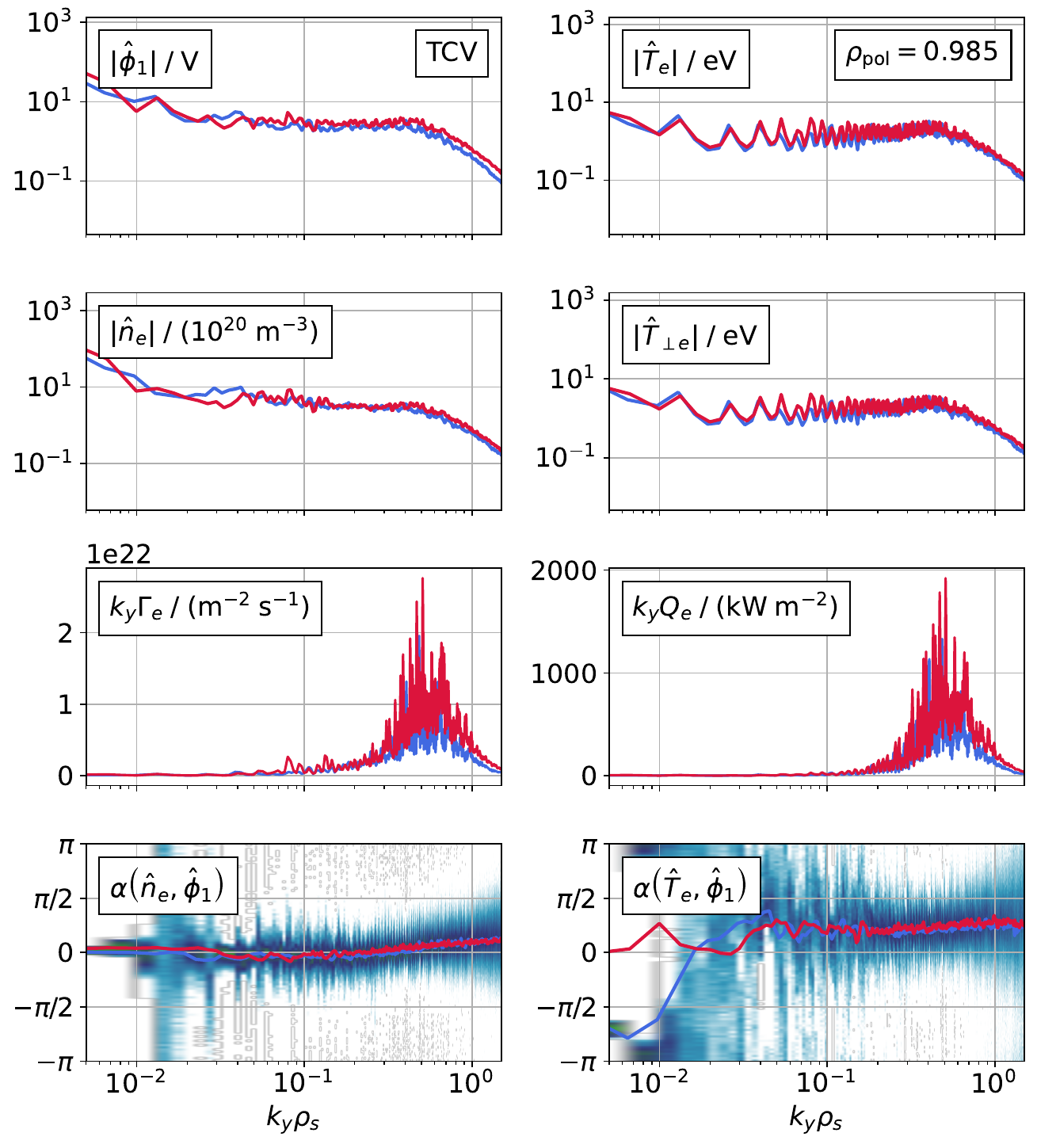}
\caption{\label{fig:fig12} Same as Fig.~\ref{fig:fig11} but for TCV.}
\end{figure}

We now perform a Fourier analysis of the nonlinear spectra on a given flux-surface following the procedure outlined in \cite{ulbl2022,frei2025vspec}. Figs.~\ref{fig:fig11} and \ref{fig:fig12} show the amplitude spectra of the electrostatic potential $\phi_1$, density $n_e$, the total and perpendicular electron temperatures, $T_e$ and $T_{\perp e}$, the electron particle and heat fluxes, $k_y \Gamma_e$ and $k_y Q_e$, as well as the cross-phases $\alpha(\hat n_e, \hat \phi_1)$ and $\alpha(\hat T_e, \hat \phi_1)$, as functions of the binormal wavenumber $k_y \rho_s$ on the $\rho_{\mathrm{pol}} = 0.985$ flux-surface. We observe that the amplitude spectra are similar between \textit{fav} and \textit{unfav} in both devices, the same being true of the cross-phases, which remain small, $\left| \alpha \right| \lesssim \pi / 4$, over the whole spectral range. In AUG, the particle flux, $k_y \Gamma_e$, is carried both by a low wavenumber component peaking at $k_y \rho_s \simeq 0.03$ and by a broad band around $k_y \rho_s \simeq 0.4$, while the heat flux is carried almost entirely by the latter. On the other hand, a single broad band centred on $k_y \rho_s \simeq 0.5$ dominates both fluxes in TCV. Consistently with the larger fluctuation levels reported in Figs.~\ref{fig:fig2} and \ref{fig:fig3}, the TCV flux spectra are larger in \textit{unfav}, by up to a factor of two at the peak, consistent with the larger transport measured in \textit{unfav} (see Section~\ref{subsec:particleandenergy}). In AUG, the \textit{unfav} spectra are only slightly higher at small $k_y$, consistent with the similar transport. The similar shape of the amplitude spectra and of the cross-phases indicates that the same instability drives the turbulence in \textit{fav} and \textit{unfav}, as confirmed by the linear analysis below.

\subsection{Fluctuation Statistics} \label{subsec:pdf}

To assess whether the drift configuration affects the character of the turbulence beyond its level (Figs.~\ref{fig:fig2} and \ref{fig:fig3}), we compute the probability density functions (PDFs) of the non-axisymmetric fluctuations at the OMP. More precisely, the relative fluctuation $\tilde f / \langle f \rangle_\varphi = (f - \langle f \rangle_\varphi)/\langle f \rangle_\varphi$, with $\langle f \rangle_\varphi$ the toroidal average taken at each time, such that the axisymmetric component (including the GAM) is removed, is accumulated over the QSS window and all poloidal planes, from which the radial profiles of its standard deviation $\sigma$, skewness $S$ and kurtosis $K$ are evaluated. We focus on the electron density, i.e. $f = n_e$. 

\begin{figure}
\centering
\includegraphics[scale=0.5]{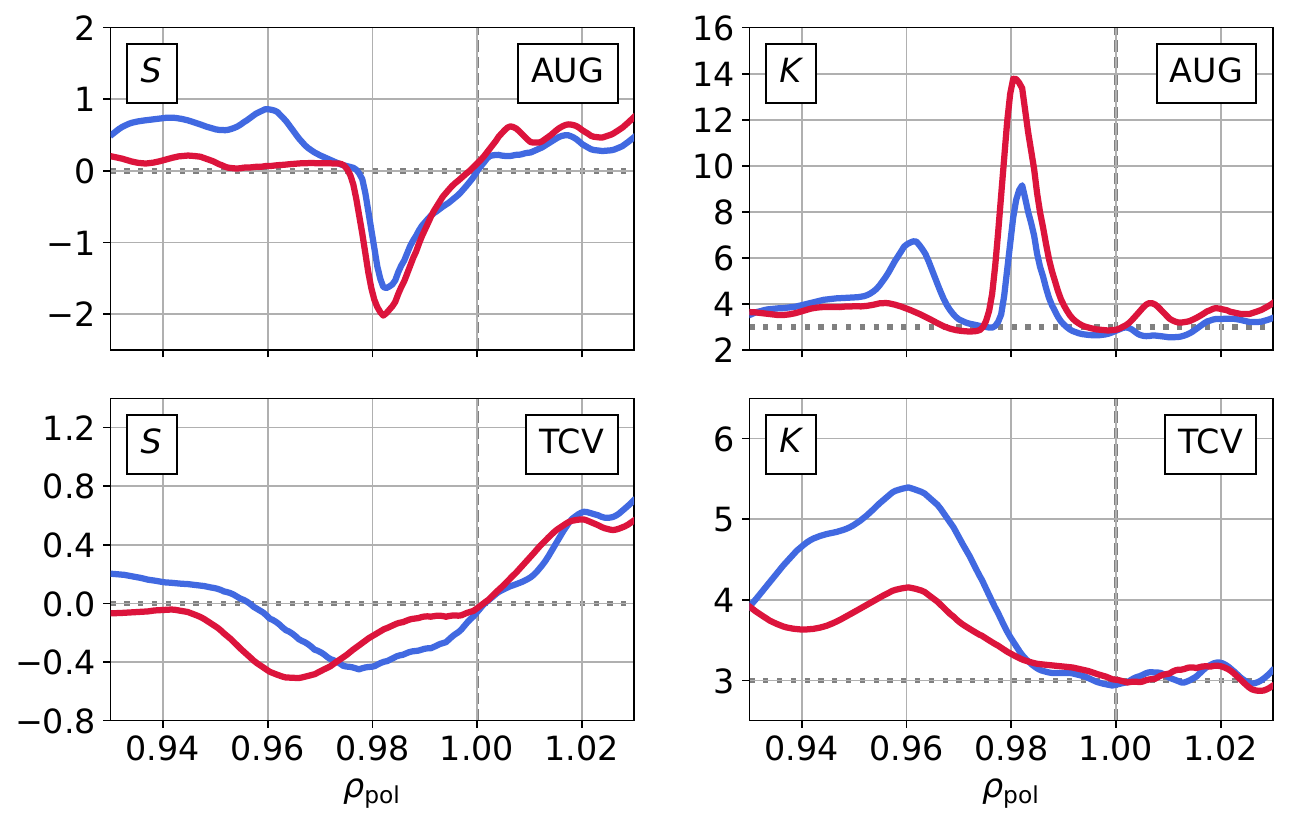}
\caption{\label{fig:skewkurt} Radial profiles of the skewness $S$ (left) and kurtosis $K$ (right) of the electron density fluctuation $\tilde n_e / \langle n_e \rangle_\varphi$ at the OMP, for AUG (top) and TCV (bottom), in \textit{fav} and \textit{unfav}. The profiles are obtained over a time window of approximately $1$~ms in AUG and $0.5$~ms in TCV. Note that $S = 0$ and $K = 3$ for a Gaussian distribution (horizontal dotted lines). The vertical grey dashed line marks the separatrix.}
\end{figure}

We find that the \textit{fav} and \textit{unfav} asymmetry lies in the amplitude of the fluctuations rather than in the shape of their distribution.  Indeed, as observed in Figs.~\ref{fig:fig2} and \ref{fig:fig3}, the amplitude is larger in \textit{unfav}, in particular in TCV, where the electron density fluctuation reaches $\sigma = 11.4\%$ against $7.4\%$ in \textit{fav} at $\rho_{\mathrm{pol}} = 0.97$, and weakly in AUG, where the difference depends on the radial position ($14.8\%$ against $11.1\%$ at the separatrix). On the other hand, once normalized to $\sigma$, the PDFs are nearly identical in TCV. Moreover, the radial profiles of $S$ and $K$, shown in Fig.~\ref{fig:skewkurt} for the electron density fluctuations, are similar between the two configurations. In all cases, $S$ changes sign at the separatrix, from negative inside (density depletions) to positive in the SOL (density excursions), while $K$ remains within about $1$ of its Gaussian value, i.e. $K = 3$, from the separatrix outwards. Larger deviations are found only inside the separatrix. In AUG, the transition from the quiescent inner region, where $\sigma \lesssim 3\%$, to the turbulent edge at $\rho_{\mathrm{pol}} \simeq 0.98$ is intermittent, $S$ and $K$ reaching $-1.6$ and $9$ in \textit{fav}, and $-2.0$ and $14$ in \textit{unfav}. In TCV, $K$ peaks at $\rho_{\mathrm{pol}} \simeq 0.96$, reaching $5.4$ in \textit{fav} against $4.2$ in \textit{unfav}. Hence, the \textit{fav} and \textit{unfav} difference in $K$ has opposite signs in the two devices, such that no systematic dependence of the shape of the PDFs on the drift configuration is found. Together with the similar nonlinear spectra and cross-phases, this indicates that the drift configuration modifies the level of the turbulence rather than its nature.

\subsection{Linear Analysis and Mode Identification} \label{subsec:linear}

We now identify the linear instabilities driving the turbulent transport whose nonlinear spectra are shown in Figs.~\ref{fig:fig11} and \ref{fig:fig12}. Since all linear modes are strongly coupled in the full-$f$ nonlinear simulations, we linearize the \texttt{GENE-X} model in a circular flux-tube geometry centred on a given flux-surface, following \cite{frei2023}. This yields a linear evolution equation for the perturbed, non-Maxwellian part of the full-$f$ spectral coefficient $\mathcal{N}_\alpha^{p j}$, retaining its parallel and perpendicular advections driven by the perturbed electromagnetic fields and background pressure gradients, and collisions through the linearized LBD operator. The linear system is integrated in time for each binormal wavenumber $k_y$ until a single mode grows exponentially. Consistent with the nonlinear runs, $6$ parallel and $4$ perpendicular spectral coefficients are retained. The analysis is performed at $\rho_{\mathrm{pol}} = 0.98$ for both AUG and TCV, with the normalized gradients, collisionality, plasma beta and other geometrical quantities taken from the local parameters of Fig.~\ref{fig:fig10}.

\begin{figure}
\centering
\includegraphics[scale=0.55]{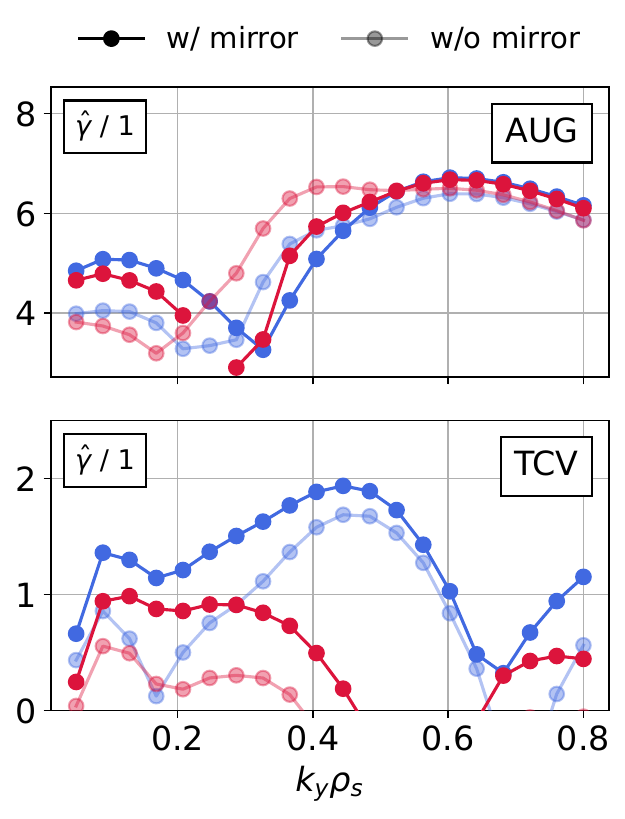}
\caption{\label{fig:fig13} Normalized (to $c_s / L_{\mathrm{ref}}$) growth rate $\hat \gamma $ as a function of the normalized binormal wavenumber $k_y \rho_s$ obtained for AUG (top) and TCV (bottom), both at $\rho_{\mathrm{pol}} = 0.98$, in \textit{fav} (blue) and \textit{unfav} (red). The opaque (faded) lines indicate $\hat \gamma$ with (without) the mirror force included.}
\end{figure}

We note that the present linear calculations ignore the effects of $\bm E \times \bm B$ shear, $\gamma_E$. The normalized growth rates, $\hat\gamma$, are shown in Fig.~\ref{fig:fig13} as a function of $k_y \rho_s$ for AUG and TCV. In AUG, the \textit{fav} and \textit{unfav} results are essentially indistinguishable, the only possible source of asymmetry in the linear calculations being a variation in the local parameters, which are nearly identical between the two configurations (see Fig.~\ref{fig:fig10}). In TCV, on the contrary, the \textit{fav} growth rate exceeds the \textit{unfav} one by about a factor of two, its maximum being displaced towards higher $k_y \rho_s$. We remark that this difference originates from the local gradients and not from the drift configuration itself: as reported in Appendix~\ref{appendix:localparameters}, the TCV \textit{fav} simulation features an electron temperature gradient $\omega_{T_e}$ about $40\%$ larger than in \textit{unfav} near $\rho_{\mathrm{pol}} = 0.98$ (see Fig.~\ref{fig:fig10}), which is precisely the drive of the DTEM identified below. We further remark that this larger linear drive in \textit{fav} is opposite to the larger fluctuation levels and turbulent fluxes measured in \textit{unfav} (Figs.~\ref{fig:fig2}, \ref{fig:fig3} and \ref{fig:fig12}), such that the differences in turbulent levels between the two configurations cannot be attributed to the background gradients driving the turbulence.

The results from AUG show two distinct peaks, at $k_y \rho_s \simeq 0.1$ and $k_y \rho_s \simeq 0.6$, both within the range of wavenumbers carrying the nonlinear fluxes of Fig.~\ref{fig:fig11}, which confirms that the $k_y \rho_s \simeq 0.6$ mode dominates the transport. In TCV, the growth rate peaks at $k_y \rho_s \simeq 0.45$ in \textit{fav}, close to the maximum of the nonlinear flux spectra of Fig.~\ref{fig:fig12}, while in \textit{unfav} the maximum is shifted to $k_y \rho_s \simeq 0.13$ and the spectrum is flatter. In all cases the real frequencies (not shown) lie in the electron diamagnetic direction and grow in magnitude with $k_y \rho_s$.

To identify the nature of these modes, we artificially disable the trapped-particle contributions in the linear equation by removing the mirror force. In TCV, removing the mirror force stabilizes the mode at all $k_y$, reducing $\hat \gamma$ by $30$--$80\%$ depending on $k_y$ and by about $50\%$ at the \textit{unfav} maximum, whereas in AUG only the long-wavelength peak is affected, the short-wavelength peak remaining unchanged. Consistent with previous linear studies, the TCV mode and the AUG long-wavelength mode are therefore TEMs, while the AUG $k_y \rho_s \sim 0.6$ mode can be identified as a resistive (collisional) eDW. Ion-temperature-gradient (ITG) modes are not expected to contribute given the weak values of $\eta_i = L_n / L_{T_i} = \omega_{T_i} / \omega_n$ (see Fig.~\ref{fig:fig10}) in both devices.

The collisionality and $\beta_e$ scans reported in Appendix~\ref{appendix:localparameters} support this identification. In TCV, the nominal collisionality lies on the dissipative branch of the TEM (DTEM), just above the minimum of $\hat \gamma$ reached near $\nu_e^* \simeq 2$--$3$, while in all cases the modes are essentially electrostatic, the KBM being destabilized only for $\beta_e$ one to two orders of magnitude above the nominal values.

We therefore conclude that, in both \textit{fav} and \textit{unfav} AUG, the turbulence is driven by an unstable resistive eDW at shorter wavelengths and a TEM at longer wavelengths, whereas in TCV it is dominated by the DTEM. Similar results are found on the other edge flux-surfaces.

\subsection{Radial Density Correlation Length} \label{subsec:corrlength}

It is well known that an $\bm E \times \bm B$ shear layer can reduce turbulent transport by shearing the turbulent eddies. This in turn reduces the radial correlation length $\ell_c$ and hence the random-walk diffusion coefficient $\sim \ell_c^2 / \tau_c$, with $\tau_c$ the eddy turnover time. Comparing $\ell_c$ between \textit{fav} and \textit{unfav} can therefore serve as a measure of how effectively the $\bm E \times \bm B$ shear $\gamma_E$ regulates turbulent transport through eddy shearing.

\begin{figure}
\centering
\includegraphics[scale=0.5]{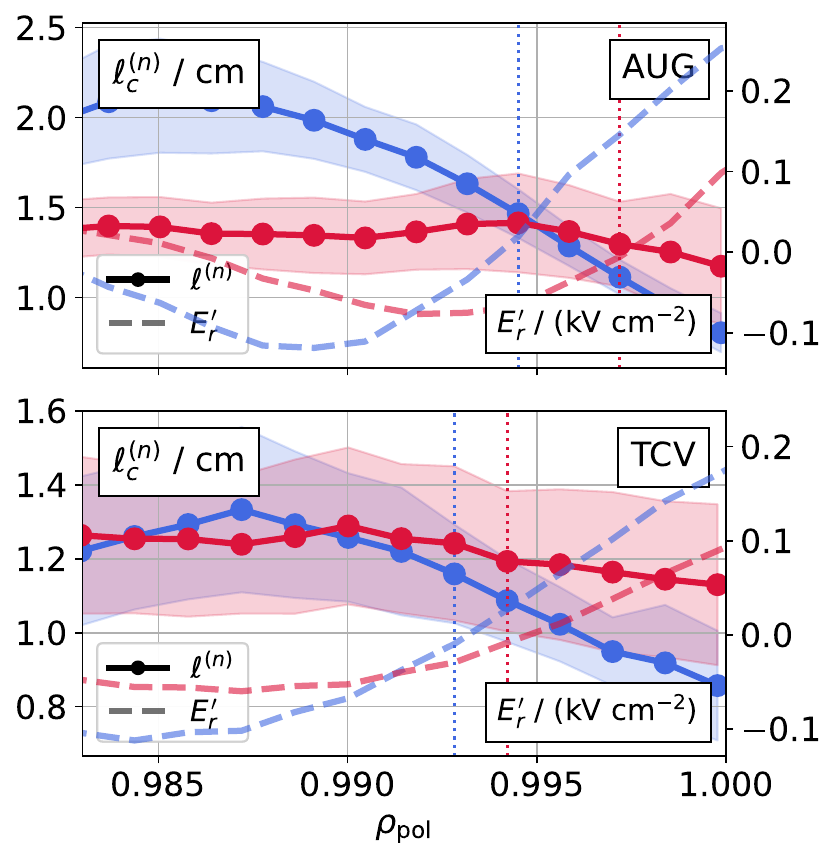}
\caption{\label{fig:fig15} Radial correlation length $\ell_c$ (solid coloured lines with markers) computed at the OMP for the AUG (top) and TCV (bottom) simulations. The shaded areas are the standard deviations associated with the time ($2$~ms) and toroidal averages. The $E_r$ shear ($E_r^\prime$) is shown on the right axis by the shaded dashed lines. The vertical dotted lines mark the transition between the inner and outer shear layers.}
\end{figure}

The correlation length $\ell_c$ measured at the OMP is shown in Fig.~\ref{fig:fig15} for both \textit{fav} and \textit{unfav} in AUG and TCV. We obtain $\ell_c$ from the $1/e$ decay of the auto-correlation of the density fluctuations ($\delta n_e$) along the OMP, averaged over time and the toroidal direction. The radial derivative of $E_r$ (proportional to $\gamma_E$) is also shown. As expected, $E_r^\prime$ is negative (positive) in the inner (outer) shear layer, passing through zero at the $E_{r,\textrm{min}}$ location. In the inner shear layer, where $E_r^\prime$ is of similar magnitude in \textit{fav} and \textit{unfav}, no clear \textit{fav}/\textit{unfav} trend in $\ell_c$ is seen: $\ell_c$ is somewhat larger in \textit{fav} than in \textit{unfav} in AUG and similar in TCV. In the outer shear layer, by contrast, $E_r^\prime$ increases and becomes larger in \textit{fav} than in \textit{unfav} by at least a factor of $2$. Correspondingly, $\ell_c$ decreases significantly across the outer shear layer in \textit{fav}, whereas no significant change is observed in \textit{unfav} between the inner and outer shear layers. This suggests that, in both AUG and TCV, $\gamma_E$ is strong enough in \textit{fav} to reduce the radial decorrelation length of the turbulent structures, while it is ineffective in \textit{unfav}. This decrease of $\ell_c$ is also consistent with the improved particle and energy confinement reported in Table~\ref{tab:1}. DBS measurements in comparable TCV Ohmic discharges yield $\ell_c \simeq 3.5$~mm near $\rho_{\mathrm{pol}} \simeq 0.96$, with no systematic difference between the two configurations \cite{rienacker2026}, consistent with the present results in the inner shear layer. Being limited to $\rho_{\mathrm{pol}} \lesssim 0.99$, these measurements do not yet test the decrease of $\ell_c$ obtained here in \textit{fav} across the outer shear layer.

\section{Mean Flow Acceleration By Turbulence} \label{sec:meanflow}

In this section, we investigate the nonlinear mechanism driving the mean (zonal) poloidal flow, which we have identified as the origin of the $E_r$ asymmetry between \textit{fav} and \textit{unfav}. We first demonstrate in Section~\ref{subsec:localvorticity} the dominant role of the electrostatic radial current in the time evolution of the vorticity $\omega$. We then establish the theory of poloidal flow acceleration within the full-$f$ formalism in Section~\ref{subsec:flowacceleration}, which identifies the total velocity stress $\Pi$ as the nonlinear drive of the poloidal flow by turbulence. Its poloidal and radial structure is analysed in Section~\ref{subsec:stressstructure}, its relation with the Reynolds stress and its decomposition in Section~\ref{subsec:velocitystress}, while the nonlinear energy transfer into the mean poloidal flow is analysed in Section~\ref{subsec:nrgtransfer}.

\subsection{Local Vorticity Dynamics Driven by the Electrostatic Current} \label{subsec:localvorticity}

\begin{figure*}
\centering
\includegraphics[scale=0.5]{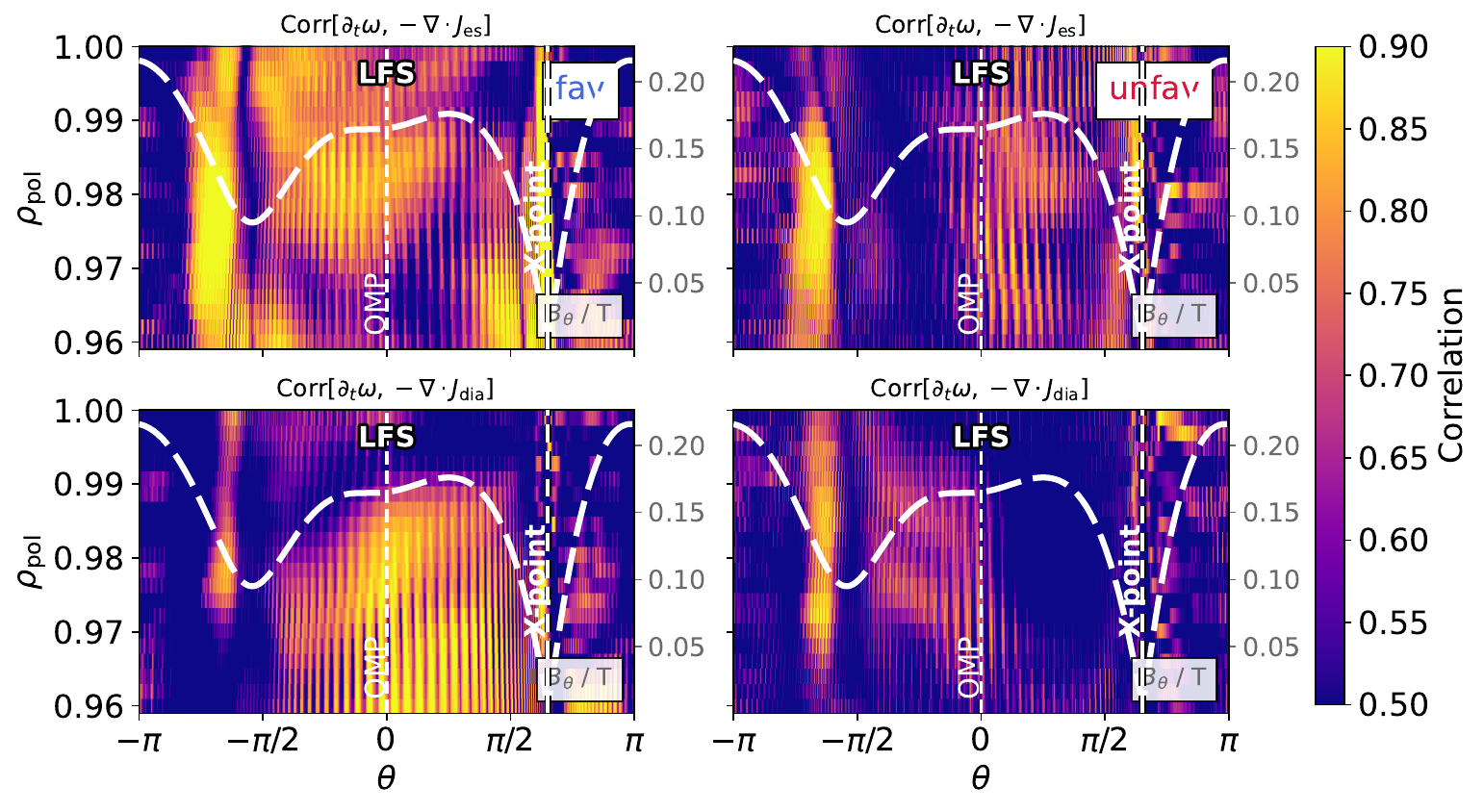}
\caption{\label{fig:fig16} Pointwise time correlation between $\partial_t \omega$ and $- \grad \cdot \bm J_\perp^{es}$ (top row) and $- \grad \cdot \bm J_\perp^{dia}$ (bottom row), see \eqnref{eq:vorticity}, as a function of the geometrical poloidal angle $\theta$ ($\theta = 0$ at the OMP and increasing towards the active X-point, see Appendix~\ref{appendix:signs}) and of $\rho_{\mathrm{pol}}$, for \textit{fav} (left) and \textit{unfav} (right) in TCV. The correlations are computed on a single toroidal plane over $0.15$~ms at QSS. The dashed white lines are the poloidal magnetic field component $B_\theta$, whose minimum locates the X-point, and LFS marks the low-field side. The vertical dashed white line at $\theta = 0$ marks the OMP position. The time traces at the OMP position ($\theta = 0$) are displayed in Fig.~\ref{fig:fig17}.}
\end{figure*}

\begin{figure*}
\centering
\includegraphics[scale=0.5]{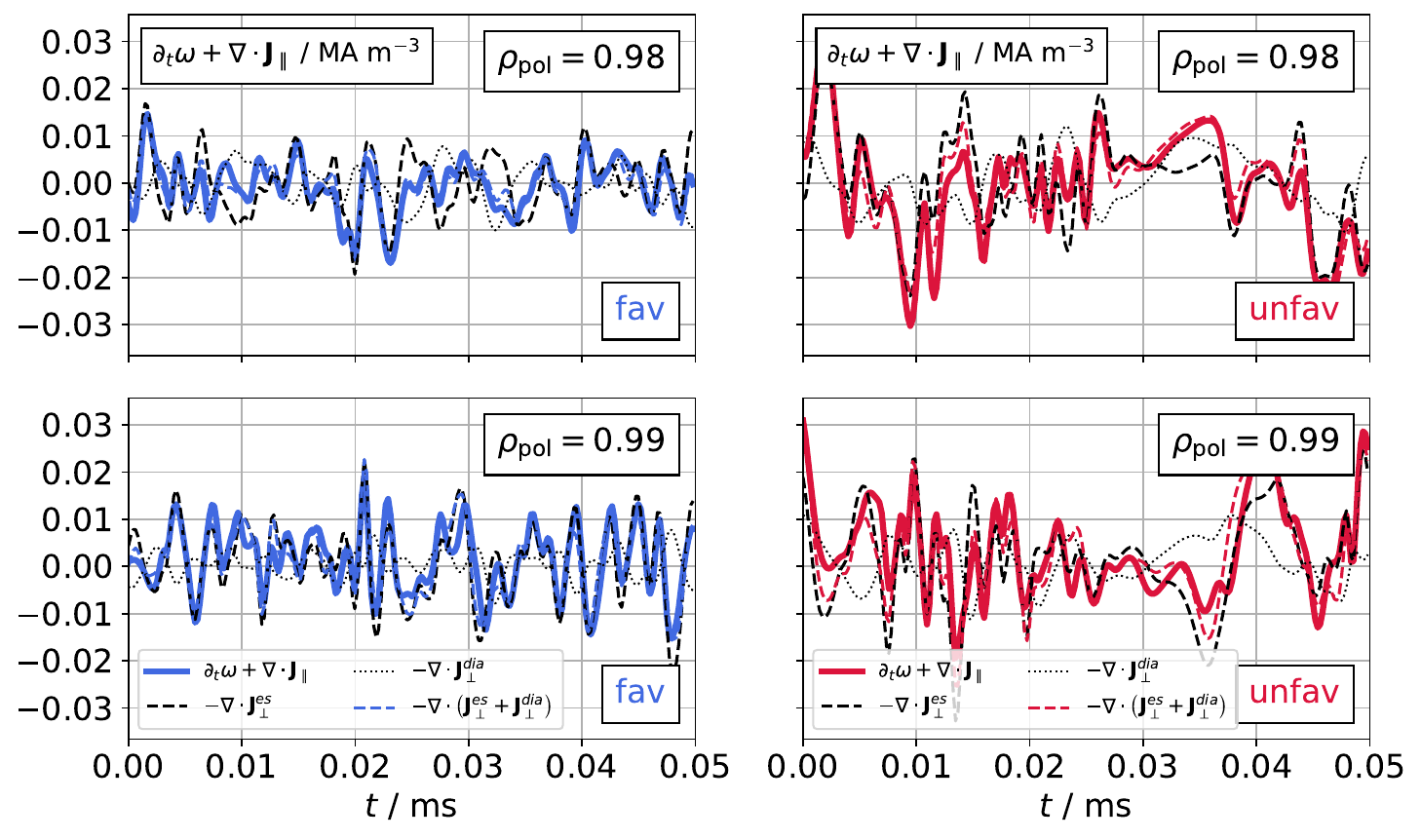}
\caption{\label{fig:fig17} Time traces of $\partial_t \omega + \grad \cdot \bm J_{\parallel}$ (thick coloured lines), $- \grad \cdot \bm J_\perp^{es}$ (dashed black lines), $- \grad \cdot \bm J_\perp^{dia}$ (dotted black lines), and their sum, $- \grad \cdot \left( \bm J_\perp^{es} + \bm J_\perp^{dia} \right)$ (dashed coloured lines), evaluated at the OMP in TCV for \textit{fav} (left column, blue) and \textit{unfav} (right column, red), outside ($\rho_{\mathrm{pol}} = 0.98$, top row) and inside ($\rho_{\mathrm{pol}} = 0.99$, bottom row) of the $E_r$ well. The traces are averaged over the toroidal angle and taken over a time window of $0.05$~ms at QSS.}
\end{figure*}

We first assess the relative importance of the electrostatic and diamagnetic perpendicular current densities entering the GK vorticity equation, \eqnref{eq:vorticity}, since they are associated with distinct physical mechanisms: the electrostatic current, $\bm J_{\perp}^{es}$, carries the nonlinear $\bm E \times \bm B$ advection of vorticity, and therefore the turbulent drive, while the diamagnetic current, $\bm J_{\perp}^{dia}$, is associated with the NC viscous damping. While the FSA of $\omega$ is related to the FSA $E_r$ (see Section~\ref{subsec:flowacceleration}), its \textit{local} evolution reflects the tilting of the turbulent structures, such that resolving the perpendicular currents poloidally locates where vorticity is generated and nonlinearly transported. We therefore compute the pointwise time correlation between $\partial_t \omega$ and the negative divergence of $\bm J_{\perp}^{es}$ and of $\bm J_{\perp}^{dia}$, evaluated on a single toroidal plane over a time window of $0.15$~ms at QSS. The poloidally and radially resolved correlations are shown in Fig.~\ref{fig:fig16} for \textit{fav} and \textit{unfav} in TCV. We remark that the parallel current density, $\bm J_\parallel$, although being the largest contribution to the local evolution of $\omega$, does not enter the FSA balance considered below, the FSA of its divergence vanishing thanks to $\bm b \cdot \grad \psi = 0$. For this reason, $\partial_t \omega$ is correlated with the perpendicular currents alone in Fig.~\ref{fig:fig16}, while $\grad \cdot \bm J_\parallel$ is subtracted from $\partial_t \omega$ in the time traces of Fig.~\ref{fig:fig17}, such that the latter isolates the perpendicular contributions and yields the lower correlation values reported below.

In \textit{fav}, the local dynamics of $\omega$ is dominated by $\grad \cdot \bm J_{\perp}^{es}$ near the separatrix. Indeed, the correlation between $\partial_t \omega$ and $- \grad \cdot \bm J_{\perp}^{es}$ reaches $0.9$ on the low-field side (LFS) in the region of the $E_r$ well ($\rho_{\mathrm{pol}} \simeq 0.995$) and is even larger close to the X-point, while the correlation with $- \grad \cdot \bm J_{\perp}^{dia}$ remains below $0.5$ there. We remark that, when all the currents are considered, the vorticity equation, \eqnref{eq:vorticity}, is verified numerically, i.e. the time correlation is unity. Since $\grad \cdot \bm J_{\perp}^{es}$ is a measure of the nonlinear interaction with turbulence, this indicates that the local vorticity dynamics is set by the nonlinear $\bm E \times \bm B$ advection close to the separatrix and, in particular, near the X-point, while the diamagnetic (NC) contribution remains weak in these regions.

In \textit{unfav}, by contrast, the correlation with $- \grad \cdot \bm J_{\perp}^{es}$ does not exceed $0.8$ in the same region and is comparable to the one obtained with $- \grad \cdot \bm J_{\perp}^{dia}$, such that the local dynamics of $\omega$ results from a balance between $\grad \cdot \bm J_{\perp}^{es}$ \textit{and} $\grad \cdot \bm J_{\perp}^{dia}$. The contribution from $- \grad \cdot \bm J_{\perp}^{es}$ is also found much smaller near the X-point in \textit{unfav} than in \textit{fav}, highlighting a clear asymmetry between the two configurations in the spatial location of the $\bm E \times \bm B$ nonlinear advection of the vorticity. We remark that both \textit{fav} and \textit{unfav} also display a large and similar correlation with $- \grad \cdot \bm J_{\perp}^{es}$ above the OMP location. Finally, we remark that the electromagnetic current is negligible in all cases thanks to the low values of $\beta_e$, and that similar results are obtained for AUG (not shown).

These observations are confirmed by the time traces shown in Fig.~\ref{fig:fig17}, which displays $\partial_t \omega + \grad \cdot \bm J_{\parallel}$ together with $- \grad \cdot \bm J_\perp^{es}$, $- \grad \cdot \bm J_\perp^{dia}$ and their sum, evaluated at the OMP inward of ($\rho_{\mathrm{pol}} = 0.98$) and within ($\rho_{\mathrm{pol}} = 0.99$) the $E_r$ well. Retaining both perpendicular currents reproduces the local evolution of $\omega$, the time correlation between $\partial_t \omega + \grad \cdot \bm J_{\parallel}$ and $- \grad \cdot \left( \bm J_\perp^{es} + \bm J_\perp^{dia} \right)$ being $0.9$ or larger in all cases, the small remainder being carried by the polarization and electromagnetic currents. In particular, the electrostatic current alone accounts for most of this correlation, which increases from $0.56$ ($0.72$) inward of the $E_r$ well to $0.81$ ($0.82$) within it in \textit{fav} (\textit{unfav}), while the diamagnetic current remains only weakly correlated with $\partial_t \omega + \grad \cdot \bm J_{\parallel}$. The correlation associated with $- \grad \cdot \bm J_\perp^{dia}$ is systematically larger in \textit{unfav} than in \textit{fav}, being $0.29$ and $0.30$ against $0.16$ and $0.16$ at $\rho_{\mathrm{pol}} = 0.98$ and $\rho_{\mathrm{pol}} = 0.99$, respectively. While the electrostatic correlation itself is comparable in the two configurations, the electrostatic contribution therefore exceeds the diamagnetic one by a factor of $3.5$ to $5$ in \textit{fav}, against a factor of about $2.5$ in \textit{unfav}, which again demonstrates the enhanced role of $- \grad \cdot \bm J_\perp^{es}$ in \textit{fav} in driving the local evolution of $\omega$ in the $E_r$ well.

\subsection{Theory of Mean Flow Acceleration} \label{subsec:flowacceleration}

The previous subsection has established the role of the electrostatic current as the main contribution to the local evolution of the vorticity. We aim now to relate the radial electrostatic current to the mean (FSA) radial electric field $E_r$, and thus to the mean poloidal flow responsible for the \textit{fav} and \textit{unfav} asymmetry.

The present derivation starts from the GK vorticity equation, given in \eqnref{eq:vorticity}. Taking its FSA, integrating over $\psi$ and developing the polarization radial current, which can be combined exactly with the electrostatic one, yields an evolution equation for the FSA poloidal momentum density. The derivation is reported in Appendix~\ref{appendix:momentum} and introduces the full-$f$ electrostatic velocity stress $\Pi$ \cite{frei2026prl}, defined by

\begin{align} \label{eq:pidef}
  \Pi  = - \frac{1}{c}\fs{\Omega^\psi u^\psi} = \fs{w M u_\theta u^\psi}.
\end{align}
being $u_\theta = \bm e_\theta \cdot \bm u$ the poloidal component of the $\bm E \times \bm B$ velocity, $u^\psi = \bm u \cdot \grad \psi$ its radial one, $w = R^2 / ( \mathcal{J} I)$ a geometrical weight factor. The velocity stress $\Pi$ represents the mean radial transport by $u^\psi$ of the poloidal momentum density $M u_\theta$. We note that $\Pi$ constitutes the full-$f$ generalization of the RS used in previous models \cite{diamond1991, smolyakov2000, diamond2005} (see Section~\ref{subsec:velocitystress}).

Writing $\fs{\Omega^\psi} = - c \fs{M w u_\theta} = - c \mathcal{P}_\theta$ (with $\mathcal{P}_\theta = \fs{w M u_\theta}$ the FSA poloidal momentum density), the FSA vorticity equation yields an equation for the poloidal flow acceleration $\partial_t \mathcal{P}_\theta$ associated with the poloidal velocity $u_\theta$, which dominates the total ion poloidal velocity $U_{i \theta}$ around the $E_r$ well (see Section~\ref{subsec:forcebalance}). More precisely, the FSA vorticity equation becomes

\begin{align} \label{eq:vort4}
c \frac{\partial}{\partial t}  \mathcal{P}_\theta
= - \frac{c}{V^\prime} \frac{\partial}{\partial \psi}
\left[ V^\prime \left( \Pi - \Pi^{em} \right) \right] +  \mathcal{C} + \fs{ J_\perp^{dia, \psi} } - \frac{c_\psi}{V^\prime},
\end{align}
where $\Pi$ is the full-$f$ electrostatic velocity stress introduced above and where we introduce the Maxwell stress $\Pi^{em} =\fs{w B_{1 \theta} B_1^\psi} / (4 \pi) $ (with $B_{1 \theta} = \bm e_\theta \cdot \bm B_1$), being $c_\psi$ a constant of integration and $\mathcal{C} = - \fs{ \bm \Omega \bm u : \grad \grad \psi } - \fs{ \Omega^\psi \grad_\perp \cdot \bm u }$ a contribution from the shaping of the flux surfaces and from the compression of the $\bm E \times \bm B$ velocity, which is smaller than the radial gradient of $\Pi$ (see Appendix~\ref{appendix:momentum}). We remark that $\Pi$ and $\Pi^{em}$ have similar expressions, with the replacement $\sqrt{M} \bm u \to \bm B_1 / \sqrt{4 \pi}$. The combination $\Pi - \Pi^{em}$ appearing in \eqnref{eq:vort4} is the familiar competition between the Reynolds and Maxwell stresses \cite{scott2005, naulin2005, diamond2005}. We note that \eqnref{eq:vort4} is exact: no ordering assumption is made and it is full-$f$ by construction.

\eqnref{eq:vort4} is an evolution equation for the mean radial electric field $E_r$ in the limit of $\phi_1 = \phi_1(\psi)$. Indeed, in this limit, the $\bm E \times \bm B$ velocity becomes $\bm u = \omega_E \left( R^2 \grad \phi - I \bm B / B^2\right)$, with $\omega_E = - c \partial_\psi \phi_1$. This implies that $w u_\theta = - \omega_E |\grad \psi|^2 / B^2$, while $u^\psi = 0$ yielding $\Pi = 0$. In this limit, the FSA poloidal momentum density reduces to $\mathcal{P}_\theta = - \fs{M} \omega_E \fs{w |\grad \psi|^2 / B^2 }$ (assuming that the total mass density is a flux-surface function, i.e. $M \simeq \fs{M}$) and \eqnref{eq:vort4} simplifies to

\begin{align} \label{eq:vort5}
\fs{w \frac{|\grad \psi|^2}{B^2} } \frac{\partial}{\partial t} \left(\fs{M} \omega_E \right)
=  - \fs{  P_{\parallel} \frac{\bm b \times \bm \kappa}{B}\cdot \grad \psi}   - \fs{P_{\perp } \frac{\bm b \times \grad B}{B^2} \cdot \grad \psi   }  + \frac{c_\psi}{c V^\prime},
\end{align}
with $\fs{\grad_\perp \cdot \bm u } \simeq 0$ and $\bm \kappa = (\bm b \cdot \grad )\bm b$ the magnetic curvature. \eqnref{eq:vort5} represents the relaxation of $\omega_E$ (or equivalently $E_r$) to the NC equilibrium due to the NC viscosity associated with $J_\perp^{dia, \psi}$.

The presence of poloidal asymmetries in $\phi_1$ (driven by higher-order NC effects and/or by turbulence) causes a finite radial transport of poloidal momentum density, such that the contribution of the velocity stress $\Pi$ to the mean poloidal flow acceleration is not negligible. Indeed, a finite radial gradient of $\Pi$ yields a transfer of energy from the small scale turbulent structures to the mean poloidal flow \cite{diamond2005}. It is via this nonlinear process that a mean (or zonal) poloidal flow spontaneously emerges from turbulence and can, in turn, regulate it in a feedback loop \cite{diamond1994}. We expect this nonlinear process to be stronger in \textit{fav}, where the local vorticity dynamics is driven by the nonlinear $\bm E \times \bm B$ advection of vorticity near the separatrix and near the X-point yielding a nonlinear build-up of the mean poloidal flow, as shown in Section~\ref{subsec:localvorticity}.

\subsection{Poloidal and Radial Structure of the Velocity Stress} \label{subsec:stressstructure}

\begin{figure}
\centering
\includegraphics[scale=0.6]{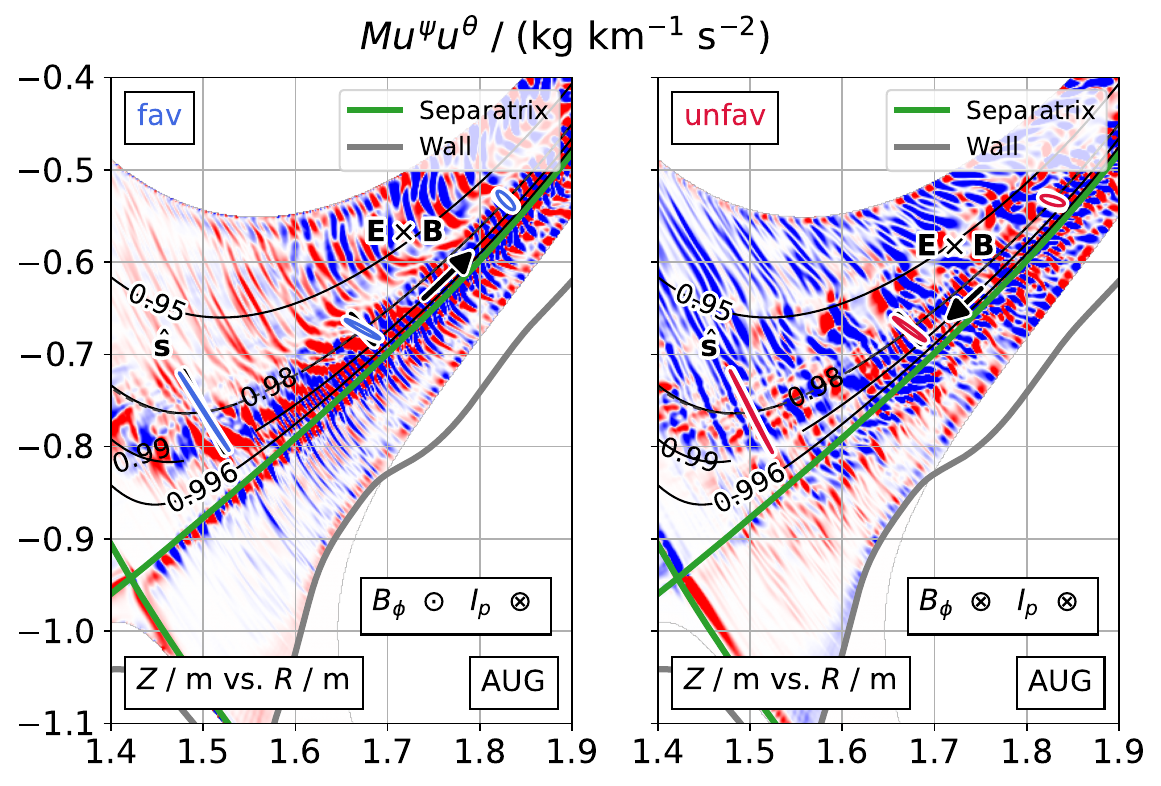}
\caption{\label{fig:fig18} Snapshots of $M u^\psi u^\theta$, whose FSA is equal to $\Pi$, obtained in \textit{fav} (left) and \textit{unfav} (right) in AUG on a single poloidal plane. The separatrix is shown by the green line, while the closed flux-surfaces $\rho_{\mathrm{pol}} = 0.95$, $0.98$, $0.99$ and $0.996$ are shown by the solid thin black lines. The black arrows show the direction of the $\bm E \times \bm B$ velocity on the $\rho_{\mathrm{pol}} = 0.99$ flux-surface. The ellipses illustrate eddies on the $\rho_{\mathrm{pol}} = 0.98$ flux-surface (dashed line), tilted by $\alpha_u^{(\hat s)}$ alone (black) and by the measured $\alpha_u$ (coloured).}
\end{figure}

The poloidal and radial structure of $\Pi$ is first inspected in Fig.~\ref{fig:fig18}, which shows snapshots of $M u^\psi u^\theta$, whose FSA is $\Pi$, in AUG on the LFS between the OMP and the X-point, where the nonlinear $\bm E \times \bm B$ advection of the vorticity dominates (see Section~\ref{subsec:localvorticity}). $M u^\psi u^\theta$ is strongly localized around the $\rho_{\mathrm{pol}} = 0.99$ flux-surface in \textit{fav}, while no clear radial localization is observed in \textit{unfav}.

The tilt of the turbulent structures in Fig.~\ref{fig:fig18} results from two competing mechanisms \cite{fedorczak2012,manz2012,fedorczak2013,schmid2017,manz2018}. The first is the magnetic shear, which distorts an eddy born near the OMP as it propagates along the field lines towards the X-point \cite{fedorczak2012,fedorczak2013,grover2024} (black ellipses in Fig.~\ref{fig:fig18}). Being set by the magnetic geometry, it is identical in \textit{fav} and \textit{unfav}. The second is the shear of the $\bm E \times \bm B$ velocity, whose direction reverses between the two configurations \cite{frei2026prl}: in the edge, the $\bm E \times \bm B$ velocity is directed away from the active X-point in \textit{fav} and towards it in \textit{unfav} (black arrows in Fig.~\ref{fig:fig18}). Since the $E_r$ well is deeper in \textit{fav}, the $\bm E \times \bm B$ shearing rate is larger than in \textit{unfav} \cite{schmid2017,cziegler2017,manz2018} and acts together with the magnetic shear to enhance the tilt of the eddies \cite{terry2000,diamond2005} (coloured ellipses in Fig.~\ref{fig:fig18}). The resulting radial localization of $M u^\psi u^\theta$ in \textit{fav} yields a larger radial gradient of $\Pi$ and, hence, a poloidal flow acceleration that reinforces the $\bm E \times \bm B$ shear, the mean flow and the turbulent stress being coupled in a self-amplifying loop \cite{diamond1994,diamond2005}.

\begin{figure}
\centering
\includegraphics[scale=0.5]{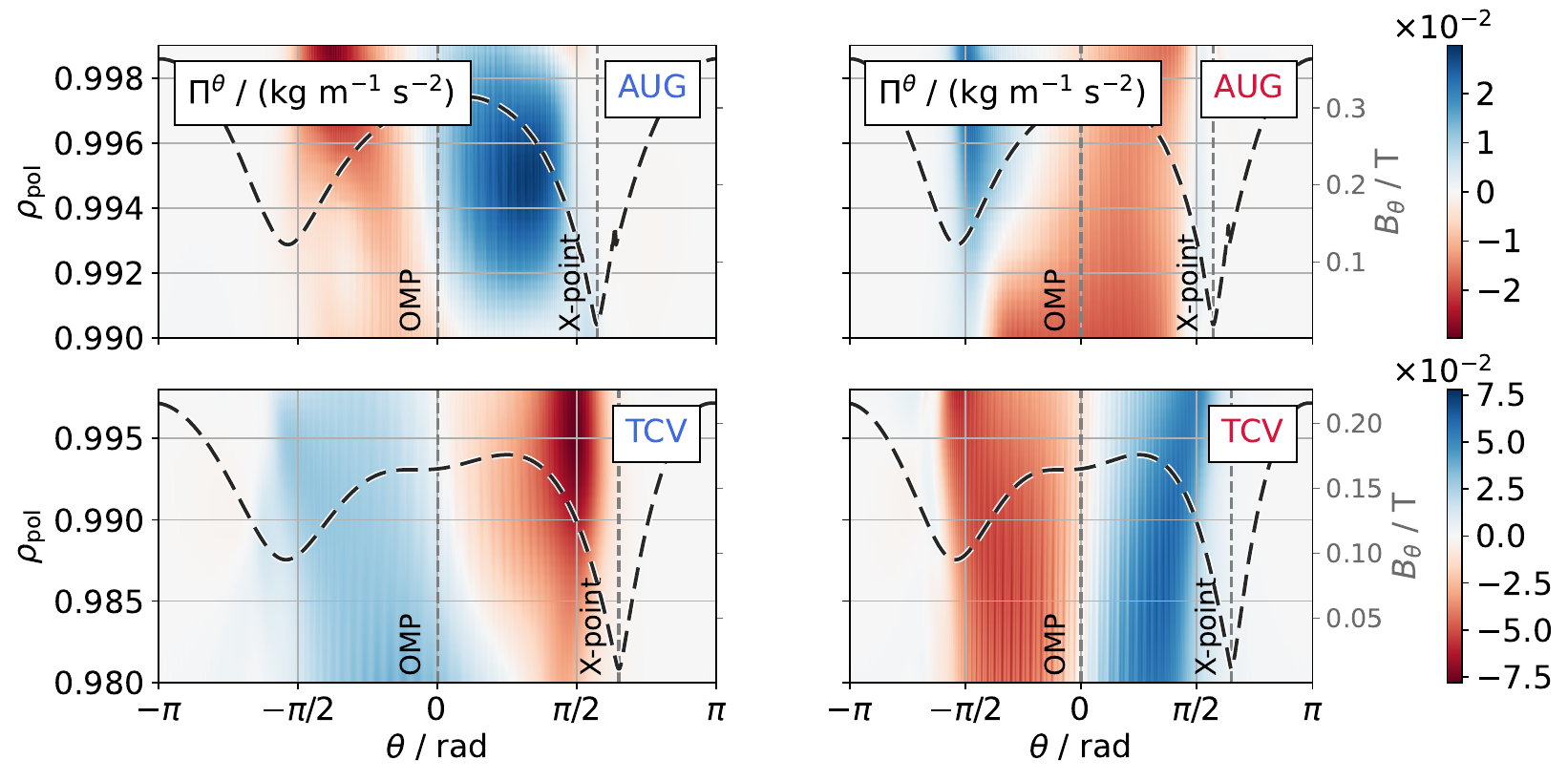}
\caption{\label{fig:fig19} Poloidally resolved velocity stress $\Pi^\theta$ (toroidally and time averaged over $0.2$~ms), whose average over $\theta$ is $\Pi$, in AUG (top row) and TCV (bottom row), in \textit{fav} (left column) and \textit{unfav} (right column). The vertical dashed lines mark the OMP position ($\theta = 0$) and the X-point. The black dashed curve on the right-hand axis is the poloidal magnetic field $B_\theta$ on the $0.9985$ flux-surface, its minimum locating the active X-point. The colour bars and the $B_\theta$ axis are shared along each row. $\Pi^\theta$ has the sign of $\Pi$ defined in \eqnref{eq:pidef} in both configurations (see Appendix~\ref{appendix:signs}).}
\end{figure}

The poloidally resolved stress $\Pi^\theta$, whose average over $\theta$ yields $\Pi$, is shown in Fig.~\ref{fig:fig19} for AUG and TCV after toroidal and time averaging over $0.2$~ms. In \textit{fav}, the magnitude of $\Pi^\theta$ increases near the separatrix between the OMP and the X-point, owing to the radially localized structures of Fig.~\ref{fig:fig18}, while $\Pi^\theta$ remains almost radially uniform in \textit{unfav}. Since the FSA weights the structures at $+\theta$ more than those at $-\theta$ (weight $\propto 1/|\grad \psi|$), this radially modulated poloidal asymmetry, present in \textit{fav} near the minimum of $B_\theta$, yields a finite radial gradient of $\Pi$ in \textit{fav} only.

\begin{figure}
\centering
\includegraphics[scale=0.5]{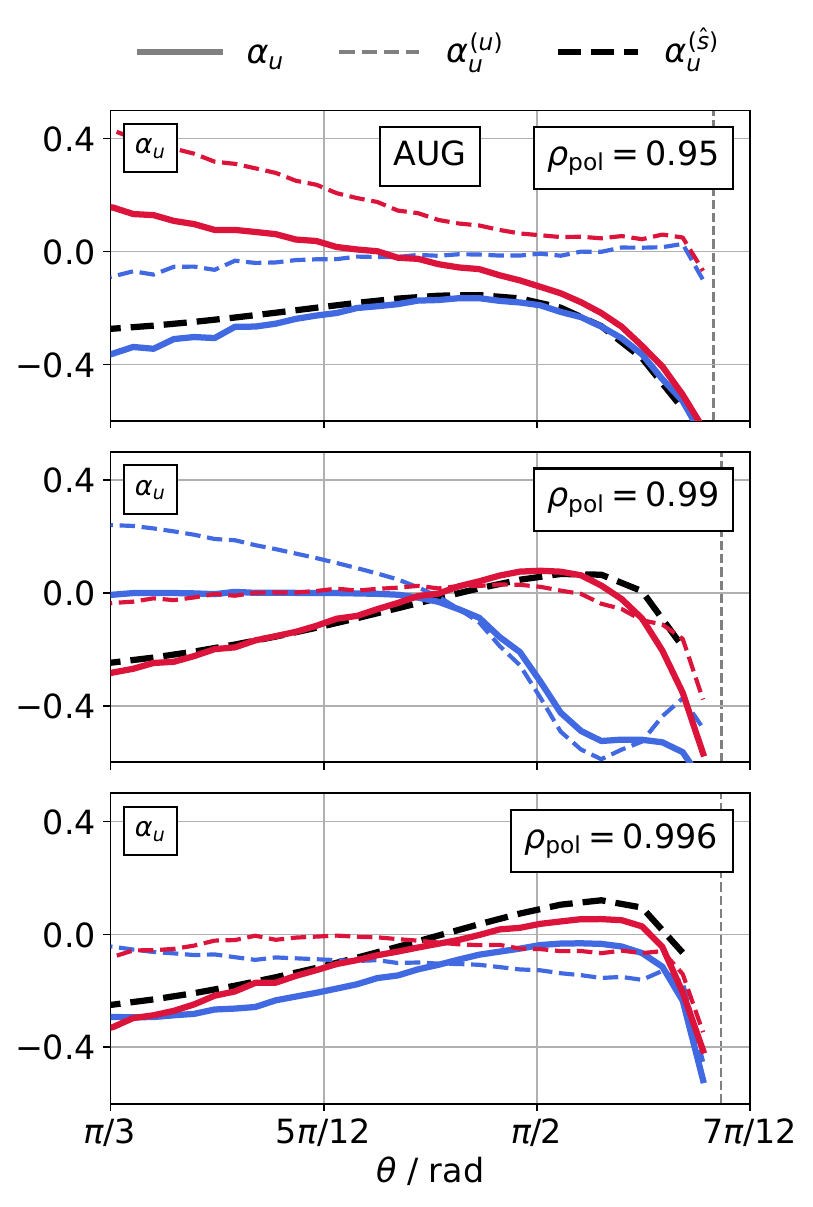}
\caption{\label{fig:tilt} Measured eddy tilt angle $\alpha_u$ (solid), flow shear contribution $\alpha_u^{(u)}$ (dashed), and magnetic shear contribution $\alpha_u^{(\hat s)}$ (thick dashed black) as a function of the poloidal angle $\theta$ on the $\rho_{\mathrm{pol}} = 0.95$, $0.99$ and $0.996$ flux-surfaces (top to bottom) in AUG, in \textit{fav} (blue) and \textit{unfav} (red). The X-point position is shown by the dashed grey line.}
\end{figure}

To quantify this competition, we introduce the eddy tilt $\alpha_u = \langle \tilde u_\theta \tilde u_\psi \rangle_{\phi, t} / \langle \tilde u_\psi^2 \rangle_{\phi, t}$ \cite{grover2024}, with $\tilde u_\psi$ and $\tilde u_\theta$ the fluctuations of the $\bm E \times \bm B$ velocity and $\langle \cdot \rangle_{\phi, t}$ the local toroidal and time average. $\alpha_u$ is the tangent of the angle between the eddy axis and the outward radial direction, measured towards increasing $\theta$, $\alpha_u < 0$ ($\alpha_u > 0$) corresponding to an anti-clockwise (clockwise) rotation of the eddy (see Appendix~\ref{appendix:signs}). Following \cite{fedorczak2012,grover2024}, we write $\alpha_u = \alpha_u^{(\hat s)} + \alpha_u^{(u)}$, where the magnetic-shear tilt $\alpha_u^{(\hat s)}$ is computed by following along the field lines a structure that is radial at the OMP until it reaches the poloidal plane at $\theta$, and $\alpha_u^{(u)} = \alpha_u - \alpha_u^{(\hat s)}$ estimates the tilt due to the sheared $\bm E \times \bm B$ velocity. These quantities are shown in Fig.~\ref{fig:tilt} between $\theta = \pi/3$ and the active X-point for AUG. Inside the edge ($\rho_{\mathrm{pol}} = 0.95$) and close to the separatrix ($\rho_{\mathrm{pol}} = 0.996$), $\alpha_u$ is dominated by $\alpha_u^{(\hat s)}$, in particular near the X-point. In the inner shear layer ($\rho_{\mathrm{pol}} = 0.99$), the two configurations instead differ. In \textit{unfav}, $\alpha_u$ follows $\alpha_u^{(\hat s)}$ over the whole poloidal range, the eddies being tilted by the magnetic shear. In \textit{fav}, the eddies remain nearly radial ($\alpha_u \simeq 0$) up to $\theta \simeq 0.45\pi$, $\alpha_u^{(u)}$ compensating $\alpha_u^{(\hat s)}$. From $\theta \simeq 0.48\pi$, the contribution from the flow shear, $\alpha_u^{(u)}$, largely dominates and tilts the eddies anti-clockwise ($\alpha_u \simeq -0.5$). The \textit{fav} and \textit{unfav} asymmetry of the eddy tilt is therefore localized inside the $E_r$ well and away from the X-point. Similar observations are obtained in TCV.

\subsection{Favre Decomposition and Radial Gradient of the Velocity Stress} \label{subsec:velocitystress}

Before quantifying the poloidal flow acceleration driven by the velocity stress $\Pi$, we first discuss its composition and, in particular, its relation with the Reynolds stress $\mathcal{R}$ (defined below). A consequence of the full-$f$ approach is the appearance of the total mass density $M$ in $\Pi$ (see \eqnref{eq:pidef}). The mass density $M$ is not present in previous local models \cite{diamond1991,diamond2005,scott2005}, which often assume $M$ to be constant (or a flux-function). However, poloidal density asymmetries and/or density fluctuations can yield a nonvanishing contribution to $\Pi$, and hence to nonlinear poloidal flow acceleration. To retain the contribution from the total mass density $M$, we introduce the Favre average, $\favre{f} = \fs{M f} / \fs{M}$ \cite{favre1965,held2018}. $\favre{f}$ reduces to the FSA if $f$ is a flux function. Introducing the weighted poloidal velocity $u^\theta = w u_\theta$, the Favre average operator allows us to write the velocity stress $\Pi$, defined in \eqnref{eq:pidef}, as

\begin{align} \label{eq:Pistressdef}
\Pi  = \fs{M} \favre{ u^\theta   u^\psi }.
\end{align}
 We remark that, using the Favre operator, the FSA poloidal momentum density can be written as $\mathcal{P}_\theta = \fs{w M u_\theta} = \fs{M} \favre{u^\theta}$. The Favre operator plays a central role in the full-$f$ formalism \cite{held2018} as it allows us to generalize the Reynolds decomposition, $\widetilde{f}= f - \fs{f}$ \cite{diamond2005}, to the Favre decomposition, $\widehat{f}= f - \favre{f}$. In particular, the Favre decomposition allows us to write

\begin{align}\label{eq:Pistress}
\Pi = \fs{M} \left( \mathcal{F} + \favre{u^\theta} \favre{u^\psi} \right),
\end{align}
where we introduce the Favre stress $\mathcal{F} = \favre{ \widehat{ u^\theta} \widehat{u^\psi}}$. The Favre stress is related to the RS $\mathcal{R} = \fs{\widetilde{u^\theta} \widetilde{u^\psi}}$, by

\begin{align} \label{eq:favrestress}
\mathcal{F}  = \mathcal{R} + \fs{ \widetilde{m} \widetilde{u^\psi} \widetilde{u^\theta}} - \favre{\widetilde{u^\psi}}\favre{\widetilde{u^\theta}},
\end{align}
with $\widetilde{m}  = \widetilde{M} / \fs{M}$ the normalized mass density fluctuations on a flux-surface. Note that $\favre{\widetilde{u^\psi}}$ (and $\favre{\widetilde{u^\theta}}$) does not vanish due to the presence of $M$ in $\favre{ \cdot }$ since $\favre{\widetilde{u^\psi}} = \fs{\widetilde{m} \widetilde{u^\psi}}$. Hence, $\mathcal{F}$ differs from $\mathcal{R}$ only through the mass density fluctuations, with $\mathcal{F} = \mathcal{R}$ if $M = \fs{M}$. The second term in \eqnref{eq:Pistress} represents the radial transport of poloidal momentum $\fs{M}\favre{u^\theta} $ by the turbulent radial flux $\fs{\widetilde{M} \widetilde{ u^\psi}}$. 

Combining \eqnref{eq:Pistress} and \eqnref{eq:favrestress}, the total velocity stress admits the natural decomposition $\Pi = \sum_i \Pi^{(i)}$, with $\Pi^{(1)} =  \fs{M} \mathcal{R}$, $\Pi^{(2)} =   \fs{M}  \fs{  \widetilde{m} \widetilde{u^\psi} \widetilde{u^\theta}} $, $\Pi^{(3)} =  \fs{M} \favre{u^\theta} \favre{u^\psi}$, and $\Pi^{(4)} = -  \fs{M} \favre{\widetilde{u^\psi}}\favre{\widetilde{u^\theta}}$. Here, $\Pi^{(1)}$ is the classical RS contribution retained in the local approaches \cite{diamond1991,diamond2005,scott2005}, while $\Pi^{(2)}$ and $\Pi^{(4)}$ are full-$f$ corrections associated with the density fluctuations, which vanish in the limit $M = \fs{M}$. We remark that $\Pi^{(3)}$ survives in this limit, such that $\Pi \to \fs{M} \left( \mathcal{R} + \fs{u^\theta} \fs{u^\psi} \right)$, the RS being retrieved only if the mean radial flow $\fs{u^\psi}$ vanishes as well. An explicit expression of the radial gradient of $\Pi$ can then be obtained, such that

\begin{align} \label{eq:gradpi}
\frac{1}{\fs{M}}\frac{d}{d \psi}   \Pi & =  \frac{d}{d \psi}   \mathcal{R}  + \left( \mathcal{F} + \favre{u^\theta} \favre{u^\psi} \right) \frac{d}{d \psi} \ln \fs{M} \nonumber \\
& + \frac{1}{\fs{M}} \frac{d}{d \psi}\fs{\widetilde{m} \widetilde{u^\psi} \widetilde{u^\theta}} +  \frac{d}{d \psi} \left(\favre{u^\theta} \favre{u^\psi} -\favre{\widetilde{u^\theta}}\favre{\widetilde{u^\psi}} \right).
\end{align}
\eqnref{eq:gradpi} demonstrates the full-$f$ modifications compared to the local theories, where the mean flow is driven solely by the radial gradient of the RS (first term). Indeed, additional nonlinear drives appear, associated with the finite FSA density gradient (second term) \cite{held2018}, with the radial gradient of the triple fluctuation product (third term), and with the turbulent flux of the mean poloidal momentum density (fourth term). We remark that density fluctuations on a flux-surface and a finite FSA density gradient therefore allow poloidal mean flow acceleration even for a radially homogeneous $\mathcal{R}$, a mechanism retained in the present simulations but absent from the local models.

\begin{figure}
\centering
\includegraphics[scale=0.5]{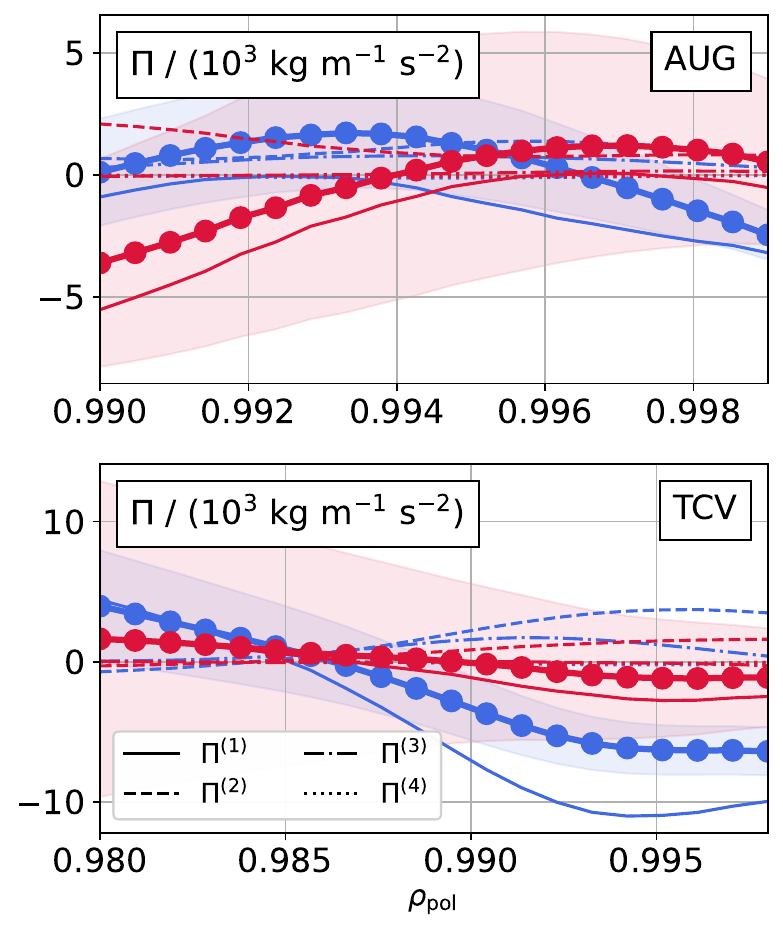}
\caption{\label{fig:fig20} Radial profiles (averaged over time) of the velocity stress, $\Pi$ defined in \eqnref{eq:Pistress} (solid coloured lines with markers), around the $E_r$ well obtained in AUG (top) and TCV (bottom). The different contributions $\Pi^{(i)}$ defined in the main text are also shown by the coloured lines without markers. The shaded coloured areas are the standard deviations associated with the time average.}
\end{figure}

The radial profiles of $\Pi$ and of its contributions $\Pi^{(i)}$ are shown in Fig.~\ref{fig:fig20} around the $E_r$ well. In TCV, $\Pi$ decreases monotonically from about $4 \times 10^3$ to $- 6 \times 10^3$~kg~m$^{-1}$~s$^{-2}$ across the well in \textit{fav}, while it remains within $\pm 2 \times 10^3$~kg~m$^{-1}$~s$^{-2}$ in \textit{unfav}, yielding a radial gradient several times larger in \textit{fav}, consistent with the localized structures of Figs.~\ref{fig:fig18} and \ref{fig:fig19}. In AUG, the amplitudes are comparable but the profiles have opposite slopes in the outer part of the well ($\rho_{\mathrm{pol}} \gtrsim 0.995$), where $\Pi_W$ is the largest (see Section~\ref{subsec:nrgtransfer}), $\Pi$ decreasing radially in \textit{fav} while it increases in \textit{unfav} over the whole radial range shown. The decomposition reveals that $\Pi^{(1)}$ dominates in all cases, the full-$f$ corrections being nevertheless not negligible. Indeed, $\Pi^{(2)}$, driven by the correlation between the density and the velocity fluctuations, is systematically of opposite sign to $\Pi^{(1)}$ and partially compensates it, the compensation being the strongest in TCV in \textit{fav} where the RS alone overestimates the total stress by nearly a factor of two. The contribution $\Pi^{(3)}$ remains smaller but becomes comparable to $\Pi^{(2)}$ near the separatrix, while $\Pi^{(4)}$ is negligible. A description of the poloidal flow acceleration based on the RS alone is therefore not sufficient. We remark, finally, that the standard deviations associated with the time average are larger in \textit{unfav}, owing to the large GAMs observed there (see Section~\ref{subsec:gam}).

\subsection{Nonlinear Energy Transfer to Mean Poloidal Flow} \label{subsec:nrgtransfer}

We now assess the efficiency of the nonlinear energy transfer from turbulence into the mean poloidal flow driven by the gradient of $\Pi$. To quantify this nonlinear process, we derive an evolution equation for a quadratic quantity whose sign depends neither on the drift configuration nor on the orientation of the coordinates (see Appendix~\ref{appendix:signs}). We therefore introduce the poloidal kinetic energy $K_\theta = \mathcal{P}_\theta^2 / 2$ and derive its evolution equation from \eqnref{eq:vort4}. Multiplying \eqnref{eq:vort4} by $\mathcal{P}_\theta / \left( c \fs{M} \right) = \favre{u^\theta} / c$, the evolution equation of the poloidal kinetic energy $K_\theta$ is \cite{manz2012}

\begin{align} \label{eq:kthetafull}
  \frac{1}{\fs{M}}\frac{\partial}{\partial t} K_\theta = - \Pi_W + \Lambda_W,
\end{align}
with $\Pi_W$ (a work per unit volume and per unit time, i.e. a power density) defined by 

\begin{align} \label{eq:workterms}
\Pi_W & =  \favre{u^\theta} \frac{1}{V^\prime} \frac{d}{d \psi} \left( V^\prime \Pi \right).
\end{align}
Each term on the right-hand side of \eqnref{eq:kthetafull} is associated with the work done by the corresponding radial current appearing in \eqnref{eq:vort4}. More precisely, $\Pi_W$ is the work done by the full-$f$ electrostatic velocity stress $\Pi$ and measures the nonlinear energy transfer between turbulence and the mean poloidal flow. On the other hand, the $\Lambda_W$ term collects the contributions from the Maxwell stress, from the flux-surface shaping and from the compression of the $\bm E \times \bm B$ flow contained in $\mathcal{C}$ (see \eqnref{eq:vort4}), the work done by the diamagnetic current (NC viscous damping of the mean poloidal flow), and the residual associated with the constant of integration $c_\psi$. In the following, we focus solely on the contribution from $\Pi_W$ associated with the nonlinear $\bm E \times \bm B$ advection of the vorticity (or equivalently with the radial transport of poloidal momentum). 

The sign convention of $\Pi_W$ is such that $\Pi_W < 0$ corresponds to a net transfer of energy from turbulence into the mean poloidal flow.
The work $\Pi_W$ can itself be decomposed into a transport and a production contribution, $\mathcal{T}$ and $\mathcal{P}$ respectively \cite{manz2012}, such that

\begin{align} \label{eq:workdecomposition}
\Pi_W = \mathcal{T} - \mathcal{P},
\end{align}
with $\mathcal{T} =  d_V \left( V^\prime \favre{u^\theta} \Pi \right)$ and $\mathcal{P} =  V^\prime \Pi d_V \favre{u^\theta}$ (with $d_V = (V^\prime)^{-1} d / d \psi$). The transport term $\mathcal{T}$ is a radial divergence and therefore redistributes $K_\theta$ radially without contributing to the volume-integrated poloidal kinetic energy. The production term $\mathcal{P}$, being the product of the velocity stress with the radial shear of the mean poloidal flow, exchanges energy between turbulence and the mean flow.

\begin{figure}
\centering
\includegraphics[scale=0.5]{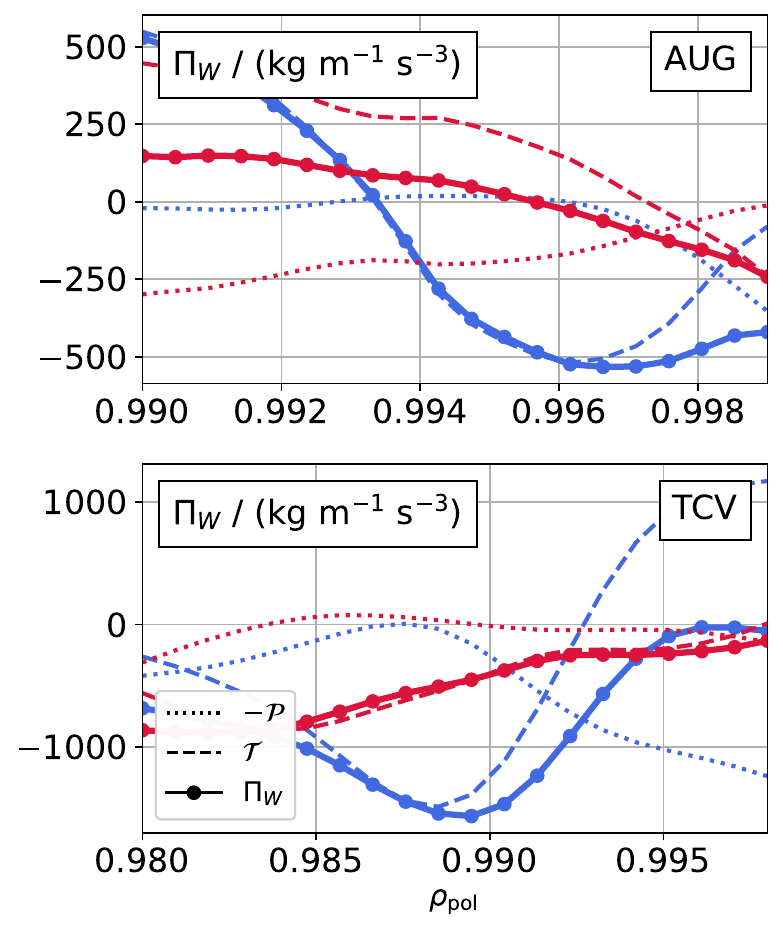}
\caption{\label{fig:fig21} Radial and time averaged (over $0.2$~ms) profiles of the total work done by the velocity stress $\Pi$, measured by $\Pi_W$ (solid coloured lines with markers) obtained in AUG (top) and TCV (bottom). The contributions from $\mathcal{T}$ (transport) and $-\mathcal{P}$ (production) are also shown for comparison by the coloured dashed and dotted lines, respectively.}
\end{figure}

We now diagnose $\Pi_W$, defined in \eqnref{eq:workterms}, in both \textit{fav} and \textit{unfav} in AUG and TCV. Fig.~\ref{fig:fig21} displays the radial and time-averaged profiles of $\Pi_W$, with $\mathcal{T}$ and $- \mathcal{P}$, obtained in AUG and TCV around the $E_r$ well. We observe that $\Pi_W$ develops a pronounced negative minimum in \textit{fav} in both AUG and TCV, located in the region of the $E_r$ well (around $\rho_\mathrm{pol} \simeq 0.996$ in AUG and $\rho_\mathrm{pol} \simeq 0.989$ in TCV). Following the sign convention of $\Pi_W$, this corresponds to turbulence accelerating the mean poloidal flow, consistently with the deeper $E_r$ well and the larger $U_{i \theta}$ reported in Section~\ref{subsec:forcebalance}.

The transfer $\Pi_W$ is systematically weaker in \textit{unfav}. In TCV, the minimum of $\Pi_W$ is reduced by roughly a factor of two. In AUG, the reduction is even more pronounced, $\Pi_W$ remaining positive over most of the $E_r$ well region and being an order of magnitude smaller than in \textit{fav}. The decomposition given in \eqnref{eq:workdecomposition} shows that $\Pi_W$ is dominated by the transport contribution $\mathcal{T}$ in the $E_r$ well in \textit{fav}, the production term $\mathcal{P}$ remaining small there. On the other hand, the latter becomes comparable to (or larger than) $\mathcal{T}$ outside the well, i.e. for $\rho_\mathrm{pol} \gtrsim 0.997$ in AUG and $\rho_\mathrm{pol} \gtrsim 0.992$ in TCV. Being a radial divergence, $\mathcal{T}$ does not create poloidal kinetic energy, such that $K_\theta$ is supplied in the $E_r$ well by the radial convergence of $V^\prime \favre{u^\theta} \Pi$, while it is produced out of turbulence in the outer shear layer in \textit{fav}.

On the other hand, in \textit{unfav}, $\mathcal{T}$ and $- \mathcal{P}$ are of opposite signs and of comparable magnitude in AUG, such that the net work $\Pi_W$ results from a strong cancellation between the radial redistribution of $K_\theta$ and its production by the mean flow shear. It is therefore remarkable that the \textit{fav} and \textit{unfav} asymmetry of the mean poloidal flow, and hence of the $E_r$ well, originates less from the amplitude of the velocity stress $\Pi$ itself than from the efficiency with which it is converted into mean poloidal kinetic energy through \eqnref{eq:kthetafull}, this efficiency being set by the radial gradient of $\Pi$.

Finally, we remark that the nonlinear energy transfer measured by $\Pi_W$ is consistent with the local vorticity dynamics reported in Section~\ref{subsec:localvorticity}. Indeed, the correlation between $\partial_t \omega$ and $- \grad \cdot \bm J_\perp^{es}$ is largest in \textit{fav} near the separatrix (see Fig.~\ref{fig:fig16} and Fig.~\ref{fig:fig17}), where $\Pi_W$ is the most negative, while in \textit{unfav} the vorticity dynamics is shared with the diamagnetic current and the transfer into the mean flow is correspondingly weaker. The larger fluctuation amplitudes and the stronger GAM activity observed in \textit{unfav} (see Fig.~\ref{fig:fig7}) therefore do not translate into a stronger mean poloidal flow, the turbulent energy being transferred less efficiently to the mean flow than in \textit{fav}.

\section{Asymmetry at High and Matched Power and Implications for H-mode Access} \label{sec:fluxdriven}

The results presented so far are obtained at the low L-mode power. We now assess whether the \textit{fav} and \textit{unfav} asymmetry survives when the power crossing the separatrix $P_{sep}$ is suddenly raised. We show that the deeper $E_r$ well found in \textit{fav}, as well as the larger $E_r$ fluctuation level and the stronger GAM activity found in \textit{unfav}, persist up to $P_{sep} \simeq 1.6$~MW in TCV, i.e. more than six times the L-mode value (see Table~\ref{tab:1}), and we discuss the implications of this persistence for the asymmetry of the L--H power threshold.

Starting from the saturated TCV simulations analysed in Section~\ref{sec:results}, volumetric heat sources are introduced close to the edge boundary, with Gaussian shape centred at $\rho_{\mathrm{pol}} = 0.86$ and normalized width $\Delta \rho_{\mathrm{pol}} = 2 \times 10^{-2}$. Their amplitudes are set to $P_e = 500$~kW on the electrons and $P_i = 800$~kW on the ions, i.e. a total injected power of $1.3$~MW. The heat sources are identical in \textit{fav} and \textit{unfav}, and both simulations are continued over about $4$~ms. The amplitude of the density source is subsequently raised from its L-mode value to about twice that value at $\Delta t = 1.2$~ms (with $\Delta t$ the time elapsed since the start of the flux-driven phase), to assess the sensitivity of $E_r$ associated with a build up of the density profile. Neumann boundary conditions are then applied in the edge to allow the temperature profiles to rise (while keeping the density fixed).

\begin{figure}
\centering
\includegraphics[scale=0.55]{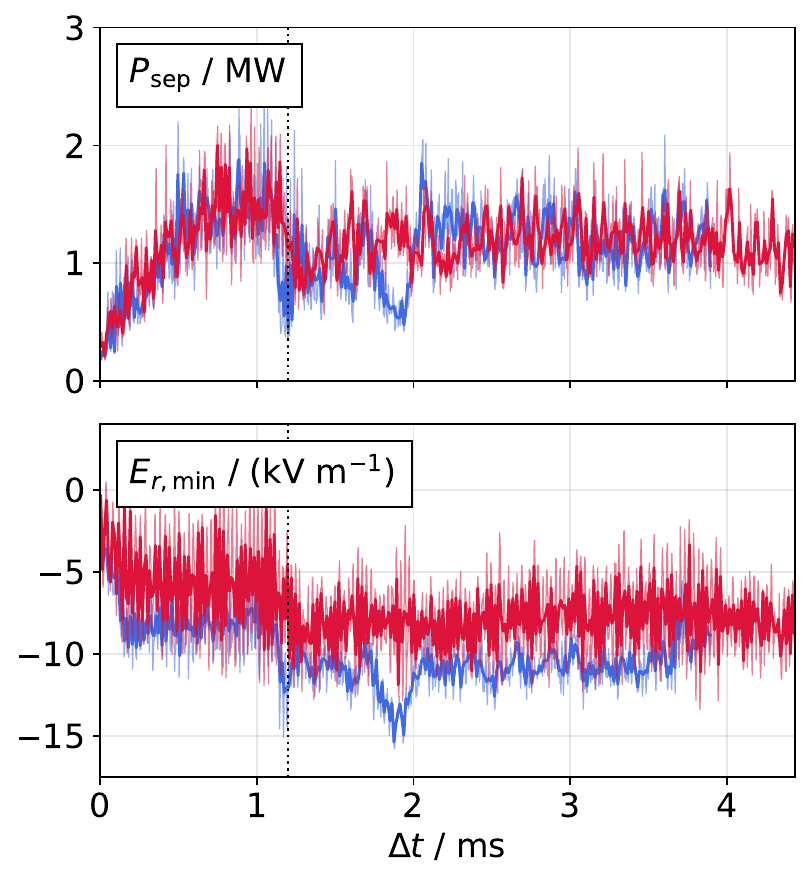}
\caption{\label{fig:fig22} Time traces of the total power ($P_{sep}$) crossing the separatrix (top) and the corresponding depth of the $E_r$ well ($E_{r, \mathrm{min}}$), measured at the OMP shown as a function of the time $\Delta t$ elapsed since the start of the power injection. The thin lines are the instantaneous traces and the thick lines their $12.6$~$\mu$s window average. The vertical dotted black line marks $\Delta t = 1.2$~ms, at which the amplitude of the density source is raised from its L-mode value to its final value. $\Delta t < 0$ corresponds to the saturated low power L-mode in TCV.}
\end{figure}

\begin{figure}
\centering
\includegraphics[scale=0.55]{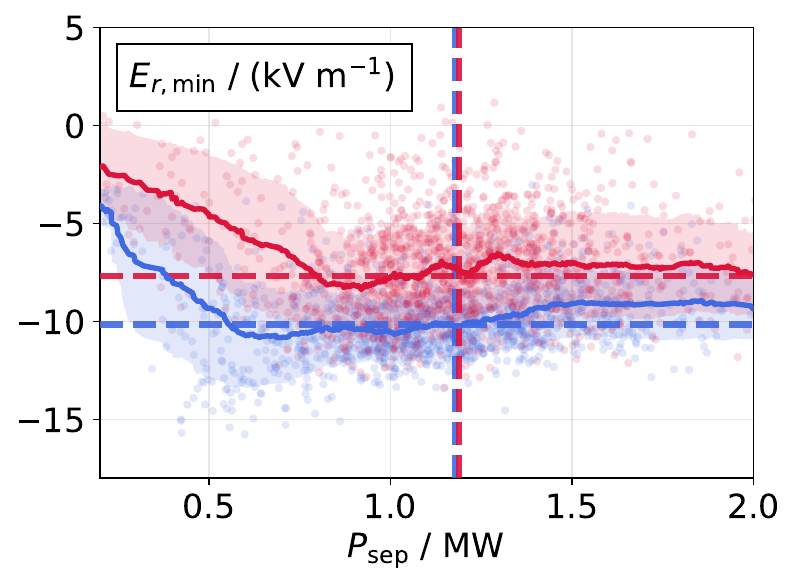}
\caption{\label{fig:fig23} Depth of the $E_r$ well ($E_{r, \mathrm{min}}$) as a function of the total power $P_{sep}$ measured along the time traces of \ref{fig:fig22}. Each point is a single snapshot, i.e. a $2.9$~$\mu$s sample. The thick solid lines are moving averages of $E_{r, \mathrm{min}}$ over a window of $161$ snapshots ordered by $P_{sep}$, evaluated over the whole range of $P_{sep}$ visited by each simulation, the axis being restricted to $0.2 \leq P_{sep} \leq 2$~MW, above which fewer than $15$ snapshots per configuration are available, and the shaded bands show the corresponding standard deviations. The dashed horizontal and vertical lines indicate respectively the mean $E_{r, \mathrm{min}}$ and the mean $P_{sep}$ evaluated in the QSS reached for $\Delta t > 3$~ms. The $E_{r, \mathrm{min}}$ asymmetry is retrieved over the whole range of $P_{sep}$.}
\end{figure}

The power crossing the separatrix, $P_{sep}$, is measured at $\rho_{\mathrm{pol}} = 0.98$ retaining both the $\bm E \times \bm B$ and diamagnetic contributions of both species. Note that $P_{sep}$ measured at the beginning of the two simulations retrieves the values reported in Table~\ref{tab:1} and is radially constant to within $2$ to $6~\%$ over $0.96 \leq \rho_{\mathrm{pol}} \leq 0.99$, indicating QSS in the low power L-mode phase. All the quantities shown below are averaged over windows of $12.6$~$\mu$s. The time traces of the measured power $P_{sep}$ and of the depth of the $E_r$ well are shown in Fig.~\ref{fig:fig22}, while Fig.~\ref{fig:fig23} displays them against each other. We first remark that the increase of the density source at $\Delta t = 1.2$~ms results in a sudden deepening of $E_{r, \mathrm{min}}$, by about $2$~kV/m in both configurations, accompanied by a decrease of $P_{sep}$ of approximately $0.4$~MW. This fast change is \textit{not} a transition resulting from a self-consistent organization of the turbulence \cite{zholobenko2026prl}, but is caused by the modification of the density profiles following the increase of the density source.

The \textit{fav} and \textit{unfav} time traces of $P_{sep}$ are very similar, both rising from the L-mode values and converging to $P_{sep} \simeq 1.2$~MW achieving matched high power at QSS. On the other hand, the $E_r$ well deepens from $-4.6$ to $-10.8$~kV/m in \textit{fav} and from $-3.2$ to $-7.9$~kV/m in \textit{unfav}. Most importantly, the asymmetry is retrieved at all powers explored here as shown in Fig.~\ref{fig:fig23}. For $\Delta t > 1.75$~ms where the measured powers agree to within $3~\%$, we obtain $E_{r, \mathrm{min}} \simeq -10.8$~kV/m in \textit{fav} against $-7.8$~kV/m in \textit{unfav}. It is noticeable that this difference is nearly independent of the power. This is only qualitatively consistent with the TCV heating experiments, where the $E_r$ well is maintained, or slightly enhanced, in \textit{fav} along NBI and ECRH power ramps while it becomes shallower in \textit{unfav}, such that the asymmetry grows with the injected power, the \textit{fav} discharge eventually accessing H-mode while the \textit{unfav} one remains in L-mode \cite{rienacker2026}. This may result from the simulations not accessing the L--H transition, from the heating being deposited mostly on the ions near the inner boundary, and from the drift-dependent sheath and neutral physics not being retained (see Section~\ref{sec:conclusions}).

The asymmetry of the fluctuations and of the GAM activity (see Section~\ref{subsec:gam}) is also retrieved at high power. Evaluating the temporal Fourier spectrum of $E_r$ (not shown here), a sharp peak at $f_{\mathrm{GAM}} \simeq 34$~kHz is observed in \textit{unfav}. In \textit{fav}, no distinct peak is found over the same frequency range, consistent with the strong reduction of the GAM activity inside the $E_r$ well reported earlier. Extracting $A_{\mathrm{GAM}}$ as in Section~\ref{subsec:gam}, we obtain $A_{\mathrm{GAM}} \simeq 2.6$~kV/m in \textit{unfav} against $1.3$~kV/m in \textit{fav}, such that the GAM-driven $\bm E \times \bm B$ shear remains comparatively more important in \textit{unfav} at high power. Finally, we note that the OMP profiles (not shown here) obtained for $\Delta t > 3$~ms are similar between \textit{fav} and \textit{unfav}. These results confirm the picture developed in Section~\ref{sec:meanflow} at a power several times larger than the L-mode values: the turbulent energy of \textit{unfav} remains in the fluctuations and in the GAM oscillations, and the mean poloidal flow and hence the $E_r$ well, remains weaker than in \textit{fav}.

These results have direct consequences on the asymmetry of the L--H power threshold $P_{LH}$. Indeed, if the L--H transition is controlled by a critical value $E_c$ of $E_{r, \mathrm{min}}$ near the separatrix \cite{cavedon2020,cavedon2024}, the systematic offset in $E_{r, \mathrm{min}}$ reported in Fig.~\ref{fig:fig23} can be translated into an offset in the power required to reach it. More precisely, evaluating the instantaneous response of the $E_r$ well depth to the injected power for $0.2 \leq P_{sep} \leq 0.5$~MW, where the $E_r$ well is still building up, yields a change of $E_{r, \mathrm{min}}$ larger in \textit{fav} than in \textit{unfav}, such that $d E_{r, \mathrm{min}} / d P_{sep} \simeq 25$ and $\simeq 6$~kV~m$^{-1}$~MW$^{-1}$ in \textit{fav} and in \textit{unfav} respectively. The \textit{unfav} well is therefore less responsive to the injected power than the \textit{fav} one.

Extrapolating the QSS values reached for $\Delta t > 3$~ms, being $P_{sep} = 1.18$~MW with $E_{r, \mathrm{min}} = -10.2$~kV/m in \textit{fav} and $-7.7$~kV/m in \textit{unfav} (dashed lines in Fig.~\ref{fig:fig23}), to a hypothetical critical value $E_c = -15$~kV/m then yields $P_{LH} \simeq 1.4$~MW in \textit{fav} against $P_{LH} \simeq 2.4$~MW in \textit{unfav}, i.e. a ratio $P_{LH}^{\mathrm{unfav}} / P_{LH}^{\mathrm{fav}} \simeq 1.7$, to be compared with the factor of about two measured experimentally \cite{ryter1996,ryter2016,plank2023}. We emphasize that this estimate of $P_{LH}$ involves no free parameter beyond the assumed value of $E_c$, and that it is obtained with the same heat sources, the same magnetic equilibrium and the same boundary conditions in the two configurations. We remark, however, that this estimate remains qualitative, $E_c$ being assumed universal and reached by neither simulation, such that $P_{LH}$ is an extrapolation.

The mechanism identified in Section~\ref{sec:meanflow} indicates where this offset in $E_{r,\mathrm{min}}$ originates. The drift configuration changes neither the free energy available to the turbulence nor the nature of the instability that it drives (see Section~\ref{subsec:linear}), but it does change the efficiency with which the turbulent velocity stress is converted into a mean poloidal flow, $\Pi_W$ being weaker in \textit{unfav}. Hence, a larger fraction of the turbulent energy remains in the fluctuations and in the GAM oscillations in \textit{unfav}, instead of being transferred to the mean flow as in \textit{fav}, such that a larger input power is required to build the same $E_r$ well. Since the fluctuation amplitudes in \textit{unfav} at matched heating power are larger \cite{cziegler2017,bielajew2023}, one expects in addition a stronger $\bm E \times \bm B$ shear than in \textit{fav} to be required in order to quench the edge turbulence \cite{biglari1990,terry2000}, the two effects acting in the same direction. This trend, however, depends on the fluctuating quantity and on the heating conditions \cite{rienacker2026}. The present picture is also consistent with the absence, in \textit{unfav}, of a single threshold value of $E_r$ at the L--H transition, which indicates that parameters other than $E_r$, such as the edge turbulence characteristics (see Section~\ref{sec:turbulence}), must change with the drift configuration \cite{hubbard2012,plank2023,plank2023b}. Since the present simulations do not access the L--H transition itself, this discussion remains to be tested by following an actual transition, as recently achieved with fluid models \cite{zholobenko2026prl,delucca2026}.

\section{Summary and Conclusions} \label{sec:conclusions}

In the present work, we have investigated the influence of the drift configuration on turbulence, transport and radial electric field in the edge and SOL of AUG and TCV, by means of first-principles, diverted, full-$f$ GK simulations performed with the \texttt{GENE-X} code \cite{michels2021,frei2025vspec}. Two matched pairs of LSN L-mode discharges, differing by the direction of the ion $\grad B$ drift, have been simulated and validated against the available experimental measurements. To identify the origin of the observed asymmetry, we have derived from the GK vorticity equation an exact, full-$f$ evolution equation for the FSA poloidal momentum density. Our main findings can be summarized as follows:

\begin{itemize}

\item \textit{The profiles are matched, but the fluctuations are not.} Consistent with experiments \cite{plank2023,rienacker2026}, \texttt{GENE-X} yields nearly identical OMP density and temperature profiles in \textit{fav} and \textit{unfav} at both AUG and TCV. In the simulations, the relative fluctuation levels near the separatrix are, on the other hand, larger in \textit{unfav}, in particular in TCV. The edge confinement times are slightly larger in \textit{fav} in AUG, while in TCV the energy confinement time is enhanced by a factor of about $1.7$ in \textit{fav}, the particle confinement time remaining similar.

\item \textit{The $E_r$ asymmetry is reproduced and is carried by the poloidal flow.} The simulations retrieve, within the experimental uncertainties, the deeper $E_r$ well measured in \textit{fav}, despite the nearly identical OMP profiles. The radial force balance shows that the diamagnetic and toroidal contributions nearly cancel at the well, leaving the poloidal flow carrying the asymmetry.

\item \textit{The GAM activity is stronger in \textit{unfav}.} The GAM amplitude is larger in \textit{unfav}, by a factor of about $2$ (AUG) to $3$ (TCV) at the $E_r$ well, where the mean $E_r$ dominates over the GAM oscillation in \textit{fav}, consistently with experiments \cite{cziegler2017,vermare2026eps,rienacker2026}.

\item \textit{The poloidal flow asymmetry is not neoclassical.} The self-consistent ion poloidal flow $U_{i \theta}$ deviates strongly in \textit{fav} from the predictions of \texttt{NCLASS} \cite{houlberg1997} and \texttt{PERFECT} \cite{landreman2014}, while remaining compatible with them in \textit{unfav}, and reconstructing $E_r$ with the NC poloidal flow removes the asymmetry altogether. Axisymmetric simulations show that removing the non-axisymmetric components of $\phi_1$ suppresses the $E_r$ asymmetry (within statistical errors), together with the GAM activity.

\item \textit{The instability is the same, but the turbulence is differently regulated.} The local plasma parameters are nearly identical in \textit{fav} and \textit{unfav}, the only exception being the normalized $\bm E \times \bm B$ shear, about twice as large in \textit{fav}. The linear analysis identifies a resistive eDW and a TEM in AUG and a DTEM in TCV in both configurations, with indistinguishable growth rates in AUG. In TCV, the growth rate is about twice as large in \textit{fav}, owing to its larger $\omega_{T_e}$, which is opposite to the larger fluctuation levels found in \textit{unfav}. Thus, the difference in the nonlinear turbulence levels cannot be attributed to linear physics. Across the outer shear layer, the radial correlation length $\ell_c$ decreases in \textit{fav} while remaining unchanged in \textit{unfav}.

\item \textit{In \textit{fav}, the nonlinear $\bm E \times \bm B$ advection dominates the vorticity dynamics and drives the mean poloidal flow.} In \textit{fav}, the local evolution of the vorticity is dominated by the electrostatic current, which carries the nonlinear $\bm E \times \bm B$ advection, on the LFS in the radially localized region of the $E_r$ well, and even more so near the X-point, where the intrinsic magnetic and $\bm E \times \bm B$ shears tilt the turbulent structures in opposite directions \cite{fedorczak2012,fedorczak2013,schmid2017,manz2018}. The exact FSA balance equation for the poloidal momentum density identifies the radial gradient of the total electrostatic velocity stress $\Pi = \fs{M} \favre{u^\theta u^\psi}$ -- generalizing the RS drive to the full-$f$ regime \cite{favre1965,held2018} -- as the nonlinear drive of the mean poloidal flow, its finite radial gradient in \textit{fav} being driven by the radially localized poloidal asymmetries generated by the eddy tilting.

\item \textit{The asymmetry originates in the efficiency of the nonlinear energy transfer.} The work $\Pi_W$ done by the velocity stress on the mean poloidal flow develops a pronounced negative minimum in the region of the $E_r$ well in \textit{fav}. In \textit{unfav}, the transfer is weaker (by a factor of about two in TCV and by an order of magnitude in AUG). The transport-production decomposition of $\Pi_W$ \cite{manz2012} in \textit{fav} shows that $\Pi_W$ is dominated by the transport contribution inside the $E_r$ well and by production in the outer shear layer, indicating a net nonlinear energy transfer from turbulence to the mean flow in this region.

\item \textit{The asymmetry persists at high power.} In flux-driven simulations at matched high power in TCV, the $E_r$ well remains deeper in \textit{fav}, while the larger fluctuation level and the stronger GAM activity of \textit{unfav} are also retrieved. The \textit{unfav} $E_r$ well is also less responsive to the injected power. Assuming that the L--H transition is controlled by a critical value of $E_{r, \mathrm{min}}$, the offset in the depth of the $E_r$ well yields an estimated ratio $P_{LH}^{\mathrm{unfav}} / P_{LH}^{\mathrm{fav}} \simeq 1.7$, close to the experimental factor of about two, this estimate remaining an extrapolation since the L--H transition is not simulated.

\end{itemize}

These results yield a consistent physical picture of the \textit{fav} and \textit{unfav} asymmetry. The direction of the ion magnetic drift with respect to the X-point does not modify the free energy available to the turbulence, nor the nature of the linear instabilities that it drives, but it does control how efficiently the turbulent stress is converted into a mean poloidal flow near the separatrix, the turbulent energy remaining instead in the fluctuations themselves and in the GAM oscillations in \textit{unfav}. This picture is consistent with the larger L--H power threshold observed in \textit{unfav} \cite{ryter1996,ryter2016,plank2023}, the deeper $E_r$ well in \textit{fav} being reached at a lower input power (see Section~\ref{sec:fluxdriven}).

Although the present simulations reproduce the key experimental findings, the physical model considered here neglects (or possibly underestimates) effects that can also be affected by the reversal of the toroidal magnetic field direction. First, the sheath boundary conditions at the divertor targets set the parallel flows and the electrostatic potential in the SOL, and therefore the outer boundary condition of the near-separatrix shear layer where the $E_r$ well develops. Since the target conditions and the in-out divertor asymmetry are sensitive to the direction of the ion $\grad B$ drift \cite{chankin1997,ahomantila2017}, the SOL can thus contribute to the \textit{fav} and \textit{unfav} asymmetry in a way that is absent here, such that the present results isolate the turbulent drive of the asymmetry rather than its full magnitude. Second, neutrals are modelled only through the localized and axisymmetric edge density source of Ref.~\cite{frei2026ppcf}, identical in \textit{fav} and \textit{unfav}. In the experiments, on the contrary, the neutral density is poloidally localized near the X-point and in the divertor, where the neutral recycling is itself set by the drift configuration. A poloidally asymmetric ionization source, together with the charge-exchange friction \cite{versloot2011}, enters directly the poloidal momentum balance derived in Section~\ref{subsec:flowacceleration}, \eqnref{eq:vort4}. Indeed, since $\Pi$ depends on the poloidal asymmetries of $M$ and of $\bm u$ (see \eqnref{eq:pidef}), a poloidally asymmetric ionization source can itself drive a finite radial gradient of $\Pi$, and hence a mean poloidal flow that differs between the two configurations, an effect that cannot be assessed here. We remark, on the other hand, that the identical source used in \textit{fav} and \textit{unfav} is what allows us to attribute the asymmetry obtained here entirely to the turbulent drive. 

The present study can be extended to more realistic conditions, in particular by extending the flux-driven simulations (see Section~\ref{sec:fluxdriven}) to follow the build-up of the $E_r$ well across the L--H transition, by improving the treatment of the sheath and of the neutral dynamics, and by performing dedicated NC calculations in X-point geometry. Finally, we remark that the same nonlinear mechanism is now retrieved in two devices (AUG and TCV) of different size, shape and collisionality. The same mechanism is also invoked in flux-driven fluid simulations \cite{delucca2026}, suggesting that it is robust across devices and models.

\ack{This work has been carried out within the framework of the EUROfusion Consortium, funded by the European Union via the Euratom Research and Training Programme (Grant Agreement No 101052200 — EUROfusion). Views and opinions expressed are however those of the author(s) only and do not necessarily reflect those of the European Union or the European Commission. Neither the European Union nor the European Commission can be held responsible for them.

The simulations presented herein were carried out on the CINECA Pitagora supercomputer under the project TSVVC26.}

\roles{\textbf{B. J. Frei}: Conceptualization, Methodology, Software, Formal analysis, Investigation, Validation, Visualization, Writing -- original draft. \textbf{R. Bilato}: Methodology, Software, Writing -- review \& editing. \textbf{W. Zholobenko}: Methodology, Software, Validation, Writing -- review \& editing. \textbf{O. Grover}: Methodology, Software, Writing -- review \& editing. \textbf{C. Angioni}: Investigation, Writing -- review \& editing. \textbf{M. Bergmann}: Software, Writing -- review \& editing. \textbf{P. Ulbl}: Methodology, Software, Writing -- review \& editing. \textbf{F. Jenko}: Supervision, Resources, Funding acquisition, Writing -- review \& editing. \textbf{S. Rien\"acker}: Investigation, Data curation, Writing -- review \& editing. \textbf{L. Vermare}: Investigation, Data curation, Supervision, Writing -- review \& editing. \textbf{P. Hennequin}: Investigation, Supervision, Writing -- review \& editing. \textbf{B. Labit}: Investigation, Resources, Writing -- review \& editing.}

\data{The data that support the findings of this study are available upon reasonable request from the authors. A link to a Zenodo repository containing the data will be added to ensure reproducibility.}

\appendix

\section{Coordinates and Magnetic Field Geometry} \label{appendix:signs}

We summarize here the magnetic field geometry and associated coordinates and sign conventions used throughout this work.

We use the right-handed cylindrical coordinates $(R, \phi, Z)$, the toroidal angle $\phi$ increasing counter-clockwise when the torus is viewed from above. The poloidal plane is shown with $R$ increasing to the right and $Z$ upwards, such that $\hat{\bm e}_\phi$ points into the page (see, e.g., Figs.~\ref{fig:fig1} and \ref{fig:fig18}). All results are presented in the LSN geometry of the experiments. The flux coordinates $(\psi, \theta, \phi)$ are right-handed, i.e. $\mathcal{J} = \left( \grad \psi \times \grad \theta \cdot \grad \phi \right)^{-1} > 0$, with $\psi$ increasing outwards and $\theta = 0$ at the OMP, $\theta$ increasing clockwise in the poloidal plane, i.e. from the OMP towards the active X-point on the LFS. The associated covariant basis vectors are $\bm e_\psi = \mathcal{J} \grad \theta \times \grad \phi$, $\bm e_\theta = \mathcal{J} \grad \phi \times \grad \psi$ and $\bm e_\phi = \mathcal{J} \grad \psi \times \grad \theta = R^2 \grad \phi$, with $\hat{\bm e}_\theta = \bm e_\theta / \left| \bm e_\theta \right|$, such that a vector field $\bm A$ has the contravariant components $A^i = \bm A \cdot \grad x^i$ and the covariant components $A_i = \bm A \cdot \bm e_i$, with $x^i = \left( \psi, \theta, \phi \right)$. The equilibrium magnetic field is written as $\bm B = I \grad \phi + \sigma_I \grad \phi \times \grad \psi$, with $I = R B_\phi$, $\sigma_B = \mathrm{sign}(B_\phi)$ and $\sigma_I = \pm 1$ the direction of the plasma current $I_p$ with respect to $\hat{\bm e}_\phi$, such that $B_\theta = \sigma_I \left| \grad \psi \right| / R$, the field-line helicity being set by $\sigma_B \sigma_I$. 

\section{Boundary and Initial Conditions, Numerical Resolution and Computational Cost} \label{appendix:A}

We report here the boundary and initial conditions and the details of the numerical resolution of the \texttt{GENE-X} simulations, complementing Section~\ref{subsec:numsetup}, and provide the computational cost of each simulation used in this work. The reference normalization, the numerical resolution and the associated computational cost are collected in Table~\ref{tab:2}.

We use the spectral approach of the \texttt{GENE-X} code \cite{frei2025vspec}, the spectral coefficients $\mathcal{N}_\alpha^{pj}$ being discretized on $N_\varphi$ equidistant poloidal planes, each of them carrying a Cartesian $(R,Z)$ grid of constant spacing $\Delta {RZ}$. The latter corresponds to approximately $1.5$ and $0.7$ times the ion sound Larmor radius $\rho_s = \sqrt{m_i T_e} / (e B_{\mathrm{ref}})$ near the separatrix in AUG and TCV, respectively. The inner boundary of the numerical domain is located at $\rho_{\mathrm{pol}} = 0.9$ in AUG and $\rho_{\mathrm{pol}} = 0.85$ in TCV. The velocity-space resolution and the constant reference temperatures $\tau_\alpha$, which set the scaling of the velocity-space basis, are based on previous investigations in both TCV \cite{frei2025} and AUG \cite{frei2026ppcf}, and yield a total number of degrees of freedom of $6.9 \times 10^8$ in AUG and $3.9 \times 10^8$ in TCV. A standard fourth-order Runge-Kutta scheme is employed with a fixed time step $\Delta t = 3 \times 10^{-4} \, t_{\mathrm{ref}} = 5.1$~ns in AUG and $\Delta t \simeq 4.5 \times 10^{-4} \, t_{\mathrm{ref}} = 4.2$~ns in TCV, being $t_{\mathrm{ref}} = L_{\mathrm{ref}} / c_{\mathrm{ref}}$ the reference time.

At the edge and SOL boundaries, where the distribution functions are Maxwellian (see Section~\ref{subsec:numsetup}), the density and the ion and electron temperatures are set in AUG to $n = 2.27 \times 10^{19}$~m$^{-3}$, $T_i = 302$~eV and $T_e = 523$~eV at the edge boundary and to $n = 0.20 \times 10^{19}$~m$^{-3}$ and $T_i = T_e = 35$~eV at the SOL boundary, while in TCV they are set to $n = 2.48 \times 10^{19}$~m$^{-3}$, $T_i = 210$~eV and $T_e = 190$~eV at the edge boundary and to $n = 0.30 \times 10^{19}$~m$^{-3}$ and $T_i = T_e = 25$~eV at the SOL boundary, the same values being used in \textit{fav} and \textit{unfav}. On the other hand, the electromagnetic fields are set to zero. To avoid artificial flow shear near the edge boundary, Neumann boundary conditions on flux variables are applied. The simulations are initialized with isotropic Maxwellian distribution functions of equal ion and electron densities, as imposed by QN, and different ion and electron temperatures. The corresponding initial profiles are sine profiles \cite{michels2021}, which interpolate the edge and SOL values over $0.92 \leq \rho_{\mathrm{pol}} \leq 0.995$ in AUG and $0.90 \leq \rho_{\mathrm{pol}} \leq 0.98$ in TCV, and are constant outside this interval. 

All four simulations are performed on the \texttt{Pitagora} supercomputer at CINECA, each of them using $8$ nodes equipped with two $128$-core AMD EPYC 9745 processors, distributed over $32$ MPI tasks ($16$ in the toroidal direction and $2$ over the species) with $64$ OpenMP threads per task. The time per timestep is $1.3$~s for TCV and $2.8$~s for AUG, the difference being essentially the ratio of the number of grid points per poloidal plane. Evolving the plasma from the initial condition through the transient relaxation of the profiles and up to the data collection for the statistical analysis performed in this work required between $1.3$ and $2.6$~ms of plasma time per simulation, which corresponds to $1.0 \times 10^3$~h of wall-clock time on $8$ nodes, and to approximately $2.1$~MCPUh (million core-hours) in total. The present computational cost neglects the contributions from the axisymmetric and flux-driven simulations.

\begin{table}[h!]
\centering
\begin{tabular}{lcc}
\hline
 & AUG & TCV \\
\hline
\hline
Discharges (\textit{fav} / \textit{unfav}) & \#36983 / \#37375 & \#86593 / \#86594 \\
$L_{\mathrm{ref}}$ / m & $1.65$ & $0.9054$ \\
$B_{\mathrm{ref}}$ / T & $2.53$ & $1.42$ \\
$T_{\mathrm{ref}}$ / eV & $100$ & $100$ \\
$n_{\mathrm{ref}}$ / $10^{19}$~m$^{-3}$ & $1$ & $1$ \\
$c_{\mathrm{ref}} = \sqrt{T_{\mathrm{ref}} / m_{\mathrm{ref}}}$ / km~s$^{-1}$ & $97.9$ & $97.9$ \\
$t_{\mathrm{ref}} = L_{\mathrm{ref}} / c_{\mathrm{ref}}$ / $\mu$s & $16.9$ & $9.3$ \\
$\rho_{\mathrm{ref}} = m_{\mathrm{ref}} c_{\mathrm{ref}} / (e B_{\mathrm{ref}})$ / mm & $0.40$ & $0.72$ \\
\hline
$\Delta {RZ}$ / mm & $1.43$ & $0.87$ \\
Grid points per poloidal plane & $4.5 \times 10^5$ & $2.5 \times 10^5$ \\
$N_\varphi$ & $32$ & $32$ \\
$N_{v_\parallel}$, $N_\mu$ & $6$, $4$ & $6$, $4$ \\
$\tau_i / T_{\mathrm{ref}}$, $\tau_e / T_{\mathrm{ref}}$ & $1.8$, $2.8$ & $1.4$, $1.4$ \\
\hline
Cores per simulation & $2048$ & $2048$ \\
Wall-clock time per time step / s & $2.8$ & $1.3$ \\
Time steps (\textit{fav} / \textit{unfav}) / $10^5$ & $5.1$ / $5.0$ & $3.6$ / $3.0$ \\
Plasma time (\textit{fav} / \textit{unfav}) / ms & $2.55$ / $2.46$ & $1.51$ / $1.26$ \\
Runtime (\textit{fav} / \textit{unfav}) / days & $16.3$ / $16.1$ & $5.5$ / $4.6$ \\
Cost (\textit{fav} / \textit{unfav}) / MCPUh & $0.80$ / $0.79$ & $0.27$ / $0.23$ \\
\hline
\end{tabular}
\caption{Reference normalization quantities, numerical resolution and computational cost of the \texttt{GENE-X} simulations presented in this work, for the AUG and TCV \textit{fav} and \textit{unfav} configurations, with $m_{\mathrm{ref}}$ the proton mass. The runtime is the wall-clock time on the $8$ nodes used by each simulation, and the cost is expressed in millions of core-hours (MCPUh) on the \texttt{Pitagora} supercomputer at CINECA.}
\label{tab:2}
\end{table}

\section{Derivation of the Flux-Surface-Averaged Poloidal Momentum Balance} \label{appendix:momentum}

We report here the derivation of the FSA poloidal momentum balance, \eqnref{eq:vort4}, used in Section~\ref{subsec:flowacceleration}, with the magnetic field representation, the flux coordinates $x^i = \left( \psi, \theta, \phi \right)$ and the notation of Appendix~\ref{appendix:signs}. The derivation starts from the GK vorticity equation, given in \eqnref{eq:vorticity}. Using the property of the FSA operator, $\fs{\grad \cdot \bm A } =  V^{\prime -1} \partial_\psi \left(V^\prime \fs{  A^\psi}\right)$ for a vector field $\bm A$, we obtain

\begin{align} \label{eq:vort1}
 \fs{\frac{\partial \omega}{\partial t}}   + \frac{1}{V^\prime} \frac{d}{d \psi} \left[  V^\prime \left(   \fs{  u^\psi \grad \cdot \bm \Omega }  +   \fs{J_\perp^{pol, \psi}} +  \fs{J_\perp^{em, \psi}}  + \fs{J_\perp^{dia, \psi}}  \right) \right] = 0,
\end{align}
with $u^\psi = \bm u \cdot \grad \psi$ (and similarly for the other radial currents). We note that the QN condition (\eqnref{eq:qn}) is used in $\grad \cdot  \bm J_{\perp}^{es}$. The first term in \eqnref{eq:vort1} can be written as $\fs{\partial_t \omega}  = \partial_t V^{\prime -1} \partial_\psi \left( V^\prime \fs{\Omega^\psi} \right)$, such that \eqnref{eq:vort1} can be integrated over $\psi$, yielding

\begin{align} \label{eq:vort2}
 \frac{\partial }{\partial t}  \fs{\Omega^\psi} +  \fs{u^\psi \grad \cdot \bm \Omega} + \fs{ J_\perp^{pol, \psi} }  +  \fs{ J_\perp^{em, \psi} } +  \fs{ J_\perp^{dia, \psi} }  = \frac{c_\psi}{V^\prime},
\end{align}
with $c_\psi$ a constant of integration, independent of $\psi$. The contravariant component of $\bm \Omega$ is related to the covariant poloidal component of $\bm u$ through $\Omega^\psi = - c M w u_\theta$. To make progress, the polarization radial current, $J_\perp^{pol, \psi}$, can be developed as follows 

\begin{align} \label{eq:jpol}
 J^{pol, \psi}_\perp  = \bm \Omega \cdot \grad u^\psi -   \bm \Omega \bm u : \grad \grad \psi  - \Omega^\psi \, \grad_\perp \cdot \bm u ,
\end{align}
with $\grad_\perp \cdot \bm u = \div \bm u + \bm u \cdot \bm \kappa$ and $\grad \grad \psi $ the Hessian of $\psi$. \eqnref{eq:jpol} is an exact analytical expression and allows us to combine the contributions of the electrostatic ($u^\psi \grad \cdot \bm \Omega$) and polarization currents, such that 

\begin{align} \label{eq:jesandjpol}
\fs{u^\psi \grad \cdot \bm \Omega} + \fs{J^{pol, \psi}_\perp} = \frac{1}{V^\prime} \frac{\partial}{\partial \psi} \left( V^\prime \fs{\Omega^\psi u^\psi} \right)
+ \mathcal{C},
\end{align}
with $\mathcal{C} = - \fs{ \bm \Omega \bm u : \grad \grad \psi } - \fs{ \Omega^\psi \grad_\perp \cdot \bm u }$. In deriving \eqnref{eq:jesandjpol}, we have used the property that $\fs{\grad \cdot \left( \bm \Omega u^\psi \right)} = V^{\prime -1} \partial_\psi \left( V^\prime \fs{\Omega^\psi u^\psi} \right)$. Here, $\Pi$ is the full-$f$ electrostatic velocity stress defined in \eqnref{eq:pidef}. We note that the first term in $\mathcal{C}$, proportional to $\grad \grad \psi$, represents the contributions of the radial and poloidal variations of the flux surface, while the second term is a finite compressional contribution arising from the compression of the $\bm E \times \bm B$ velocity in toroidal geometry, as $\grad_\perp \cdot \bm u \simeq - \bm u \cdot \grad \ln B$. We however notice that the contributions from the latter terms are smaller than the contribution from the radial gradient of $\Pi$. Indeed, we can estimate that $V^{\prime -1} \partial_\psi V^{\prime} \Pi  \sim \Pi / L_c$, with $L_c$ of the order of the radial correlation length $\ell_c$ (see Fig.~\ref{fig:fig15}), while $\fs{ \bm \Omega \bm u : \grad \grad \psi }$ and $ \fs{ \Omega^\psi  \grad_\perp \cdot \bm u } $ vary on a scale comparable to $R_0$, with $\ell_c / R_0 \ll 1$.

Finally, the contribution from the electromagnetic radial current can be written as 

\begin{align} \label{eq:maxwellgrad}
\fs{J^{em, \psi}_\perp} = \frac{1}{V^\prime} \frac{\partial}{\partial \psi}
\left( \frac{c  V^\prime }{4 \pi} \fs{ w  B_{1 \theta} B_1^\psi } \right).
\end{align}
Combining \eqnref{eq:jesandjpol} and \eqnref{eq:maxwellgrad} and writing $\fs{\Omega^\psi} = - c \fs{M w u_\theta} = - c \mathcal{P}_\theta$, with $\mathcal{P}_\theta = \fs{w M u_\theta}$ the FSA poloidal momentum density, finally yields \eqnref{eq:vort4}.

\section{Gyrokinetic GAM Dispersion Relation} \label{appendix:gam}

In this appendix, we derive the GAM dispersion relation based on the GK formalism considered in \texttt{GENE-X}. Comprehensive reviews of the kinetic theory of GAMs and of their experimental observation in magnetic confinement devices can be found in \cite{qiu2018} and \cite{conway2021}, respectively.

Assuming that $\phi_1$ (and $E_1$) is a flux-surface function, such that $u^\psi = 0$, and developing the radial component of the diamagnetic current $J^{dia, \psi}$ in the absence of turbulence, \eqnref{eq:vort2} reduces to \eqnref{eq:vort5}, whose two pressure terms represent the NC damping of the poloidal flow, so that it describes its relaxation. In particular, the decay of $\phi_1$ through $\Omega^\psi$ couples to the axisymmetric $m=1$ pressure sidebands. Owing to toroidal effects, the non-vanishing divergence of $\bm u$ counteracts this decay and instead generates GAM oscillations \cite{winsor1968}. A dispersion relation for these oscillations can be derived by considering the evolution of the parallel and perpendicular $m=1$ pressure sidebands, which we carry out below within the GK framework considered here.

The evolution equations of the parallel and perpendicular pressures due to $\bm E \times \bm B$ advection are given by

\begin{subequations} \label{eq:pressurebalance}
\begin{align}
 \frac{d }{d t}  \ln \left( \frac{P_{\parallel \alpha}}{B^2} \right) & -  \frac{2 c }{B} \bm b \times \bm \kappa   \cdot \bm E_1  = 0 , \\
 \frac{d }{d t} \ln \left( \frac{P_{\perp \alpha}}{B^2}\right)  & -    \frac{ c}{B} \bm b \times  \grad \ln B \cdot \bm E_1 = 0,
\end{align}
\end{subequations}
 with $d_t = \partial_t + \bm u \cdot \grad$ and where we use $\grad \cdot \bm u \simeq - 2 \bm u \cdot \grad \ln B$ in the low-beta limit. Assuming a small inverse aspect ratio expansion, the evolution of the parallel and perpendicular pressure $m=1$ sidebands can be obtained from \eqnref{eq:pressurebalance}. Together with the vorticity equation, the GAM dispersion relation reads

\begin{align}
\omega_{GAM}^2   & =  \frac{4 B^2 \fs{ P_{\parallel}} }{\fs{M}} \left[ c_\kappa   \left(   c_B +   c_\kappa \right) +  \frac{3}{2} \lambda c_B^2  \right]
\label{eq:omegagamgen}
\end{align}
where $\lambda = \fs{ P_{\perp}} / \fs{ P_{\parallel}} $ measures the total pressure anisotropy, with $P_{\parallel} = \sum_\alpha P_{\parallel \alpha}$ and $P_{\perp} = \sum_\alpha P_{\perp \alpha}$ the total parallel and perpendicular pressures. Here, we introduce $c_\kappa = \fs{  \bm b \times \bm \kappa / B \cdot \grad \psi \sin \theta }$ and $c_B = \fs{  \bm b \times \grad B / B^2 \cdot \grad \psi   \sin \theta  }$, which, in the low inverse aspect ratio limit, simplify to $c_\kappa = c_B = -  1 / (2R B)$. In this case, \eqnref{eq:omegagamgen} reduces to

\begin{align} \label{eq:omegagam}
\omega_{GAM}^2 = \frac{\fs{ P_{\parallel}}}{\fs{M}}   \frac{1}{2 R^2}\left( 4 +  3  \lambda  \right).
\end{align}
The GAM dispersion relation given in \eqnref{eq:omegagam} retains the pressure anisotropy, but neglects the radial variation of the profiles as well as finite-orbit-width and wave-particle resonance effects, which can modify $\omega_{GAM}$ and are responsible for its collisionless damping \cite{qiu2018}. We use \eqnref{eq:omegagam} to diagnose the GAM frequency in our simulations in Section~\ref{subsec:gam}.

\section{Local Plasma Parameters and Linear Parameter Scans} \label{appendix:localparameters}

In this appendix, we report the local plasma parameters computed from the \texttt{GENE-X} simulations and the dependence of the linear growth rates on the electron collisionality and plasma beta.

\begin{figure}
\centering
\includegraphics[scale=0.52]{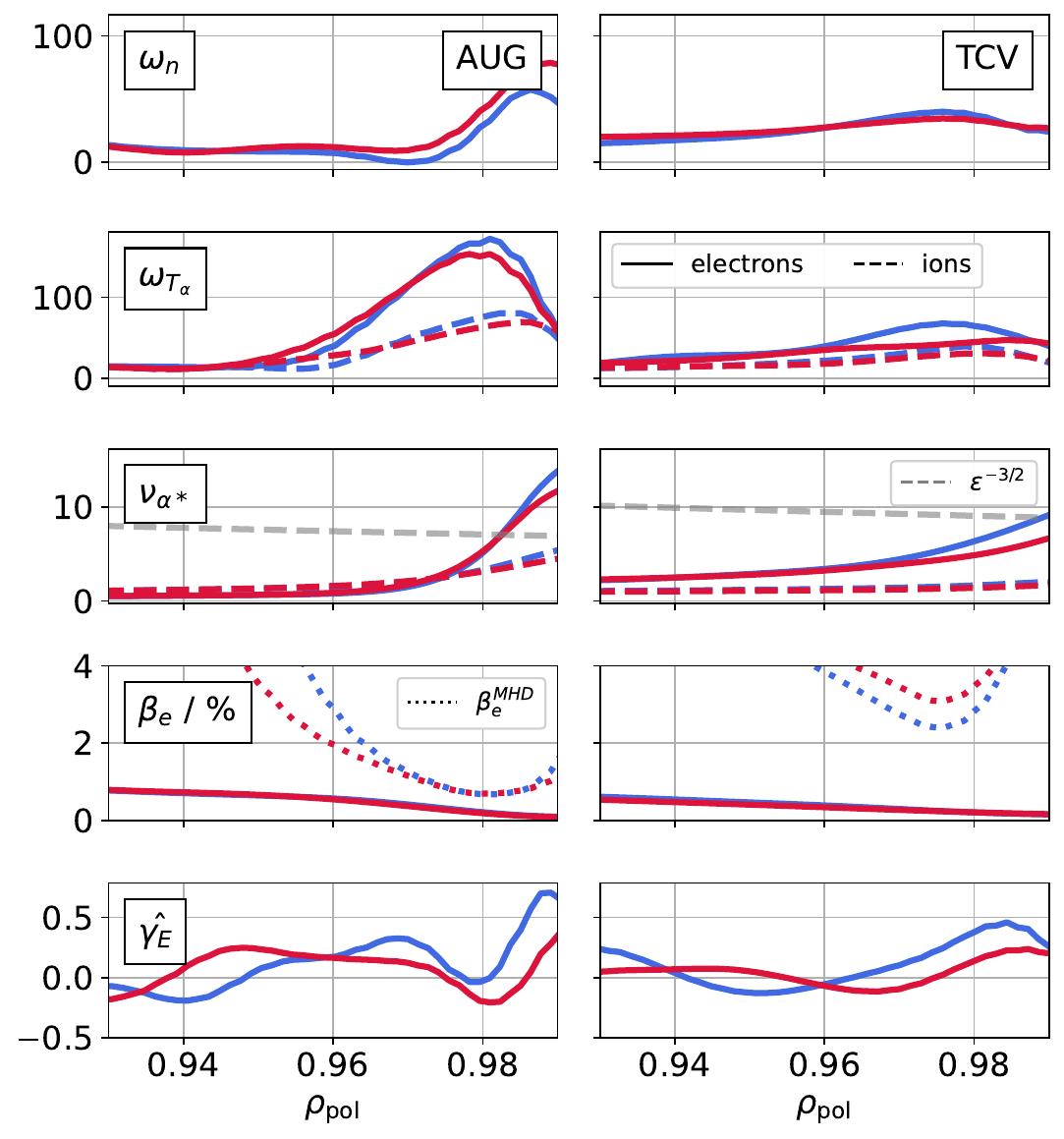}
\caption{\label{fig:fig10} Local parameters of AUG (left column) and TCV (right column). From top to bottom: normalized density gradient ($\omega_n = L_{\mathrm{ref}} / L_n$ with $L_f = - d_V (V^\prime f ) / f$), normalized temperature gradient  ($\omega_{T_\alpha} = L_{\mathrm{ref}} / L_{T_\alpha}$), normalized collisionality ($\nu_\alpha^* = q  R_0 \nu_\alpha / (\epsilon^{3/2} v_{th \alpha})$ with $\nu_{\alpha}$ the collision frequency) with the Pfirsch-Schl\"uter and banana transition ($\sim \epsilon^{-3/2}$), electron plasma beta ($\beta_e$) with the MHD critical value ($\beta_e^{\mathrm{MHD}}$), and normalized $\bm E \times \bm B$ shear ($\hat \gamma_E = - r / q d_r \left( E_r / B\right) L_{\mathrm{ref}} / c_{\mathrm{ref}}$). Note here that $\epsilon$ is the local inverse aspect ratio and $q$ the FSA safety factor.}
\end{figure}

The local plasma parameters are shown and defined in Fig.~\ref{fig:fig10}: the normalized density and temperature gradients ($\omega_n$ and $\omega_{T_\alpha}$), the normalized collisionality ($\nu_\alpha^*$), the electron plasma beta ($\beta_e$), and the normalized $\bm E \times \bm B$ shear ($\hat\gamma_E$). We first observe small variations in the normalized gradients $\omega_n$ and $\omega_{T_\alpha}$ between \textit{fav} and \textit{unfav}, although the TCV \textit{fav} simulation features a larger $\omega_{T_e}$ near $\rho_{\mathrm{pol}} = 0.98$, exceeding the \textit{unfav} value by about $40\%$. The normalized collisionalities show that the electrons enter the Pfirsch-Schl\"uter regime in AUG for $\rho_{\mathrm{pol}} \gtrsim 0.99$, where $\nu_e^*$ exceeds $\epsilon^{-3/2}$, while the ions remain in the plateau regime, whereas both species stay in the banana and plateau regimes in TCV, suggesting that the trapped-particle contribution is larger in TCV. The $\beta_e$ profiles indicate that electromagnetic effects are small in both cases and remain below the critical magnetohydrodynamic (MHD) value $\beta_e^{MHD}$ for kinetic ballooning mode (KBM) onset. We note that $\beta_e$ is just below $\beta_e^{MHD}$ in AUG around $\rho_{\mathrm{pol}} \simeq 0.98$. Contrary to the other local parameters, the bottom panel reveals a clear difference in the normalized $\bm E \times \bm B$ shear between \textit{fav} and \textit{unfav}: because of the deeper $E_r$ well in \textit{fav}, $\hat\gamma_E$ (defined in Fig.~\ref{fig:fig10}) around the well is about twice as large in \textit{fav} as in \textit{unfav}, peaking at $\hat\gamma_E \simeq 0.9$ against $0.4$ in AUG and $0.6$ against $0.3$ in TCV. Overall, the close similarity of the local parameters shows that possible differences in edge turbulence between \textit{fav} and \textit{unfav} cannot be explained by differences in the linear theory, as demonstrated in Section~\ref{subsec:linear}.

\begin{figure}
\centering
\includegraphics[scale=0.55]{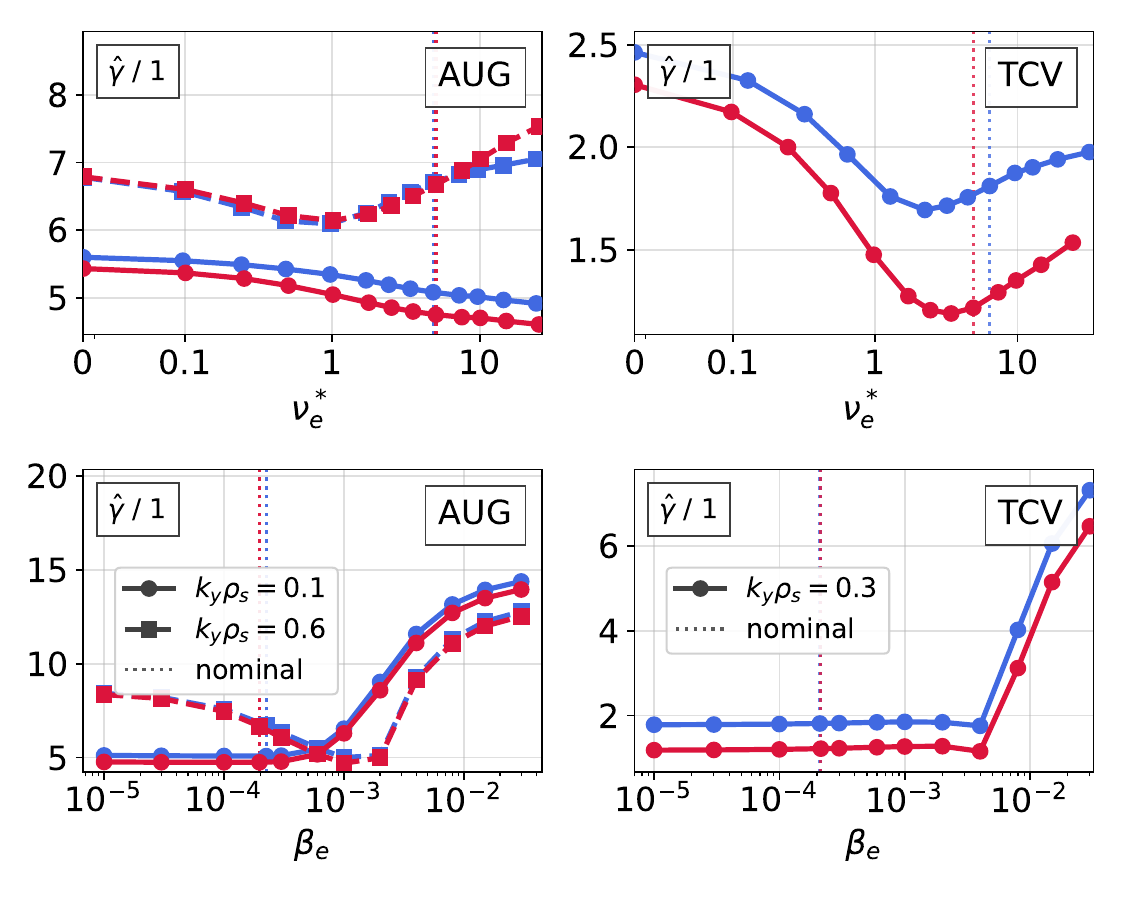}
\caption{\label{fig:fig14} Normalized growth rate $\hat \gamma$ as a function of the electron collisionality $\nu_e^*$ (top row) and of the electron plasma beta $\beta_e$ (bottom row), for AUG (left column) and TCV (right column), both at $\rho_{\mathrm{pol}} = 0.98$, in \textit{fav} (blue) and \textit{unfav} (red). The AUG scans are performed at the two unstable peaks of Fig.~\ref{fig:fig13}, $k_y \rho_s = 0.1$ (circles, solid lines) and $k_y \rho_s = 0.6$ (squares, dashed lines), and the TCV scans at $k_y \rho_s = 0.3$ (circles). The nominal values are marked by the vertical dotted lines of the corresponding colour.}
\end{figure}

We now report the collisionality and $\beta_e$ scans of the linear growth rates of Section~\ref{subsec:linear}, performed at $\rho_{\mathrm{pol}} = 0.98$ about the nominal local parameters of Fig.~\ref{fig:fig10}. Varying the collisionality about its nominal value (see the top row of Fig.~\ref{fig:fig14}) reveals a non-monotonic dependence. In AUG, the long-wavelength peak is stabilized by collisional detrapping, while the short-$k_y$ mode is destabilized through the non-adiabatic electron response. In TCV, the nominal collisionality lies on the DTEM branch: the growth rate is first damped by collisions with respect to the collisionless limit, reaching a minimum near $\nu_e^* \simeq 2$--$3$, and is then destabilized, the nominal values ($\nu_e^* \simeq 6.4$ in \textit{fav} and $4.9$ in \textit{unfav}) lying just above this minimum on the rising branch. Consistently, Fig.~\ref{fig:fig10} indicates that the TCV electrons are well inside the plateau regime at $\rho_{\mathrm{pol}} = 0.98$, with $\nu_e^*$ still below $\epsilon^{-3/2}$, i.e. collisional enough to sit on the dissipative branch but far from the collisionless TEM limit.

Reducing $\beta_e$ below its nominal value (see the bottom row of Fig.~\ref{fig:fig14}) has only a weak destabilizing effect in all cases, confirming that the modes are essentially electrostatic and are stabilized by finite $\beta_e$. The KBM is eventually destabilized when $\beta_e$ is raised to $\gtrsim 2 \times 10^{-3}$ in AUG and $\gtrsim 8 \times 10^{-3}$ in TCV, i.e. one to two orders of magnitude above the nominal values, with the real frequency shifting toward the ion diamagnetic direction. The critical $\beta_e^{MHD}$ is also shown in Fig.~\ref{fig:fig10}, where AUG is found to lie closer to the linear threshold than TCV.

\bibliography{library}

\end{document}